\documentclass[twocolumn]{aastex62}

\usepackage{savesym}
\savesymbol{tablenum}
\usepackage{siunitx}
\restoresymbol{SIX}{tablenum}

\usepackage{graphicx,epsf}
\usepackage{subfigure}
\usepackage{graphicx}	
\usepackage{amsmath}	
\usepackage{amssymb}	
\usepackage{units}
\usepackage{newtxtext,newtxmath}
\usepackage{siunitx}
\usepackage{footnote}

\newcommand{\Msun}{{\rm M}_\odot}
\newcommand{\Rsun}{{\rm R}_\odot}
\newcommand{\Lsun}{{\rm L}_\odot}
\newcommand{\kms}{\textrm{km}\,\textrm{s}^{-1}}

\def\arcsec{\hbox{$^{\prime\prime}$}}

\def\ca{{CaST}}
\def\cas{{CaSTs}}

\shorttitle{Ca-rich Transient SN~2019ehk}
\shortauthors{Jacobson-Gal\'{a}n et al.}

\begin{document}

\title{SN~2019ehk: A Double-Peaked Ca-rich Transient with Luminous X-ray Emission and Shock-Ionized Spectral Features}

\correspondingauthor{Wynn Jacobson-Gal\'{a}n (he, him, his)}
\email{wynn@u.northwestern.edu}

\author[0000-0002-3934-2644]{Wynn V. Jacobson-Gal\'{a}n}
\affil{Department of Physics and Astronomy, Northwestern University, 2145 Sheridan Road, Evanston, IL 60208, USA}
\affil{Center for Interdisciplinary Exploration and Research in Astrophysics (CIERA), 1800 Sherman Ave, Evanston, IL 60201, USA}
\affil{Department of Astronomy and Astrophysics, University of California, Santa Cruz, CA 95064,
USA}

\author[0000-0003-4768-7586]{Raffaella Margutti}
\affil{Department of Physics and Astronomy, Northwestern University, 2145 Sheridan Road, Evanston, IL 60208, USA}
\affil{Center for Interdisciplinary Exploration and Research in Astrophysics (CIERA), 1800 Sherman Ave, Evanston, IL 60201, USA}

\author{Charles~D.~Kilpatrick}
\affil{Department of Astronomy and Astrophysics, University of California, Santa Cruz, CA 95064,
USA}

\author[0000-0002-1125-9187]{Daichi Hiramatsu}
\affil{Department of Physics, University of California, Santa Barbara, CA 93106-9530, USA }
\affil{Las Cumbres Observatory, 6740 Cortona Dr, Suite 102, Goleta, CA 93117-5575, USA
}

\author{Hagai Perets}
\affil{Technion - Israel Institute of Technology, Physics department, Haifa Israel 3200002
}

\author{David Khatami}
\affil{Department of Astronomy and Theoretical Astrophysics Center, University of California, Berkeley, CA 94720, USA}

\author{Ryan J. Foley}
\affil{Department of Astronomy and Astrophysics, University of California, Santa Cruz, CA 95064,
USA}

\author[0000-0002-7868-1622]{John Raymond}
\affil{Center for Astrophysics \textbar{} Harvard \& Smithsonian, 60 Garden Street, Cambridge, MA 02138, USA}

\author{Sung-Chul Yoon}
\affil{Department of Physics and Astronomy, Seoul National University, 08826, Seoul, South Korea
}
\affil{Center for Theoretical Physics (CTP), Seoul National University, 08826, Seoul, South Korea}

\author[0000-0002-4674-0704]{Alexey Bobrick}
\affil{Lund University, Department of Astronomy and Theoretical physics, Box 43, SE 221-00 Lund, Sweden
}

\author{Yossef Zenati}
\affil{Technion - Israel Institute of Technology, Physics department, Haifa Israel 3200002}

\author{Llu\'is Galbany}
\affil{Departamento de F\'isica Te\'orica y del Cosmos, Universidad de Granada, E-18071 Granada, Spain}

\author{Jennifer Andrews}
\affil{Department of Astronomy/Steward Observatory, 933 North Cherry Avenue, Rm. N204, Tucson, AZ 85721-0065, USA}

\author{Peter J. Brown}
\affil{George P. and Cynthia Woods Mitchell Institute for Fundamental Physics and Astronomy, Department of Physics and Astronomy, Texas A\&M University, College Station, TX, 77843, USA}

\author{R\'{e}gis Cartier}
\affil{Cerro Tololo Inter-American Observatory, National Optical Astronomy Observatory, Casilla 603, La Serena, Chile}

\author[0000-0001-5126-6237]{Deanne L. Coppejans}
\affil{Department of Physics and Astronomy, Northwestern University, 2145 Sheridan Road, Evanston, IL 60208, USA}
\affil{Center for Interdisciplinary Exploration and Research in Astrophysics (CIERA), 1800 Sherman Ave, Evanston, IL 60201, USA}

\author{Georgios~Dimitriadis}
\affil{Department of Astronomy and Astrophysics, University of California, Santa Cruz, CA 95064,
USA}

\author{Matthew Dobson}
\affil{Astrophysics Research Centre, School of Mathematics and Physics, Queen's University Belfast, BT7 1NN, UK}

\author{Aprajita Hajela}
\affil{Department of Physics and Astronomy, Northwestern University, 2145 Sheridan Road, Evanston, IL 60208, USA}
\affil{Center for Interdisciplinary Exploration and Research in Astrophysics (CIERA), 1800 Sherman Ave, Evanston, IL 60201, USA}

\author[0000-0003-4253-656X]{D. Andrew Howell}
\affil{Department of Physics, University of California, Santa Barbara, CA 93106-9530, USA }
\affil{Las Cumbres Observatory, 6740 Cortona Dr, Suite 102, Goleta, CA 93117-5575, USA
}

\author{Hanindyo Kuncarayakti}
\affil{Department of Physics and Astronomy, University of Turku, FI-20014 Turku, Finland}
\affil{Finnish Centre for Astronomy with ESO (FINCA), FI-20014 University of Turku, Finland.}

\author[0000-0002-0763-3885]{Danny Milisavljevic}
\affil{Department of Physics and Astronomy, Purdue University, 525 Northwestern Avenue, West Lafayette, IN 47907, USA}

\author{Mohammed Rahman}
\affil{The Thacher School, 5025 Thacher Rd, Ojai, CA 93023}

\author[0000-0002-7559-315X]{C\'{e}sar~Rojas-Bravo}
\affil{Department of Astronomy and Astrophysics, University of California, Santa Cruz, CA 95064,
USA}

\author[0000-0003-4102-380X]{David J. Sand}
\affil{Department of Astronomy/Steward Observatory, 933 North Cherry Avenue, Rm. N204, Tucson, AZ 85721-0065, USA}

\author{Joel Shepherd}
\affil{Seattle Astronomical Society}

\author{Stephen J. Smartt}
\affil{Astrophysics Research Centre, School of Mathematics and Physics, Queen's University Belfast, BT7 1NN, UK}

\author{Holland Stacey}
\affil{The Thacher School, 5025 Thacher Rd, Ojai, CA 93023}

\author{Michael Stroh}
\affil{Department of Physics and Astronomy, Northwestern University, 2145 Sheridan Road, Evanston, IL 60208, USA}
\affil{Center for Interdisciplinary Exploration and Research in Astrophysics (CIERA), 1800 Sherman Ave, Evanston, IL 60201, USA}

\author[0000-0002-9486-818X]{Jonathan J. Swift}
\affil{The Thacher School, 5025 Thacher Rd, Ojai, CA 93023}

\author{Giacomo Terreran}
\affil{Department of Physics and Astronomy, Northwestern University, 2145 Sheridan Road, Evanston, IL 60208, USA}
\affil{Center for Interdisciplinary Exploration and Research in Astrophysics (CIERA), 1800 Sherman Ave, Evanston, IL 60201, USA}

\author{Jozsef Vinko}
\affil{CSFK Konkoly Observatory, Konkoly-Thege ut 15-17, Budapest, 1121, Hungary}
\affil{Department of Optics and Quantum Electronics, University of Szeged, Domter 9, Szeged, 6720, Hungary} 
\affil{ELTE E\"{o}tv\"{o}s Lor\'{a}nd University, Institute of Physics, P\'{a}zm\'{a}ny P. s\'{e}t\'{a}ny 1/A, Budapest, 1117 Hungary}

\author{Xiaofeng Wang}
\affil{Physics Department, Tsinghua University, Beijing, 100084}
\affil{Beijing Planetarium, Beijing Academy of Science and Technology, Beijing, 100044}

\author{Joseph P. Anderson}
\affil{European Southern Observatory, Alonso de C\'ordova 3107, Casilla 19 Santiago, Chile}

\author{Edward A. Baron}
\affil{Homer L. Dodge Department of Physics and Astronomy, University of Oklahoma, 440 W. Brooks, Rm. 100, Norman, OK 73019-2061, USA}

\author{Edo Berger}
\affil{Center for Astrophysics \textbar{} Harvard \& Smithsonian, 60 Garden Street, Cambridge, MA  02138, USA}

\author[0000-0003-0526-2248]{Peter K. Blanchard}
\affil{Department of Physics and Astronomy, Northwestern University, 2145 Sheridan Road, Evanston, IL 60208, USA}
\affil{Center for Interdisciplinary Exploration and Research in Astrophysics (CIERA), 1800 Sherman Ave, Evanston, IL 60201, USA}

\author{Jamison Burke}
\affil{Department of Physics, University of California, Santa Barbara, CA 93106-9530, USA }
\affil{Las Cumbres Observatory, 6740 Cortona Dr, Suite 102, Goleta, CA 93117-5575, USA
}

\author{David~A.~Coulter}
\affil{Department of Astronomy and Astrophysics, University of California, Santa Cruz, CA 95064,
USA}

\author[0000-0003-4587-2366]{Lindsay DeMarchi}
\affil{Department of Physics and Astronomy, Northwestern University, 2145 Sheridan Road, Evanston, IL 60208, USA}
\affil{Center for Interdisciplinary Exploration and Research in Astrophysics (CIERA), 1800 Sherman Ave, Evanston, IL 60201, USA}

\author{James M. DerKacy}
\affil{Homer L. Dodge Department of Physics and Astronomy, University of Oklahoma, 440 W. Brooks, Rm. 100, Norman, OK 73019-2061, USA}

\author{Christoffer Fremling}
\affil{Division of Physics, Mathematics and Astronomy, California Institute of Technology, Pasadena, CA 91125, USA}

\author[0000-0001-6395-6702]{Sebastian Gomez}
\affil{Center for Astrophysics \textbar{} Harvard \& Smithsonian, 60 Garden Street, Cambridge, MA 02138, USA}

\author{Mariusz Gromadzki}
\affil{Astronomical Observatory, University of Warsaw, Al. Ujazdowskie 4, 00-478 Warszawa, Poland}

\author[0000-0002-0832-2974]{Griffin Hosseinzadeh}
\affil{Center for Astrophysics \textbar{} Harvard \& Smithsonian, 60 Garden Street, Cambridge, MA 02138, USA}

\author{Daniel Kasen}
\affil{Department of Astronomy and Theoretical Astrophysics Center, University of California, Berkeley, CA 94720, USA}
\affil{Nuclear Science Division, Lawrence Berkeley National Laboratory, 1 Cyclotron Road, Berkeley, CA 94720}

\author{Levente Kriskovics}
\affil{CSFK Konkoly Observatory, Konkoly-Thege ut 15-17, Budapest, 1121, Hungary}
\affil{ELTE E\"{o}tv\"{o}s Lor\'{a}nd University, Institute of Physics, P\'{a}zm\'{a}ny P. s\'{e}t\'{a}ny 1/A, Budapest, 1117 Hungary} 

\author[0000-0001-5807-7893]{Curtis McCully}
\affil{Department of Physics, University of California, Santa Barbara, CA 93106-9530, USA }
\affil{Las Cumbres Observatory, 6740 Cortona Dr, Suite 102, Goleta, CA 93117-5575, USA
}

\author{Tom\'as E. M\"uller-Bravo}
\affil{School of Physics and Astronomy, University of Southampton, Southampton, Hampshire, SO17 1BJ, UK
}

\author{Matt Nicholl}
\affil{Birmingham Institute for Gravitational Wave Astronomy and School of Physics and Astronomy, University of Birmingham, Birmingham B15 2TT, UK}
\affil{Institute for Astronomy, University of Edinburgh, Royal Observatory, Blackford Hill, EH9 3HJ, UK}

\author{Andr\'as Ordasi}
\affil{CSFK Konkoly Observatory, Konkoly-Thege ut 15-17, Budapest, 1121, Hungary}

\author[0000-0002-7472-1279]{Craig Pellegrino}
\affil{Department of Physics, University of California, Santa Barbara, CA 93106-9530, USA }
\affil{Las Cumbres Observatory, 6740 Cortona Dr, Suite 102, Goleta, CA 93117-5575, USA}

\author{Anthony~L.~Piro}
\affil{The Observatories of the Carnegie Institution for Science, 813 Santa Barbara Street, Pasadena, CA 91101, USA}

\author{Andr\'as P\'al}
\affil{CSFK Konkoly Observatory, Konkoly-Thege ut 15-17, Budapest, 1121, Hungary}
\affil{ELTE E\"{o}tv\"{o}s Lor\'{a}nd University, Institute of Physics, P\'{a}zm\'{a}ny P. s\'{e}t\'{a}ny 1/A, Budapest, 1117 Hungary}
\affil{ELTE E\"{o}tv\"{o}s Lor\'{a}nd University, Department of Astronomy, P\'{a}zm\'{a}ny P. s\'{e}t\'{a}ny 1/A, Budapest, 1117 Hungary}

\author{Juanjuan Ren}
\affil{National Astronomical Observatory of China, Chinese Academy of Sciences, Beijing, 100012, China}

\author{Armin Rest}
\affil{Space Telescope Science Institute, Baltimore, MD 21218}
\affil{Department of Physics and Astronomy, The Johns Hopkins University, Baltimore, MD 21218}

\author{R. Michael Rich}
\affil{Department of Physics and Astronomy, University of California at Los Angeles, PAB 430 Portola Plaza, Los Angeles, CA, 90095-1547 United States}

\author{Hanna Sai}
\affil{Physics Department, Tsinghua University, Beijing, 100084}

\author{Kriszti\'an S\'arneczky}
\affil{CSFK Konkoly Observatory, Konkoly-Thege ut 15-17, Budapest, 1121, Hungary}

\author[0000-0002-9632-6106]{Ken J. Shen}
\affil{Department of Astronomy and Theoretical Astrophysics Center, University of California, Berkeley, CA 94720, USA}

\author{Philip Short}
\affil{Institute for Astronomy, University of Edinburgh, Royal Observatory, Blackford Hill, EH9 3HJ, UK}

\author{Matthew R. Siebert}
\affil{Department of Astronomy and Astrophysics, University of California, Santa Cruz, CA 95064,
USA}

\author{Candice Stauffer}
\affil{Department of Physics and Astronomy, Northwestern University, 2145 Sheridan Road, Evanston, IL 60208, USA}
\affil{Center for Interdisciplinary Exploration and Research in Astrophysics (CIERA), 1800 Sherman Ave, Evanston, IL 60201, USA}

\author{R\'obert Szak\'{a}ts}
\affil{CSFK Konkoly Observatory, Konkoly-Thege ut 15-17, Budapest, 1121, Hungary}

\author{Xinhan Zhang}
\affil{Physics Department, Tsinghua University, Beijing, 100084}

\author{Jujia Zhang}
\affil{Yunnan Astronomical Observatory of China, Chinese Academy of Sciences, Kunming, 650011, China}

\author{Kaicheng Zhang}
\affil{Physics Department, Tsinghua University, Beijing, 100084}

\begin{abstract}
We present panchromatic observations and modeling of the Calcium-rich supernova (SN) 2019ehk in the star-forming galaxy M100 (d$\approx$16.2 Mpc) starting 10 hours after explosion and continuing for $\sim300$ days. SN\,2019ehk shows a double-peaked optical light curve peaking at $t = 3$ and $15$ days. The first peak is coincident with luminous, rapidly decaying \textit{Swift}-XRT discovered X-ray emission ($L_x\approx10^{41}~\rm{erg~s^{-1}}$ at $3$ days; $L_x \propto t^{-3}$), and a Shane/Kast spectral detection of narrow H$\alpha$ and \ion{He}{ii} emission lines ($v \approx500\,\kms$) originating from pre-existent circumstellar material (CSM). We attribute this phenomenology to radiation from shock interaction with extended, dense material surrounding the progenitor star at $r<10^{15}~\rm{cm}$ and the resulting cooling emission. We calculate a total CSM mass of $\sim$ $7\times10^{-3}$~$\Msun$ ($M_{\rm He}/M_{\rm H}$~$\approx$~6) with particle density $n\approx10^{9}\,\rm{cm^{-3}}$. Radio observations indicate a significantly lower density $n < 10^{4}\,\rm{cm^{-3}}$ at larger radii $r>(0.1-1)\times10^{17}\,\rm{cm}$. The photometric and spectroscopic properties during the second light curve peak are consistent with those of Ca-rich transients (rise-time of $t_r =13.4\pm0.210$ days and a peak $B$-band magnitude of $M_B =-15.1\pm0.200$ mag). We find that SN~2019ehk synthesized $(3.1\pm0.11)\times10^{-2} ~ \Msun$ of ${}^{56}\textrm{Ni}$ and ejected $M_{\rm ej} = (0.72\pm 0.040)~\Msun$ total with a kinetic energy $E_{\rm k}=(1.8\pm0.10)\times10^{50}~\rm{erg}$. Finally, deep \emph{HST} pre-explosion imaging at the SN site constrains the parameter space of viable stellar progenitors to massive stars in the lowest mass bin ($\sim10 \ \Msun$) in binaries that lost most of their He envelope \emph{or} white dwarfs (WDs). The explosion and environment properties of SN~2019ehk further restrict the potential WD progenitor systems to low-mass hybrid HeCO WD + CO WD binaries.

\end{abstract}


\section{Introduction} \label{sec:intro}

Calcium-rich (Ca-rich) transients are a new class of faint, rapidly evolving astronomical transients that has been identified in the past two decades \citep{filippenko03, perets10, kasliwal12}. Observationally, Ca-rich transients are characterized by peak magnitudes of $-14$ to $-16.5$, rise-times $t_r<15$ days, and strong calcium features in photospheric and nebular phase spectra \citep{taubenberger17}. The majority of these objects exhibit low ejecta and ${}^{56}\textrm{Ni}$ masses of $\lesssim 0.5~\Msun$ and $\lesssim 0.1~\Msun$, respectively. Ca-rich transients do not necessarily have larger than average Ca mass but rather are ``rich'' in [\ion{Ca}{ii}] emission during the nebular phase. Consequently, Ca-rich spectra typically exhibit minimal [\ion{O}{i}] $\lambda\lambda6300, 6364$ emission and contain an integrated [\ion{Ca}{ii}]/[\ion{O}{i}] flux ratio greater than $\sim$2. 

The ``Ca-rich'' naming convention was reinforced by the Ca and O abundances of 0.135 and 0.037~$\Msun$ derived from the nebular spectrum of prototypical event, SN~2005E \citep{perets10}. However, subsequent modeling of Ca-rich transient nebular spectra using optical and near-infrared data highlight uncertainty in this estimate and suggest that chemical abundances may vary widely between events  \citep{milisavljevic17}. A potential explanation for the prominence of \ion{Ca}{ii} emission relative to other species is that the distribution of $^{56}$Ni throughout the SN ejecta over-excites calcium ions \citep{polin2019b}. Because of this, we choose to adopt the label suggested by \cite{shen19} and refer to these objects as ``Calcium-Strong Transients'' (\cas) from this point forward.

Early-time spectra of ``gold sample'' \cas \ \citep{shen19} resemble that of core-collapse type Ib SNe (SNe Ib) with detectable photospheric \ion{He}{i} and no observed H$\alpha$ emission. However, the large fraction of objects found in old stellar environments on the outskirts of early-type galaxies disfavors a massive star origin for most \cas \ \citep{perets11, kasliwal12}. Parenthetically, \cas \ tend to occur in group or cluster environments of early-type elliptical galaxies with no evidence for local star formation or globular clusters \citep{perets10, lyman14, foley15,lunnan17}. \cite{Perets2014} finds the location of CaSTs to be mostly consistent with older stellar populations, with many of these objects having large separations from early-type host galaxies known to have large stellar halos. \cite{shen19} also find that the radial distribution of \cas \ is consistent with old (>5 Gyr), metal-poor stellar populations. However, a non-negligible fraction of \cas \ were found in disk-shaped galaxies \citep{perets10,Perets2014, de20}.

A variety of progenitor scenarios have been proposed to explain the observed properties of \cas \ and their environments. \cite{shen19} outline the three scenarios that are most consistent with current observations. First, ultra-stripped-envelope SNe could reproduce the low ejecta and ${}^{56}\textrm{Ni}$ masses and rapidly evolving light curves, but cannot reconcile the lack of star formation at most \ca \ explosion sites. Similar discrepancies disfavor the second scenario wherein a WD is tidally disrupted by a neutron star (NS) or an intermediate-mass black hole (IMBH) \citep{rosswog08, metzger12, macleod14, sell15, margalit16,Bobrick2017,Zenati2019, zenati2019b}. While such a system is likely to occur in dense stellar systems like globular or super star clusters, there is currently no evidence for \cas \ occurring in such environments. However, SN kicks may push such systems outside of their typical cluster environments and still allow NS/BH + WD systems to reside at \ca \ explosion sites. Furthermore, NS+WD binaries occur at only 0.3-3\% of the type Ia SN (SN Ia) rate for similar age populations, which is much less than the \ca \ rate of 10-94\% with respect to SNe~Ia \citep{perets10,Frohmaier18,toonen18, de20}. Lastly, the detonation of a helium shell on the surface of a WD remains a viable option for \cas \ since its application in the study of SN~2005E \citep{perets10, waldman11, woosley11}. In this case, the detonation of the He-shell could lead a partial second detonation of the C/O core for low mass WDs and still match the \ca \ observables. A complete second detonation of a near-Chandrasekhar mass WD would otherwise result in a SN~Ia \citep{nomoto82b, woosley86, fink10, waldman11, polin19, townsley19, perets19, zenati2019b}. Given the proper conditions needed for helium shell detonations, this explosion scenario can successfully produce heightened Ca abundances through which the ejecta can effectively cool and subsequently produce the prominent \ion{Ca}{ii} emission lines seen in \cas \ \citep{holcomb13, polin2019b}. 

Despite attempts to find a singular progenitor scenario, some diversity is observed amongst SNe that display unusually large [\ion{Ca}{ii}]/[\ion{O}{i}] flux ratios. This then suggests that \cas \ might be a heterogeneous class of objects with different physical origins. For example, the large inferred ejecta mass ($\sim$ $2-4$~$\Msun$) for iPTF15eqv is difficult to reconcile with other homogeneous properties of \cas \ \citep{milisavljevic17}. However, iPTF15eqv was only observed after optical peak, and its light curve is consistent with being more luminous than any of the known  CaSTs. Together with its prominent H$\alpha$ emission during nebular phase (also shown by the CaST PTF09dav, \citealt{sullivan11}), these findings might imply that iPTF15eqv is unrelated to the general sample of CaSTs, thus demonstrating the existence of different explosion channels responsible for Ca-rich emission at late times in SNe. An additional outlier amongst \cas \ is ``Calcium-strong'' SN~2016hnk \citep{galbany19, wjg19}, which fits observationally within the class based on its peak luminosity, rise-time and [\ion{Ca}{ii}]/[\ion{O}{i}] ratio, yet has a slowly decaying light curve as well as similarities to ``SN~1991bg-like'' SNe. This object may represent the extremes of the ``Ca-rich'' classification while still remaining consistent with the helium shell detonation scenario that is now considered to be a feasible explosion mechanism for \cas \ \citep{de20}.

While the [\ion{Ca}{ii}]/[\ion{O}{i}] flux ratio is the common metric for classifying new \cas, it is now clear that there is a substantial spread in this ratio amongst events: some objects such as SN~2003dg, PTF09dav and PTF10iuv have negligible [\ion{O}{i}] emission, while SN~2012hn has an oxygen composition comparable to type IIb/IIP SNe \citep[e.g.,][]{valenti14}. Furthermore, type Iax SNe (SNe~Iax) are also thermonuclear explosions that are rich in [\ion{Ca}{ii}] emission at nebular times, yet do not belong to the typical \ca \ class \citep{foley09, foley16}.

Similar to other transients in the ``thermonuclear zoo'', \cas \ have never been detected in X-ray observations \citep{sell15, milisavljevic17, de18, sell18, prentice19}. The earliest X-ray follow-up of a \ca \ was at $t \approx 26$ days after explosion by \cite{sell18} who was testing a progenitor scenario involving the tidal disruption of a WD by an IMBH for SN~2016hnk. However, the fact that multiple other studies have also found X-ray non-detections in \cas \ suggests that either (i) these objects resulted from progenitor environments where X-ray production is not possible or (ii) X-ray emission occurs in \cas \ at yet un-observed phases i.e., extremely early-times, $\sim$ $0 - 25$~days after explosion. Furthermore, no \ca \ has been detected in radio observations \citep{chomiuk16}. Progenitor mass-loss rates of $\leq 7 \times 10^{-5}$ and $\leq 2 \times 10^{-6}~\Msun\rm{yr^{-1}}$ were derived from radio non-detections in iPTF15eqv and iPTF16hgs, respectively \citep{milisavljevic17, de18}. 

In this paper we present, analyze and model multi-wavelength observations (X-ray to radio) of a new \ca, SN~2019ehk, discovered by astronomer Joel Shepherd on 2019 April 29 (MJD 58602.24) using a TEC 140mm APO refracting telescope and Atik 460 EX Mono camera with an SDSS $g$ filter. SN~2019ehk has a discovery apparent magnitude of 17.1~mag and is located at $\alpha = 12^{\textrm{h}}22^{\textrm{m}}56.13^{\textrm{s}}$, $\delta = +15^{\circ}49'33.60^{\prime \prime}$. The last non-detection of SN~2019ehk was on 2019 April 28 (MJD 58601.25), with a reported limiting $r$-band apparent magnitude of $>17.9$ mag. We fit a power law to the early-time data and derive a time of explosion of MJD $58601.8 \pm 0.1$.

We first classified SN~2019ehk as a young core-collapse SN with a blue, featureless continuum and strong \ion{Na}{i} D absorption \citep{dimitriadis19}. Later observations of SN~2019ehk suggested a SN~Ib classification with strong calcium features present in the photospheric spectra. However, the spectral time series of SN~2019ehk, coupled with its light curve evolution, indicated that it belonged to the \ca \ class. 

SN~2019ehk is located 17.4\arcsec\:east and 13.9\arcsec\:north of the nucleus of the SAB(s)bc galaxy M100 (NGC 4321). In this paper, we use a redshift-independent host-galaxy distance of 16.2 Mpc reported by \cite{Folatelli10}, which is consistent with the Cepheid distance estimated by \cite{freedman01}. However, it should be noted that there is a significant spread in reported distances for M100, which has influence on derived SN parameters. We use a redshift $z = 0.00524$ and standard $\Lambda$CDM cosmology ($H_{0}$ = 72 km s$^{-1}$ Mpc$^{-1}$, $\Omega_M = 0.27$, $\Omega_{\Lambda} = 0.73$). The main parameters of SN~2019ehk and its host-galaxy are displayed in Table \ref{tbl:params}.

SN~2019ehk presents a remarkable opportunity to advance our understanding of this class of objects. Our observational coverage of this SN includes constraining pre-explosion \textit{Hubble Space Telescope (HST)} imaging combined with a double-peaked light curve wherein the first peak is temporally consistent with luminous X-ray emission and ``flash-ionized'' Balmer series and \ion{He}{ii} spectral features. In  \S\ref{Sec:PREobservation} we outline the reduction and analysis of archival \textit{HST}, \textit{Spitzer} and \textit{Chandra} observations of the SN~2019ehk explosion site. In \S\ref{Sec:POSTobservation} we describe all optical, IR, UV, radio and X-ray observations of SN~2019ehk. In \S\ref{sec:host} we present metallicity and star formation estimates for the explosion site in M100. In \S\ref{sec:LC_analysis} and \S\ref{sec:spectro_analysis} we present analysis and comparisons of SN~2019ehk's optical photometric and spectroscopic properties, respectively, with chemical abundances of the SN and circumstellar medium derived spectroscopically in \S\ref{subsec:FS} and \S\ref{subsec:nebular_gas}. In \S \ref{sec:flare} we describe and model the first peak of the optical light curve of SN~2019ehk, while in \S\ref{Sec:Radio_Xray_Modeling} we infer properties of the explosion's immediate environment using X-ray and radio observations. Finally, in \S\ref{Sec:discussion} we discuss the possible progenitor systems responsible for SN~2019ehk. Conclusions are drawn in \S\ref{Sec:conclusion}. All uncertainties are quoted at the 68\% confidence level (c.l.) unless otherwise stated.

\section{Pre-explosion observations and data analysis}\label{Sec:PREobservation}

\begin{figure*}[t]
\centering
\includegraphics[width=\textwidth]{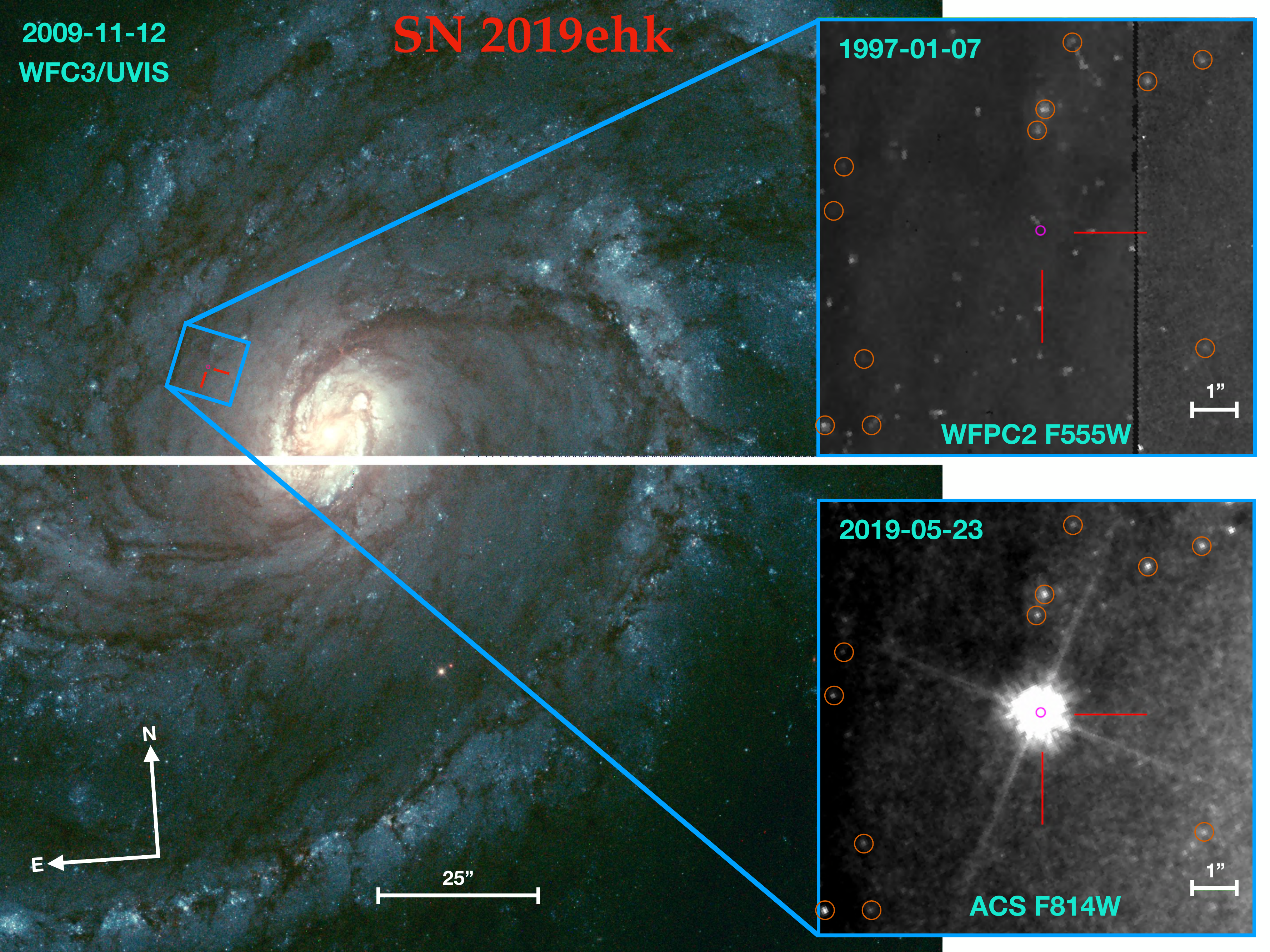} \caption{\textit{Left:} False color, \textit{HST} RGB pre-explosion image of host galaxy M100. \textit{Right Top Panel:} Zoomed-in pre-explosion image with WFPC2. \textit{Right Bottom Panel:} Post-explosion image of SN~2019ehk with ACS. Common sources between pre-/post-explosion epochs have been marked by orange circles. SN location is marked by red lines and the alignment uncertainty (at 200$\sigma$) is indicated by pink ellipses. \label{fig:hst}}
\end{figure*}

\subsection{HST observations} \label{SubSec:PreHST}

We analyze archival \textit{HST} images of M100 from the Mikulski Archive for Space Telescopes (MAST) to search for the progenitor system of SN~2019ehk. These observations span from 31 December 1993 to 12 November 2009 and include a variety of filters on the Wide Field and
Planetary Camera 2 (WFPC2) and the Wide Field Camera 3 (WFC3). Post-explosion Advanced Camera for Surveys (ACS) F814W imaging of SN~2019ehk was obtained under \textit{HST} program PID-15645 \citep{sandhst} on 23 May 2019. We follow the procedure outlined in \cite{kilpatrick18} to reduce all \textit{HST} data with the \texttt{astrodrizzle} \citep{astrodrizzle} reduction package.\footnote{\url{https://github.com/charliekilpatrick/hst123}} 

We perform a fine alignment between pre- and post-explosion images in order to accurately look for a coincident progenitor source. For this we use the ACS F814W image of SN~2019ehk on 23 May 2019 and the deepest WFPC2 archival image in F555W taken on 7 January 1997. These specific images are presented in Figure \ref{fig:hst} for reference. We first run \texttt{sextractor} \citep{sextractor} on both images to determine common sources to be used in the alignment process, with cuts made based on an individual sources' full width at half maximum (FWHM) and relative flux. We find 220 common sources between pre- and post-explosion images.

We then performed image registration on the ACS image with IRAF\footnote{IRAF, the Image Reduction and Analysis Facility, is distributed by the
National Optical Astronomy Observatory, which is operated by the Association of Universities for Research in Astronomy (AURA) under cooperative
agreement with the National Science Foundation (NSF).} tasks \texttt{ccmap} and \texttt{ccsetwcs}. We used a fourth order polynomial in \texttt{ccmap} to fit pixel coordinates of all common sources in the WFPC2 image to the tangent plane projection of the right ascensions and declinations of the same sources in the ACS image. We then adjusted the WCS solution of the WFPC2 image with \texttt{ccsetwcs}. We calculate an astrometric uncertainty of $\sigma_{\alpha} =  4.05\times 10^{-4}\arcsec$ and $\sigma_{\delta} = 2.71\times 10^{-4}\arcsec$ on the explosion site of SN~2019ehk in pre- and post-explosion images.

We apply the WCS solution from our fine alignment to all pre-explosion images and run \texttt{dolphot} to search for a source at the location of SN~2019ehk. We find no detectable source in any pre-explosion images within the uncertainty range of the relative astrometry. We then calculate 3$\sigma$ upper limits on a possible source coincident with SN~2019ehk by injecting fake stars and performing PSF photometry on these sources with \texttt{dolphot}. We present the upper limits in apparent magnitude (Vega system) for each pre-explosion \textit{HST} filter in Table \ref{tbl:hst_table} and flux limits with respect to filter functions in Figure \ref{fig:hst_sed} of the Appendix. 

All \textit{HST} limiting magnitudes are used to constrain the luminosity and temperature of the SN~2019ehk stellar progenitor. First, we use \texttt{pysynphot} to generate a grid of luminosities ($10^{-2} - 10^{8}$~L$_{\odot}$) and temperatures ($100-10000$~K) assuming a blackbody stellar model. Each blackbody luminosity is normalized using the SN distance and uncertainty. For each luminosity and temperature in our grid, we convolve the associated spectrum with each \textit{HST} filter in order to calculate the expected apparent magnitude. Then, in each filter, we cross-match the synthetic magnitude against the limit derived from fake star injection. If every synthetic magnitude is smaller than the pre-explosion limits then the luminosity/temperature grid point is rejected from the SN~2019ehk progenitor parameter space. We present the allowed/ruled out regions of pre-explosion parameter space (\S\ref{subsec:HR_progenitors}) and discuss its implications for the progenitor of SN~2019ehk on the Hertzsprung-Russell diagram. 

\subsection{Spitzer observations} \label{SubSec:PreSpitzer}

We perform a similar analysis of \textit{Spitzer} pre-explosion imaging as in \S\ref{SubSec:PreHST}. We collect archival data of M100 from the Spitzer Heritage Archive that included multi-channel observations from 21 August 2014 to 12 April 2019 \citep{2013sptz.prop10136K, 2014sptz.prop11063K, 2016sptz.prop13053K, 2018sptz.prop14089K}. For the fine alignment, we utilized explosion imaging of SN~2019ehk taken on 11 May 2019 under \textit{Spitzer} program DD-14089 \citep{2018sptz.prop14089K}. As in \S\ref{SubSec:PreHST}, we perform relative astrometry with IRAF and use \texttt{dolphot} to measure photometry of all detected sources. Upon inspection, we detect no pre-explosion source coincident with the location of SN~2019ehk. We then perform fake star injection with \texttt{dolphot} to estimate the limiting magnitudes of the SN~2019ehk progenitor. We report our 3$\sigma$ limits in the AB magnitude system in Table \ref{tbl:spitzer_table} and flux limits with respect to filter functions in Figure \ref{fig:hst_sed} of the Appendix. While the limits are not as constraining as those derived from \textit{HST} imaging, we discuss implications of these observations in the context of dusty progenitors in \S\ref{subsec:HR_progenitors}.  

\subsection{CXO observations} \label{SubSec:PreCXO}

The \emph{Chandra X-ray Observatory} (CXO) observed the location of SN\,2019ehk with ACIS-S on multiple occasions between  	1999 November 6  and 2012 February 16, for a total exposure time of 149.3 ks (observation IDs 400, 6727, 9121, 12696, 14230; PIs Garmire, Immler, Patnaude). We followed standard ACIS-S data reduction routines within \texttt{CIAO v.4.12} employing the latest calibration files. Specifically, we reprocessed the data with \texttt{chandra\_repro} and generated a merged event file from the individually re-projected files; this action also created a merged exposure map and a combined exposure map weighted PSF file. Running the source detection algorithm \texttt{wavdetect} on the merged event file using the exposure-map weighted PSF file we find no evidence for statistically significant X-ray emission from a point source at the location of SN\,2019ehk. Adopting Poisson statistics we infer a 0.5-8 keV count-rate upper limit of  $7\times 10^{-5}\rm{c\,s^{-1}}$ ($3\,\sigma$ c.l.), which translates into an unabsorbed flux limit in the range $F_x<( 1.7-4.0)\times 10^{-15} \rm{erg\,s^{-1}cm^{-2}}$ (0.3-10 keV) for a power-law spectrum with index $\Gamma=2$, Galactic absorption $2\times 10^{20}\,\rm{cm^{-2}}$ \citep{Kalberla05}, and intrinsic absorption $NH_{\rm int}=(1-10^2)\times 10^{20}\,\rm{cm^{-2}}$. For a  blackbody spectrum with $kT=(0.1-10)$ keV and $NH_{\rm int}=(1-10^2)\times 10^{20}\,\rm{cm^{-2}}$ the flux limit is  $F_x<(1-10)\times 10^{-15} \rm{erg\,s^{-1}cm^{-2}}$ (0.3-10 keV).

\section{Post-explosion observations and data analysis}\label{Sec:POSTobservation}

\begin{table}[ht]
\begin{center}
\caption{Main parameters of SN\,2019ehk and its host galaxy \label{tbl:params}}
\vskip0.1in
\begin{tabular}{lccccccccc}
\hline
\hline
Host Galaxy &  &  & &  & & & &  &  M100 (NGC~4321) \\ 
Galaxy Type &  &  & &  & & & &  &  SAB(rs)c \\
Galactic Offset &  &  & &  & & & &  &  $23\arcsec (1.8 ~ \rm kpc)$ \\
Redshift &  &  & &  & & & &  &  $0.005 \pm 0.0001$\\  
Distance &  &  & &  & & & &  &  $16.2 \pm 0.400$~Mpc\footnote{\cite{freedman01,Folatelli10}}\\ 
Distance Modulus, $\mu$ &  &  & &  & & & &  &  $31.1 \pm 0.100$~mag\\ 
$\textrm{RA}_{\textrm{SN}}$ &  &  & &  & & & &  &  $12^{\textrm{h}}22^{\textrm{m}}56.15^{\textrm{s}}$\\
$\textrm{Dec}_{\textrm{SN}}$ &  &  & &  & & & &  & $+15^{\circ}49'34.18^{\prime \prime}$\\
Time of Explosion (MJD) &  &  & &  & & & &  &  58601.8 $\pm$ 0.1\\ 
$E(B-V)_{\textrm{MW}}$ &  &  & &  & & & &  &  0.02 $\pm$ 0.001~mag\\
$E(B-V)_{\textrm{host}}$ &  &  & &  & & & &  &  0.47 $\pm$ 0.10~mag\\
$E(B-V)_{\textrm{host, \ion{H}{ii}}}$ &  &  & &  & & & &  &  0.34 $\pm$ 0.14~mag\footnote{Based on Balmer decrement of \ion{H}{ii} region at SN location.}\\
$m_{B}^{\mathrm{peak}}$ &  &  & &  & & & &  & $18.0 \pm 0.0150$~mag\\
$M_{B}^{\mathrm{peak}}$ &  &  & &  & & & &  & $-15.1 \pm 0.0210$~mag\footnote{Extinction correction applied.}\footnote{Relative to second $B$-band light curve peak}\\
$\Delta$m$_{15}$ &  &  & &  & & & &  & $1.7 \pm 0.014$~mag$^{\rm{d}}$\\
\hline
\end{tabular}
\end{center}
\label{table:Observations}
\tablecomments{No extinction corrections have been applied to the presented apparent magnitudes.}
\end{table}

\subsection{UV/Optical photometry}\label{SubSec:Phot}
\begin{figure*}[t]
\centering
\includegraphics[width=\textwidth]{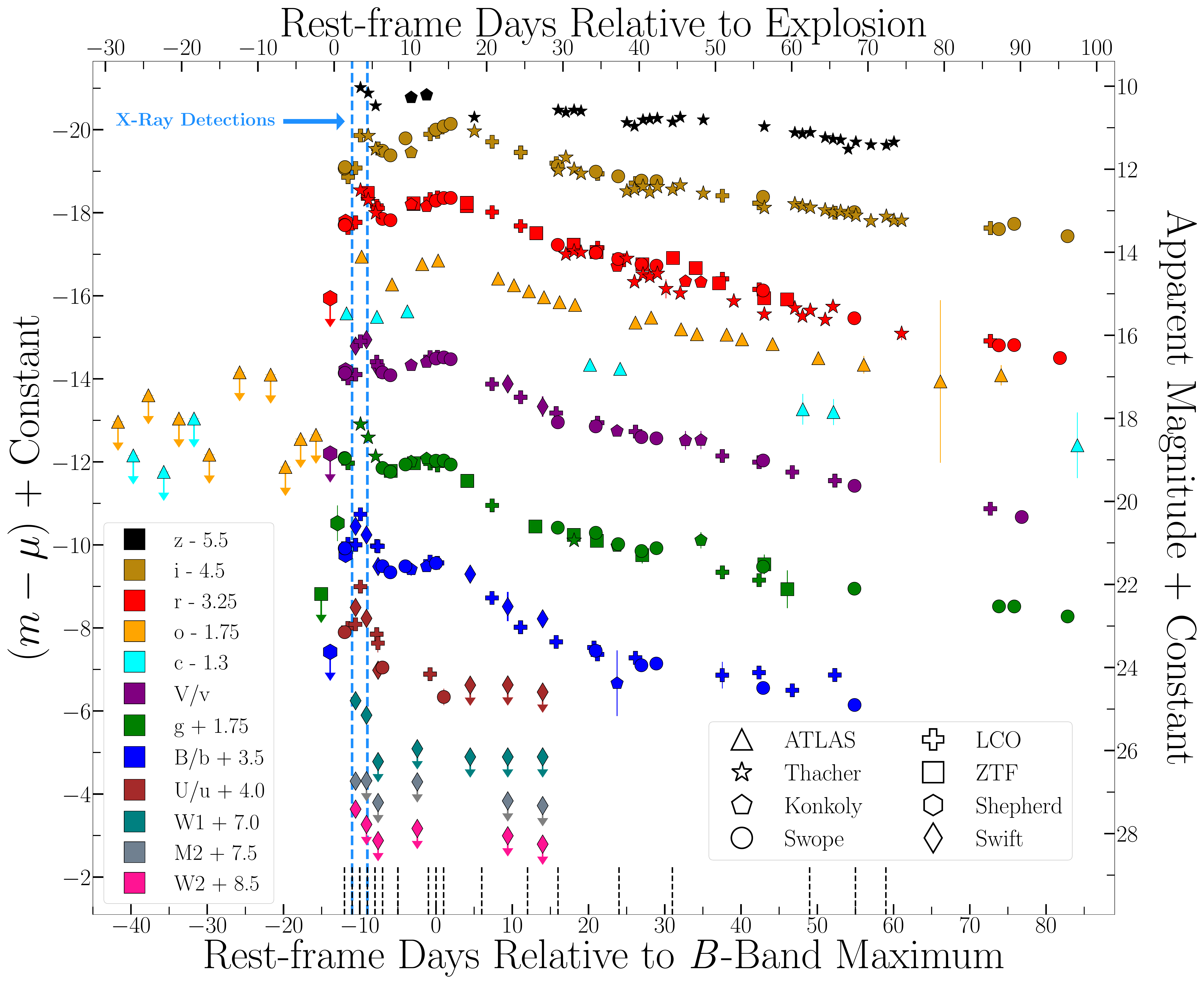}
\caption{UV/Optical light curve of SN~2019ehk with respect to second $B$-band maximum. Observed photometry presented in AB magnitude system. ATLAS data/3$\sigma$ upper limits are presented as triangles, Swope as circles, LCO as plus signs, Thacher as stars, ZTF as squares, Konkoly as polygons and J. Shepherd as hexagons. The epochs of our spectroscopic observations are marked by black dashed lines. Grey vertical dashed lines mark the time of the X-ray detections  of SN\,2019ehk \label{fig:optical_LC} }
\end{figure*}

We started observing SN\,2019ehk with the Ultraviolet Optical Telescope (UVOT; \citealt{Roming05}) onboard the Neil Gehrels \emph{Swift} Observatory \citep{Gehrels04} on 1 May 2019 until 26 May 2019 ($\delta t=$ 2.8 -- 27.3 days since explosion). We performed aperture photometry with a 3$\arcsec$ region with \texttt{uvotsource} within HEAsoft v6.26 (and corresponding calibration files), following the standard guidelines from \cite{Brown14}. We detect UV emission from the SN at the time of the first optical peak (Figure \ref{fig:optical_LC}) until $t\approx5$ days after explosion. Subsequent non-detections in $U, W1, M2, W2$ bands indicate significant cooling of the photosphere.  

SN~2019ehk was imaged between 30 April 2019 and 1 August 2019 ($\delta t =$ 1.2 -- 94.2 days since explosion) with the Direct camera on the Swope 1-m telescope at Las Campanas Observatory, Chile and the PlaneWave CDK-700 0.7m telescope at Thacher Observatory in Ojai, California. Observations were performed in Johnson \textit{BV} and Sloan \textit{ugriz} filters. For these data, we performed bias-subtraction and flat-fielding, stitching, registration, and photometric calibration using {\tt photpipe} \citep{Rest+05}.  For our photometric calibration, we used stars in the PS1 DR1 catalog \citep{Flewelling+16} transformed from \textit{gri} magnitudes to the \textit{uBVgri} Swope natural system following the Supercal method \citep{Scolnic+15}.  Difference imaging in \textit{gri} bands was performed using PS1 templates.  Final photometry was performed in the difference images with DoPhot \citep{Schechter+93}. 

Las Cumbres Observatory (LCO) $UBVgri$-band data of SN~2019ehk were obtained with the Sinistro cameras on the 1m telescopes at Sutherland (South Africa), CTIO (Chile), Siding Spring (Australia), and McDonald (USA), through the Global Supernova Project. Using {\tt lcogtsnpipe}\footnote{\tt \url{https://github.com/svalenti/lcogtsnpipe}}, a PyRAF-based photometric reduction pipeline, PSF fitting was performed. Reference images were obtained after the SN faded, and image subtraction was performed using PyZOGY \citep{guevel17}, an implementation in Python of the subtraction algorithm described in \cite{zackay16}. $UBV$-band data were calibrated to Vega magnitudes \citep{stetson00} using standard fields observed on the same night by the same telescope. $gri$-band data were calibrated to AB magnitudes using the Sloan Digital Sky Survey (SDSS, \citealt{sdss17}).

SN~2019ehk was also observed with ATLAS, a twin 0.5m telescope system installed on Haleakala and Mauna Loa in the Hawai'ian islands that robotically surveys the sky in cyan (\textit{c}) and orange (\textit{o}) filters \citep[][]{2018PASP..130f4505T}. The survey images are processed as described in \cite{2018PASP..130f4505T} and photometrically and astrometrically calibrated immediately \citep[using the RefCat2 catalogue;][]{2018ApJ...867..105T}. Template generation, image subtraction procedures and identification of transient objects are described
in \cite{smith20}. Point-spread-function photometry is carried out on the difference images and all sources greater than 5$\sigma$ are recorded and all sources go through an automatic validation process that removes spurious objects \citep{smith20}. Photometry on the difference images (both forced and non-forced) is from automated point-spread-function fitting as documented in \cite{2018PASP..130f4505T}. The photometry presented here are weighted averages of the nightly individual 30\,sec exposures, carried out with forced photometry at the position of SN2019ehk. We searched for pre-explosion outbursts in archival ATLAS observations of the SN explosion site from MJD $57400 - 58599$ ($1201 - 2$ days before explosion). We assume Gaussian errors on the flux and test different phase-dependent binning combinations of pre-explosion data but do not find any photometric detections at > 3$\sigma$ significance.

Additional follow-up photometry on SN~2019ehk was gathered at the Konkoly Observatory, Hungary, using the 0.8m RC80 telescope equipped with a $2048 \times 2048$ FLI Proline 23042-1 back-illuminated CCD camera and $B V g' r' i' z'$ filters. The frames are geometrically registered to a common pixel position then median-combined to create a deeper frame in each filter; transformation to the WCS was done by applying {\tt astrometry.net} \citep{lang10}. Using IRAF tasks, image subtraction was applied using PS1 $g r i z$ frames as templates, after pixel resampling, geometric registration, PSF- and flux-matching transformations of the template images. We then applied the publicly available {\tt Yoda} code \citep{drory03} to get simple aperture photometry on both the SN and the local comparison stars. Transformation to the standard photometric system was done using the standard magnitudes of the local comparison stars from the PS1-catalog \citep{flewelling16}. Uncertainties on the final magnitudes are computed by combining the photometric errors as given by {\tt Yoda} and the residuals of the photometric zero points derived from the local comparison stars. 

The Milky Way $V$-band extinction and color excess  along the SN line of site is $A_{V} = 0.070$~mag and \textit{E(B-V)} = 0.0227~mag \citep{schlegel98, schlafly11}, respectively, which we correct for using a standard \cite{fitzpatrick99} reddening law (\textit{$R_V$} = 3.1). In order to estimate the effect of host galaxy extinction, we use a spectroscopic observation at the SN location from the Multi Unit Spectroscopic Explorer (MUSE), which observed M100 before the SN explosion on 28 April 2019 through ESO program PID 1100.B-0651 (PI Schinnerer). We apply a $0.77\arcsec$ aperture (equal to the underlying \ion{H}{ii} region) to the MUSE data cube in order to extract a host spectrum. After accounting for the stellar absorption with  Single Stellar Population (SSP) modeling within \texttt{STARLIGHT} \citep{fernandes05} as in \cite{galbany16}, we measure a H$\alpha$ and H$\beta$ line flux ratio of 4.23 and estimate the Balmer decrement through standard assumptions of Case B recombination \citep{osterbrock06} and \cite{fitzpatrick99} extinction law ($R_V = 3.1$). We derive a line of sight host galaxy reddening of $E(B-V)= 0.339 \pm 0.135$~mag.

In addition to the color excess derived from Balmer decrement in M100, there appears to be significant host galaxy extinction in the local SN environment. All photospheric spectra show prominent \ion{Na}{i} D absorption with Equivalent Width EW $\sim$3~\AA, at the host-galaxy redshift. We attempt to use Equation 9 in \cite{Poznanski12} to convert the \ion{Na}{i} EW to  an intrinsic \textit{E(B-V)}, but the empirical relation shown in their Figure 9 becomes tenuous for EW $\gtrapprox$ 1.5~\AA. Consequently, in order to  derive an appropriate host extinction, we compare the \textit{r-i} color to a sample of type Ic SNe (SNe Ic) (see \S\ref{subsec:phot_properties}). We find that \textit{E(B-V)}=0.47\,mag (corresponding to EW = 1.3~\AA) is a reasonable estimate for host-galaxy extinction because it represents an average between the large extinction needed to match SNe Ic colors and a negligible extinction that is consistent with the observed color evolution in other \cas. 

The complete light curve of SN~2019ehk is presented in Figure \ref{fig:optical_LC} and reference photometric observations are listed in Appendix Table \ref{tbl:phot_table}. In addition to our observations, we include photometry from the Zwicky Transient Facility (ZTF; \citealt{bellm19, graham19}) public data stream. The most notable feature of the light curve is the presence of two peaks at $\delta t\approx3$ days and $\delta t\approx15$ days after explosion. Potential power sources of the first peak are presented in \S\ref{sec:flare} while the luminosity of the later peak is considered to be derived from standard ${}^{56}\textrm{Ni}$ decay modeled in \S\ref{subsec:bol_LC}. 

\subsection{Optical/NIR spectroscopy}\label{SubSec:Spec}

\begin{figure*}
\centering
\includegraphics[width=\textwidth]{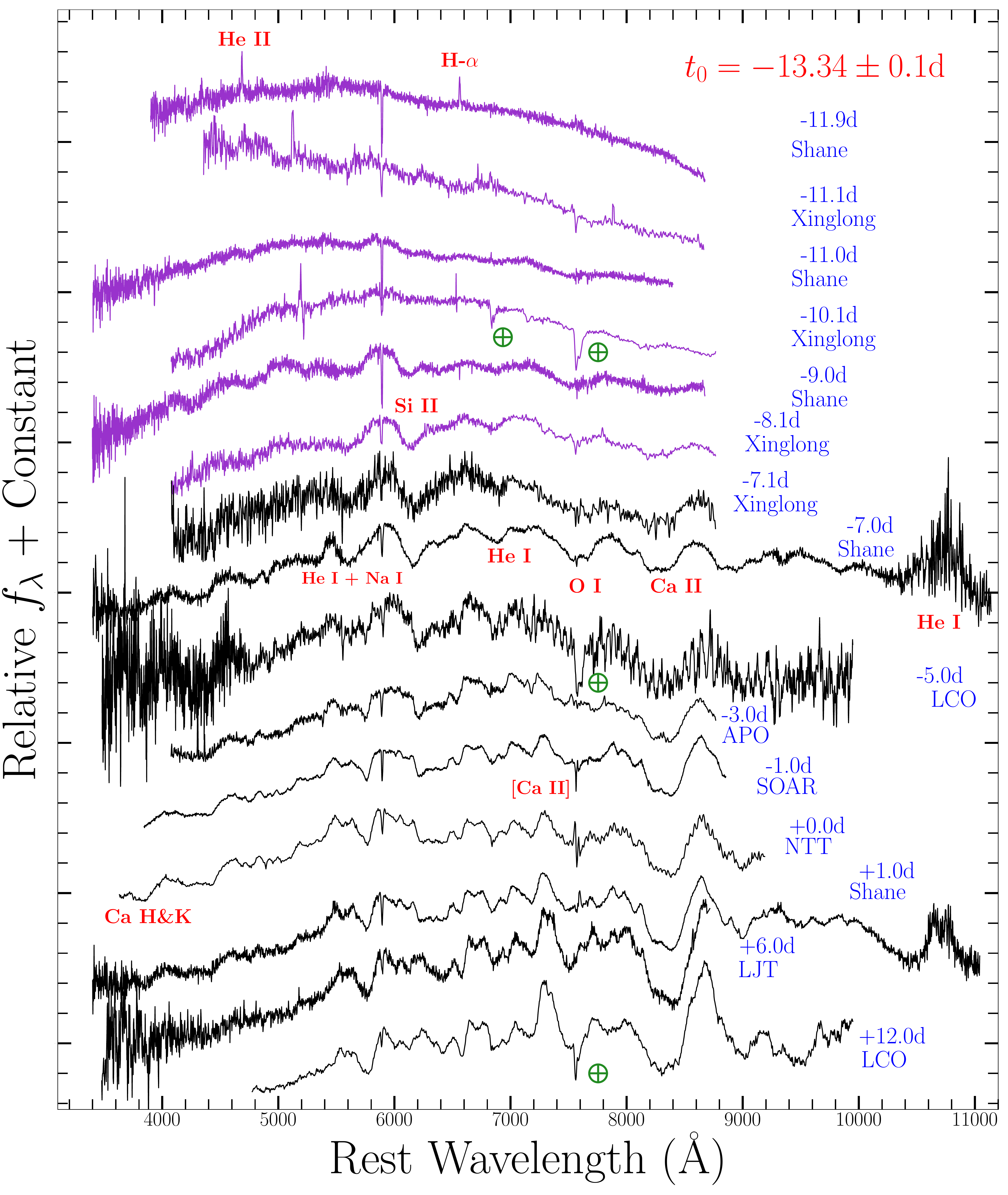}
\caption{Spectral observations of SN~2019ehk with phases (blue) marked with respect to second $B$-band maximum. Spectra during the first light curve peak are plotted in purple. Green circles with a plus indicate telluric absorption. As shown in the extremely early-time epochs, flash-ionized Balmer series and \ion{He}{ii} emission lines are only detected until $\delta t \approx 2$ days after explosion.    \label{fig:spectral_seriesA}  }
\end{figure*}

\begin{figure*}
\centering
\includegraphics[width=\textwidth]{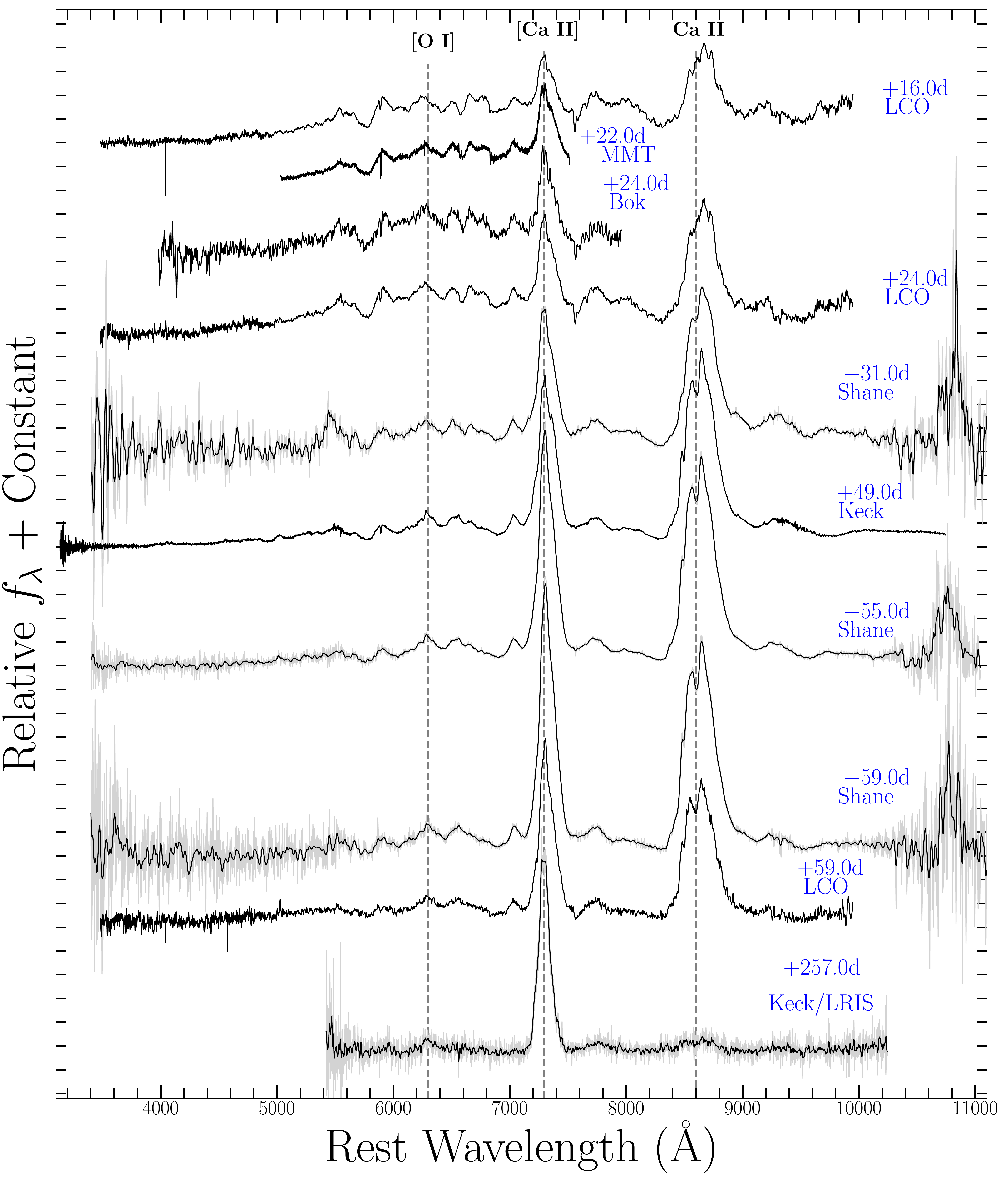}
\caption{Spectral observations of SN~2019ehk with phases (blue) marked with respect to $B$-band maximum. Raw spectra are shown in gray, and smoothed spectra with black lines.  \label{fig:spectral_seriesB}}
\end{figure*}

In Figures \ref{fig:spectral_seriesA} and \ref{fig:spectral_seriesB}, we present the complete series of optical spectroscopic observations of SN~2019ehk from -12 to +257d relative to the second $B$-band maximum ($\delta t = 1.34-270$ days relative to explosion). A full log of spectroscopic observations is presented in Appendix Table \ref{tab:spec_table}. 

SN~2019ehk was observed with Shane/Kast (Miller \& Stone 1993), SOAR/Goodman \citep{clemens04} and Keck/LRIS \citep{oke95} between -12d and +257d relative to the second light curve peak. For all these spectroscopic observations, standard CCD processing and spectrum extraction were accomplished with IRAF. The data were extracted using the optimal algorithm of \citet{1986PASP...98..609H}.  Low-order polynomial fits to calibration-lamp spectra were used to establish the wavelength scale, and small
adjustments derived from night-sky lines in the object frames were applied. We employed custom IDL routines to flux calibrate the data and remove telluric lines using the well-exposed continua of the
spectrophotometric standard stars \citep{1988ApJ...324..411W, 2003PASP..115.1220F}. Details of these spectroscopic reduction techniques are described in \citet{2012MNRAS.425.1789S}.

SN 2019ehk was observed using EFOSC2 \citep{buzzoni84} at the 3.58 m ESO New Technology Telescope (NTT) on 13 May 2019 through the ePESSTO+ program \citep{smartt15, nicholl19}. Grism \#13 was used, with spectral coverage of 3500-9300 \AA\, and resolution of 21 \AA. The exposure time was 1500 s. Standard data reduction processes were performed using the PESSTO pipeline \citep{smartt15}\footnote{\url{https://github.com/svalenti/pessto}}. The reduced spectrum was then extracted, and calibrated in wavelength and flux.

Las Cumbres Observatory (LCO) optical spectra were taken with the FLOYDS spectrographs mounted on the 2m Faulkes Telescope North and South at Haleakala (USA) and Siding Spring (Australia), respectively, through the Global Supernova Project. A $2\arcsec$ slit was placed on the target at the parallactic angle. One-dimensional spectra were extracted, reduced, and calibrated following standard procedures using the FLOYDS pipeline\footnote{\url{https://github.com/svalenti/FLOYDS\_pipeline}} \citep{valenti14b}.

One low resolution optical spectrum was taken with the 300 l/mm grating on the Boller \& Chivens (B\&C) spectrograph mounted on the 2.3m Bok telescope on Kitt Peak using a 1.5 arcsec slit width on 5 June 2019. Additionally, one moderate resolution spectrum was taken with the Binospec spectrograph \citep{binospec} on the MMT using the 600 l/mm grating and 1" slit on 3 June 2019.
Both the B\&C and Binospec spectra were reduced using standard techniques in IRAF, including bias subtraction, flat-fielding, and sky subtraction.  Flux calibration was done with spectrophotometric standard star observations taken on the same night at similar airmass.  

The spectroscopic observations of SN~2019ehk were also collected using the Xinglong 2.16-m telescope (+BFOSC), and the Lijiang 2.4-m telescope 
(+YFOSC) \citep{fan15} in China. The SN was observed between -11 to -7 days relative to second $B$-band maxiumum. All the spectra were reduced using routine tasks within IRAF and the flux was calibrated with spectrophotometric standard stars observed on the same nights. Telluric lines are removed from all of these spectra whenever possible.

\begin{figure*}
\centering
\includegraphics[width=\textwidth]{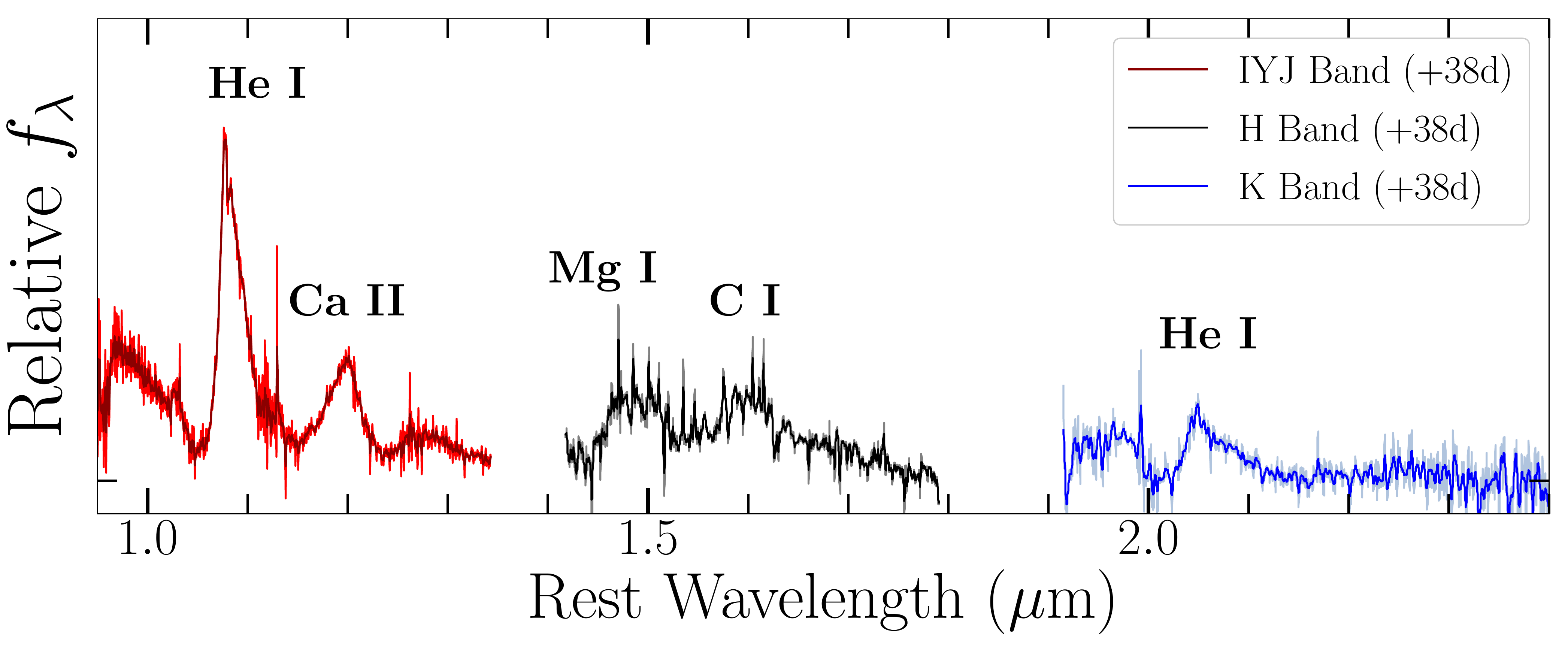}
\caption{SOAR Triple Spec NIR spectrum of SN~2019ehk on MJD 58655.9 or +38d relative to second $B$-band peak. Prominent line transitions are marked in black.  \label{fig:IR_spectrum}}
\end{figure*}

On 22 June 2019 (MJD 58656.0), we used the Triple-Spec instrument at SOAR to obtain a set of 3 ABBA observations of SN~2019ehk. We used the Spextool IDL package \citep{cushing04} to reduce the Triple-Spec data, we subtracted consecutive AB pairs to remove the sky and the bias level, then we flat fielded the science frames dividing by the normalized master flat. We calibrated 2D science frames in wavelength by using comparison lamps obtained in the afternoon before the observations. To correct for telluric features and to flux calibrate our SN spectra, we observed the A0V telluric standard HD~111744 after the SN and at a similar airmass. Finally, we extracted the SN and the telluric star spectra from the 2D wavelength calibrated frames. After the extraction of the individual spectra, we used the \texttt{xtellcorr} task (Vacca et al. 2003) included in the Spextool IDL package \citep{cushing04}, to perform the telluric correction and the flux calibration of the spectra of SN~2019ehk. We combined individual observations of SN~2019ehk in a single spectrum shown in Figure \ref{fig:IR_spectrum}.

\begin{figure}
\centering
\includegraphics[width=0.45\textwidth]{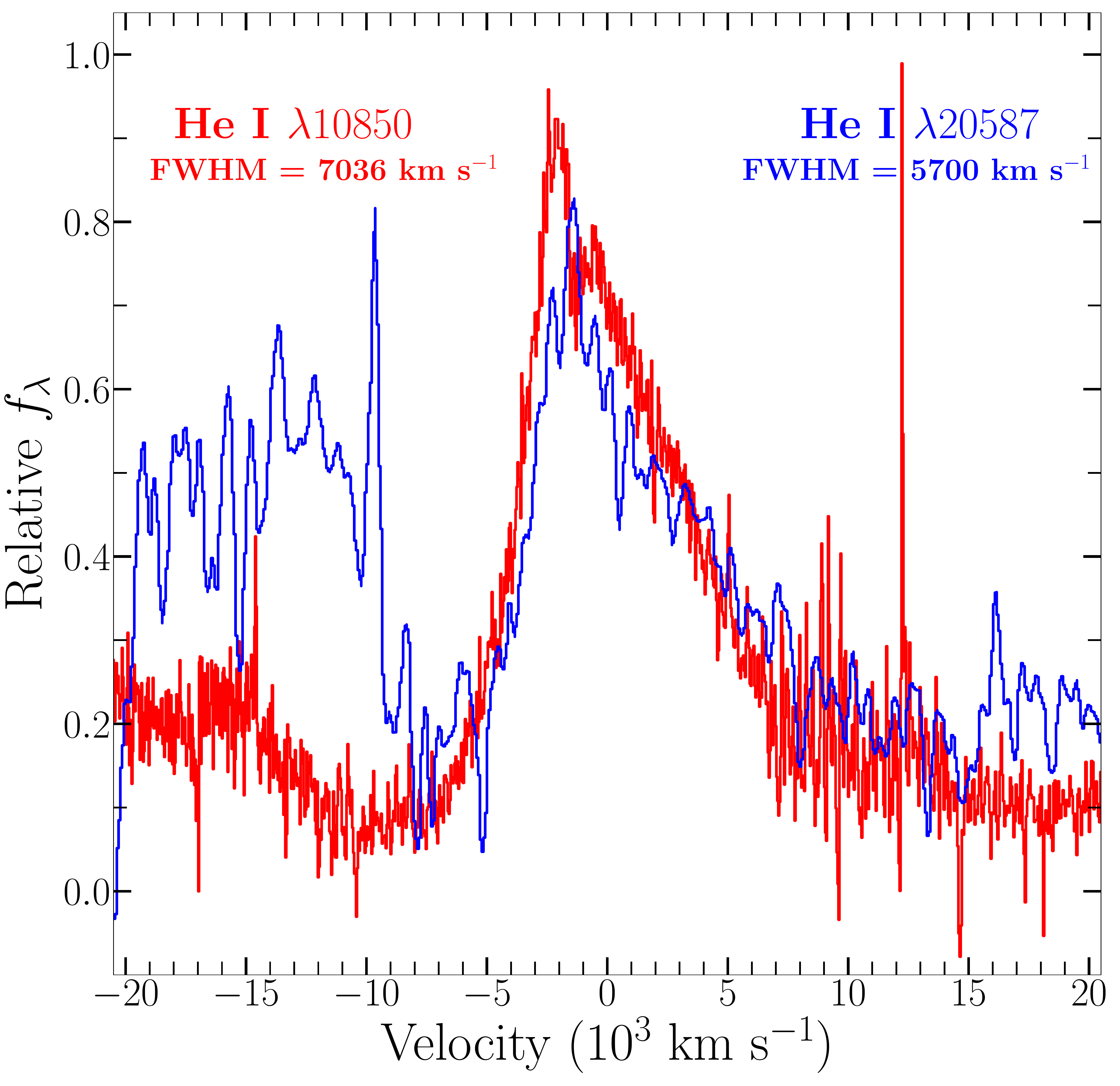}
\caption{NIR \ion{He}{i} $\lambda$10850 and $\lambda$20587 line velocity profiles (Fig. \ref{fig:IR_spectrum}). The P-Cygni line profile indicates that the helium is photospheric and expanding up to $\sim$7000\,$\kms$. However, the broad emission feature may either indicate a detached ejecta component of helium in the circumstellar material (CSM) or a blending of spectral features near 1$\mu$m. The profile of the \ion{He}{i} $\lambda$20589 line at +38 days after second $B$-band maximum shows that it becomes optically thin at lower velocity than does the \ion{He}{i} $\lambda$10830 line, presumably because of a lower population of the 1s2s $^1$S metastable levels, which results  from its much higher Einstein A value.
\label{fig:IR_spectrum-v}}
\end{figure}

\subsection{X-ray observations with \emph{Swift}-XRT and Chandra}\label{SubSec:XRT}
The X-Ray Telescope (XRT, \citealt{burrows05}) on board the \emph{Swift} spacecraft \citep{Gehrels04} started observing the field of SN\,2019ehk on 01 May 2019, until 25 May 2019 ($\delta t\approx 3-24$ d since explosion with a total exposure time of 11.4 ks, IDs 11337 and 11339). We analyzed the data using HEASoft v 6.22 and followed the prescriptions detailed in \cite{margutti13}, applying standard filtering and screening. A bright source of X-ray emission is clearly detected with significance of $>5\sigma$ against the background. Visual inspection reveals the presence of extended emission from the host galaxy at the location of the SN. Using Poisson statistics we find that X-ray emission from SN\,2019ehk is detected with significance $>3\,\sigma$ at $t\le 4.2$ d since explosion. No X-ray emission is detected above the host-galaxy level at later times.

We used \emph{Swift}-XRT pre-explosion data acquired in 2005-2006 to estimate the level of emission from the host galaxy at the SN location (IDs 35227 and 30365). Merging all the available pre-explosion observations (exposure time of $\sim$59.1 ks), and extracting a spectrum from a 20$\arcsec$ region centered at the SN location,  we find that the host-galaxy emission  is well modeled by a power-law spectrum with photon index $\Gamma=2.1\pm0.1$, corresponding to a 0.3-10 keV unaborbed flux $F_x=(1.0\pm0.1)\times 10^{-13}\,\rm{erg\,s^{-1}cm^{-2}}$. The Galactic neutral hydrogen column density along our line of sight is $\rm{NH_{MW}}=2.0\times 10^{20}\,\rm{cm^{-2}}$ \citep{Kalberla05}. We use this model to account for the contribution of the host galaxy in the two post-explosion epochs where an excess of X-ray emission from SN\,2019ehk is detected (at $t=2.8$ d and $4.2$ d). 

\begin{figure*}
\centering
\subfigure[]{\includegraphics[width=0.46\textwidth]{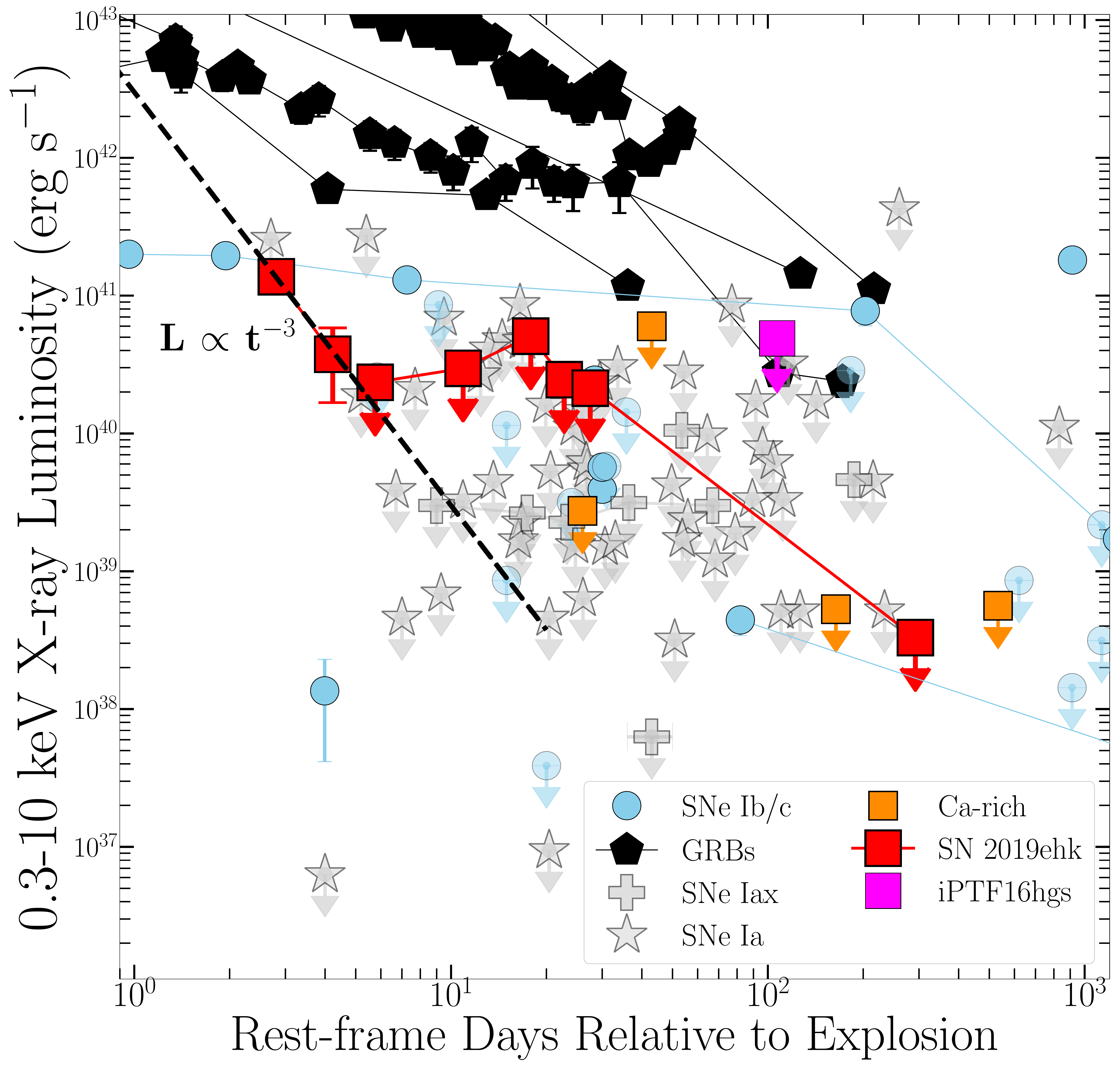}}
\subfigure[]{\includegraphics[width=0.46\textwidth]{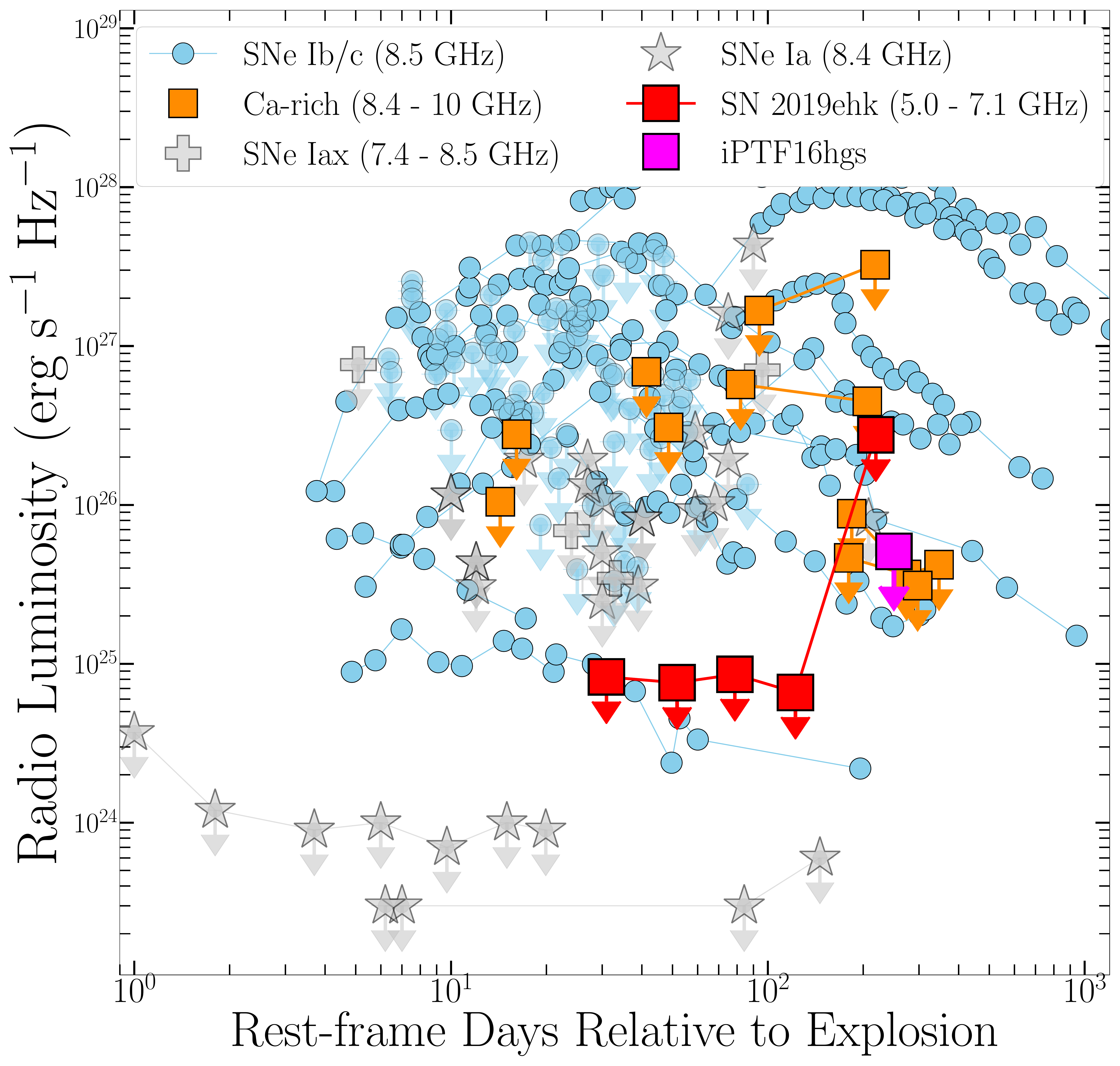}}
\caption{(a) X-Ray light curve of SN~2019ehk (red squares) and other thermonuclear transients e.g., SNe~Iax (grey plus signs), SNe Ia (grey stars) and \cas \ (orange squares). Core-collapse SNe~Ib/c are shown as light blue circles and GRBs are displayed as black polygons. The decline rate of SN~2019ehk's X-ray emission ($L_x\propto t^{-3}$) is shown as a black dashed line. (b) Radio non-detections of SN~2019ehk (red squares) compared to non-detection limits of thermonuclear SNe and SNe~Ib/c.   \label{fig:xray_radio_LC} }
\end{figure*}

For each of these two epochs we extracted a spectrum using a 20$\arcsec$ region centered at the location of the SN.  We find that the X-ray spectrum of the SN emission has a best-fitting photon index $\Gamma=0.1\pm0.3$ and $\Gamma=0.2\pm 0.9$ for the first and second epoch, respectively, corresponding to an unabsorbed 0.3-10 keV flux of $F_x=4.4 \times 10^{-12}\,\rm{erg\,s^{-1}cm^{-2}}$ and $F_x=1.3 \times 10^{-12}\,\rm{erg\,s^{-1}cm^{-2}}$. No evidence for intrinsic neutral hydrogen absorption is found ($NH_{int}<4\times 10^{22}\,\rm{cm^{-2}}$ at $3\,\sigma$ c.l. from the first epoch). We use the best-fitting spectral parameters inferred from the second epoch of observations to flux-calibrate the count-rate upper limits derived for the following epochs (Table \ref{tab:xray_obs}). At the distance of SN\,2019ehk these measurements indicate a steeply decaying, large X-ray luminosity with $L_x\propto t^{-3}$ and $L_x\ge 3\times 10^{40}\,\rm{erg\,s^{-1}}$ at very early times $t\le4.2$ d  (Figure \ref{fig:xray_radio_LC}). The very luminous X-ray emission from SN\,2019ehk at $t\approx2.8$~d  $L_x\approx10^{41}\,\rm{erg\,s^{-1}}$ rivals that of GRB\,980425. Since no other \ca \ has been observed in the X-rays a few days since explosion, it is unclear if this luminous X-ray display is a common trait of the class.

The hard 0.3-10~keV X-ray spectrum of SN\,2019ehk is suggestive of thermal bremsstrahlung emission with temperature $T>10$~keV. Fitting the SN contribution with a bremsstrahlung spectral model with $T=10-200$ keV  the inferred emission measure $EM=\int n_e n_I dV$ is $EM=(7-10)\times 10^{63}\,\rm{cm^{-3}}$ (at $\delta t=2.8$ d) and $EM=(2-3)\times 10^{63}\,\rm{cm^{-3}}$ (at $\delta t=4.2$ d), where $n_e$ and $n_I$ are the number densities of electrons and ions, respectively.

The location of SN\,2019ehk was serendipitously observed by the Chandra X-ray Observatory (CXO) on   15 February, 2020 ($\delta t=292.2$ d since explosion, exposure time of 9.95ks, ID 23140, PI Stroh) as part of follow-up observations of another supernova, SN\,2020oi, that exploded in the same host galaxy. We analyzed the data with the \texttt{CIAO} software package v4.12 and corresponding calibration files. We find no evidence for X-ray emission at the location of SN\,2019ehk and we place a 3$\sigma$ count-rate upper limit of $3.01\times 10^{-4}\rm{c\,s^{-1}}$ (0.5-8 keV, pure Poisson statistics). We adopt the spectral parameters from the latest epoch of \emph{Swift}-XRT observations that led to a detection and we infer an unabsorbed 0.3-10 keV flux limit $F_x<1.07\times 10^{-14}\,\rm{erg\,s^{-1}cm^{-2}}$, which corresponds to $L_x<3.3\times 10^{38}\,\rm{erg\,s^{-1}}$. This is the deepest limit on the late-time X-ray luminosity of a \ca\, to date (Figure \ref{fig:xray_radio_LC}).  

\subsection{Radio observations with the VLA}\label{SubSec:VLA}
We acquired deep radio observations of SN\,2019ehk with the Karl G. Jansky Very Large Array (VLA) at $\delta t=30.5-219.7$
days post explosion through project VLA/19A-271 (PI D. Coppejans).  All observations have been obtained at 6.05 GHz (C-band) with 2.048 GHz bandwidth in standard phase referencing
mode, with 3C286 as a bandpass and flux-density calibrator
and QSO J1224+21 (in A and B configuration) and QSO J1254+114 (in D configuration) as complex gain calibrators. The data have been calibrated using the VLA pipeline in the Common  Astronomy  Software  Applications  package (CASA, \citealt{McMullin07}) v5.4.1 with additional flagging. Briggs weighting with a robust parameter of 2 was used to image. SN\,2019ehk is not detected in our observations. We list the inferred flux densities in Appendix Table \ref{Tab:radio} and show how these measurements compare to radio observations of thermonuclear transients and core-collapse SNe in Figure \ref{fig:xray_radio_LC}(b).

\section{Host Galaxy and Explosion Site}\label{sec:host}

\subsection{Metallicity}\label{SubSec:metallicity}

We determine an oxygen abundance 12 + log(O/H) at the explosion site by using a MUSE spectroscopic observation taken on 28 April 2019 (PI Schinnerer). Data were reduced and analyzed following the prescriptions outlined in \cite{galbany16}. The spectrum was corrected for a host galaxy reddening of $E(B-V) = 0.34$~mag and stellar absorption is accounted for with a SSP model (e.g., see \S\ref{SubSec:Phot}). Using a combination of line flux ratios ([\ion{O}{iii}] / H$\beta$ and [\ion{N}{ii}]/H$\alpha$) into Equation 3 of \cite{pettini04}, we determine a host metallicity of 12 + log(O/H) = $8.70 \pm 0.12$~dex ($1.03 \pm 0.120$~Z$_{\odot}$). We obtain a similar value if we use the method presented in \cite{dopita16}: 12 + log(O/H) = $9.04\pm 0.20$~dex ($1.46 \pm 0.290$~Z$_{\odot}$). The $\sim$0.3~dex difference between methods is expected given known offsets amongst calibrators \citep{kewley08}. Furthermore, both values are similar to the metallicities calculated by \cite{pohlen10} from the radial distribution of gas across M100.

\subsection{Star Formation Rate}\label{SubSec:star_formation}

We utilize the same pre-explosion MUSE spectrum to determine a star formation rate at the location of SN~2019ehk. We calculate a total H$\alpha$ emission line luminosity of $L_{\textrm{H$\alpha$}} = (1.16 \pm 0.37 )\times 10^{37}$~erg~s$^{-1}$ from a 0.7$\arcsec{}$ aperture that encompasses the local \ion{H}{ii} region at the SN location. We then use Equation 2 from \cite{Kennicutt98} to estimate a star formation rate of SFR = $(9.2 \pm 2.9 )\times 10^{-5} \ \Msun$ yr$^{-1}$ at the explosion site. We also derive an effective star formation rate of $(5.3 \pm 1.7 )\times 10^{-3} \ \Msun$ yr$^{-1}$ kpc$^{-2}$. These SFR values are reasonable considering the lack of observed star formation found at most \ca \ explosion sites. Our inferred rate is consistent with the low observed SFR values derived from $\gtrsim 90 \%$ of \ca \ explosion sites. With regards to core-collapse SNe, the H$\alpha$ luminosity at the explosion site of SN~2019ehk is only consistent with the \ion{H}{ii} region luminosity at the location of $\sim 20-30 \%$ of H-stripped SNe (e.g., \citealt{galbany18, Kuncarayakti18}).

\section{Optical Light Curve Analysis}\label{sec:LC_analysis}

\subsection{Photometric Properties}\label{subsec:phot_properties}

\begin{figure*}
\centering
\subfigure[]{\includegraphics[width=0.48\textwidth]{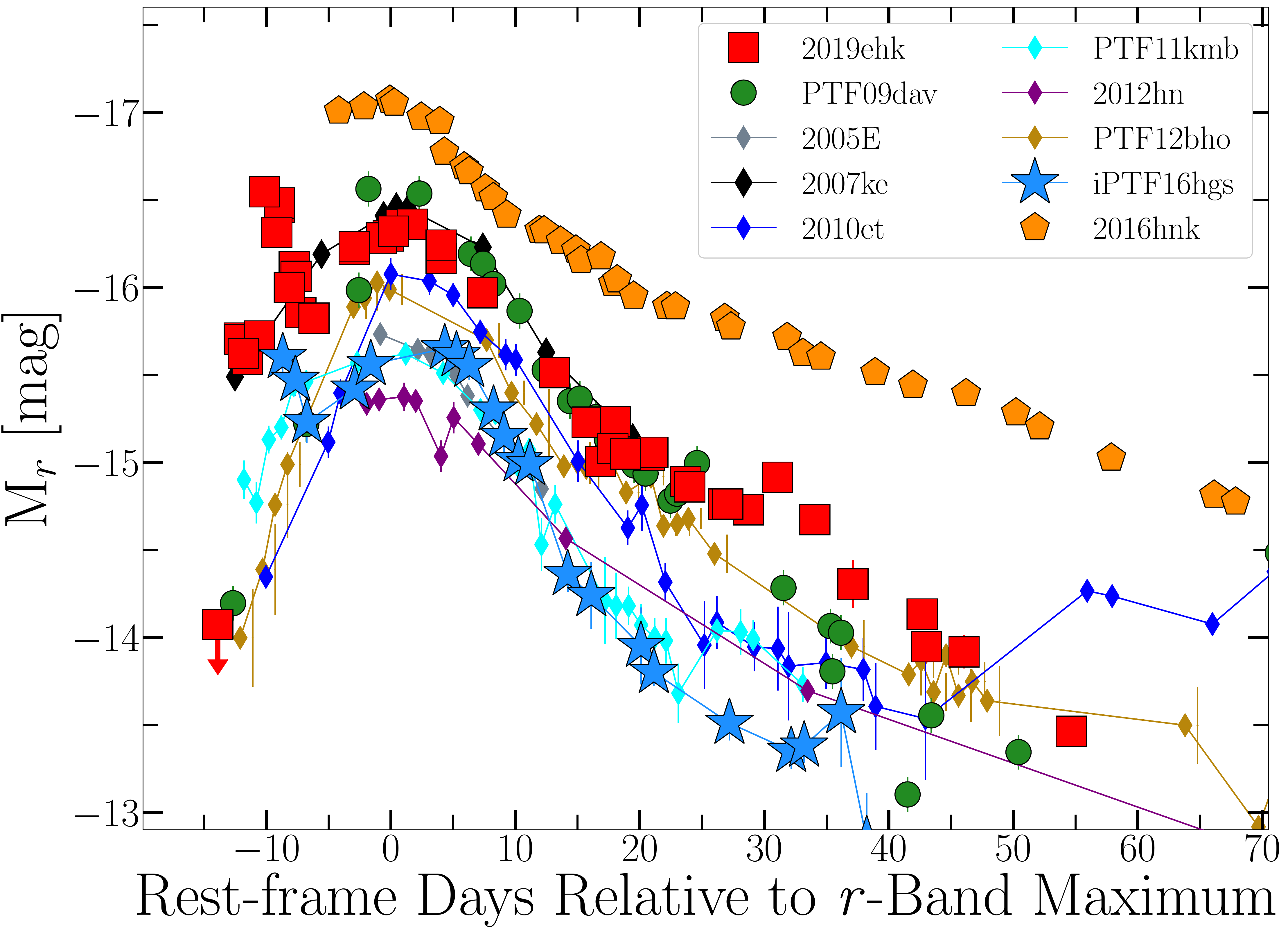}}
\subfigure[]{\includegraphics[width=0.48\textwidth]{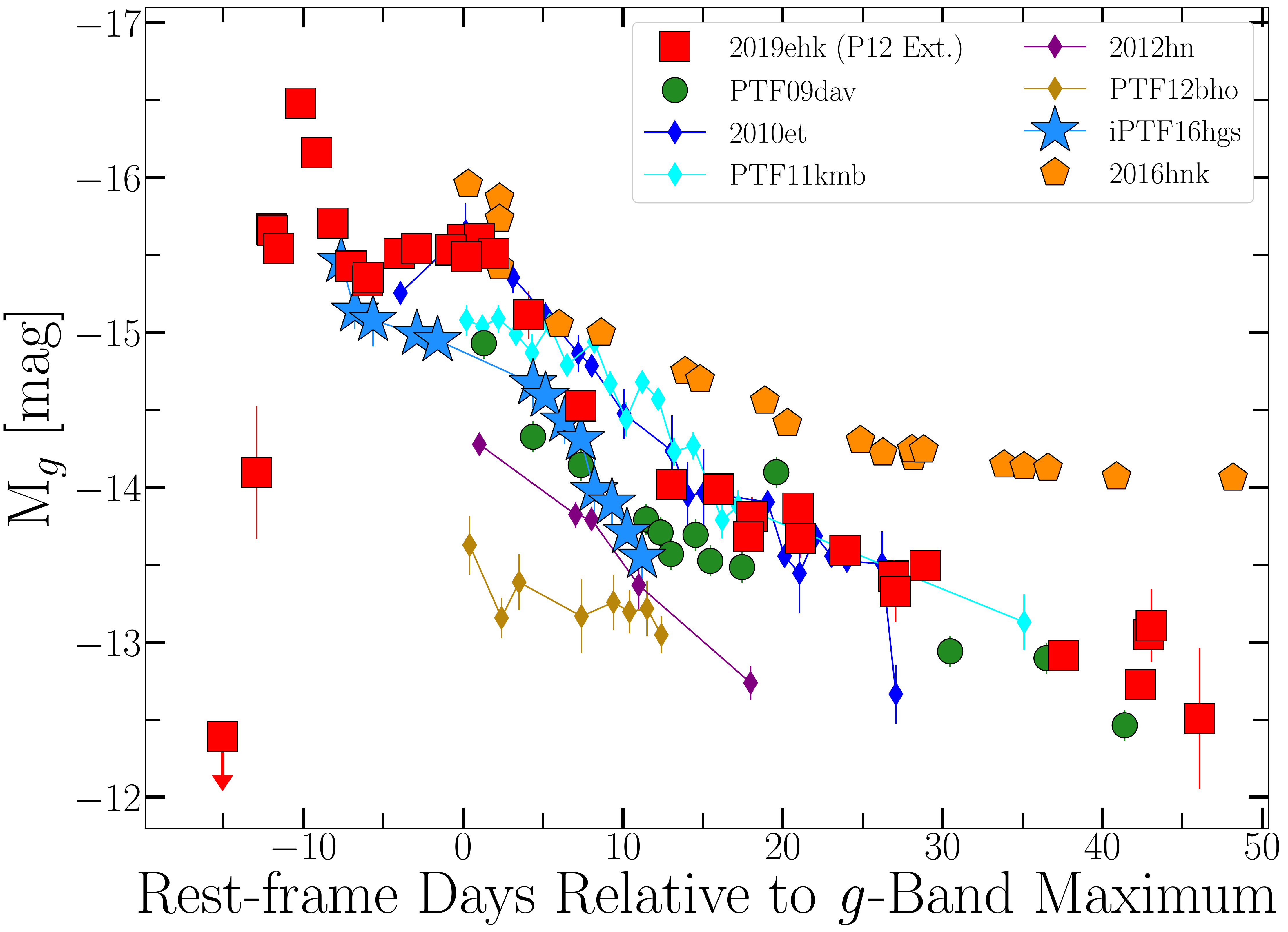}}
\caption{(a) Early-time $r-$band comparison of SN~2019ehk (red squares) and classified \cas. The peculiar, ``calcium-strong'' SN~2016hnk also presented for reference (orange polygons). SN~2019ehk is now the second object in this class to show a double-peaked light curve, iPTF16hgs (light blue stars) being the first. (b) $g-$band comparison of SN~2019ehk (red squares) and classified \cas. \label{fig:ca_rich_compare}}
\end{figure*}

SN~2019ehk is the third observed \ca \ with a double-peaked optical light curve (the others being iPTF~16hgs, \citealt{de18} and SN~2018lqo, \citealt{de20}). Consequently, we define its phase relative to the secondary, ``Nickel-powered'' peak and discuss the potential power sources of the first peak in \S\ref{sec:flare}. We fit a low-order polynomial to the SN~2019ehk light curve to find best fit $B$- and $r$-band peak absolute magnitudes of $M_B = -15.1 \pm 0.0210$~mag at MJD $58615.15\pm0.1$ and $M_r = -16.36 \pm 0.01$~mag at MJD $58616.18\pm0.2$, respectively. We calculate a \cite{phillips93} decline parameter value of $\Delta \rm{m}_{15}(B) = 1.71 \pm 0.0140$~mag from our $B$-band light curve fits. We calculate a rise-time of $t_r = 13.4 \pm 0.210$~days using the adopted times of explosion and $B$-band peak.  

\begin{figure}[h]
\centering
\includegraphics[width=0.45\textwidth]{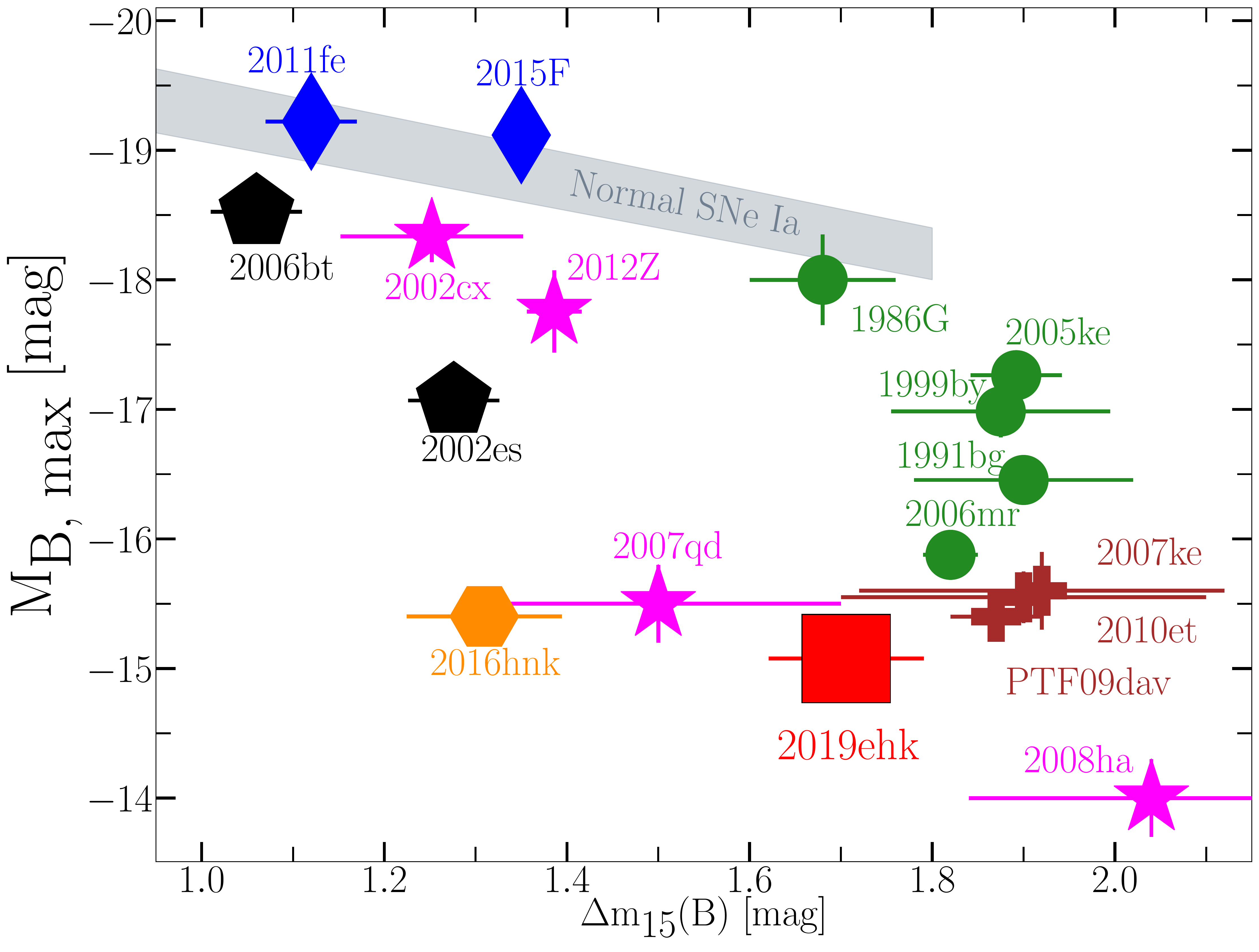} \caption{ $\Delta$m$_{15}$ vs. M$_{B,\textrm{max}}$ for SN~2019ehk (red square), normal SNe Ia (diamonds + grey region), 91bg-like SNe Ia (circles), SNe Iax (stars), 02es-like SNe Ia (pentagons), other \cas\, (plus signs), peculiar thermonuclear SN~2006bt (pentagon) and ``Calcium-strong'' SN~2016hnk (hexagon). Some uncertainities on M$_{B, \textrm{max}}$ are smaller than plotted marker size.  \label{fig:dm15}} 
\end{figure}

We present $r-$ and $g-$band light curve comparisons of SN~2019ehk and \cas \ in Figures \ref{fig:ca_rich_compare}(a)/(b). Overall, SN~2019ehk has comparable light curve evolution to other confirmed \cas: $t_r<15$~days and declines in luminosity at a similar rate. SN~2019ehk is less luminous in $r-$band than "Calcium-strong" SNe~2016hnk \citep{galbany19, wjg19} and PTF09dav \citep{sullivan11}, but has a similar light curve evolution to the next most luminous \ca \ SN~2007ke \citep{lunnan17}. Furthermore, its $r-$band evolution is consistent with iPTF16hgs \citep{de18}, the only other \ca \ with a confirmed double-peaked light curve. This duplicate first light curve peak may indicate similar underlying physics between the two objects despite the fact that SN~2019ehk is $\sim$1~mag more luminous than iPTF16hgs and declines at a slower rate. Additionally, we present $\Delta \rm{m}_{15}(B)$ vs. $M_B$ for SN~2019ehk with respect to \cas \ and other thermonuclear varieties in Figure \ref{fig:dm15}. From this comparison, SN~2019ehk is broadly consistent with the \ca \ class due to its $B-$band light curve evolution from peak out to 15 days. SN~2019ehk is clearly distinct from normal and sub-luminous SNe Ia/Iax, but has a comparable \cite{phillips93} decline parameter value to 91bg-like SNe~Ia. 

\begin{figure*}
\centering
\subfigure[]{\includegraphics[width=0.49\textwidth]{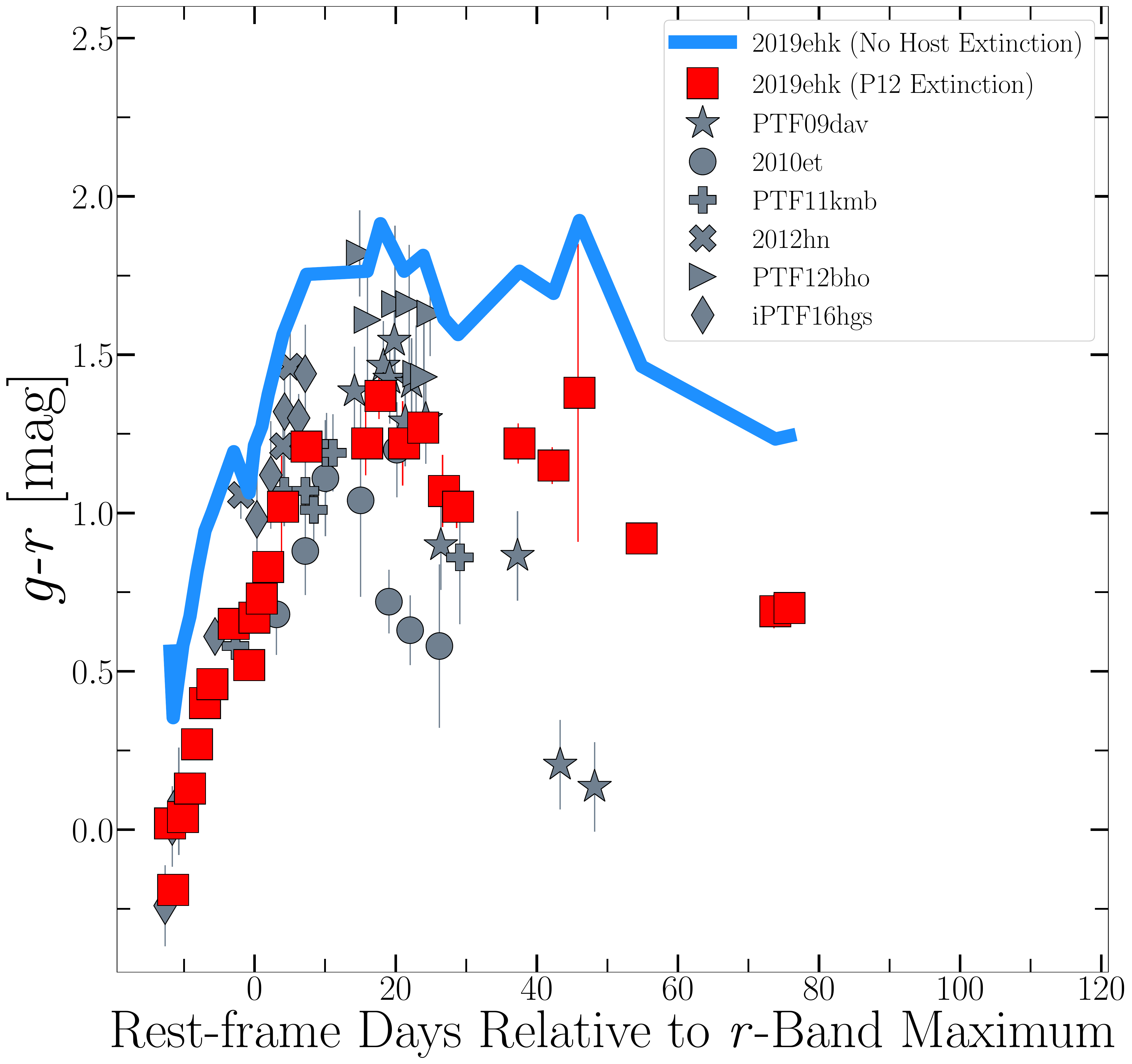}}
\subfigure[]{\includegraphics[width=0.49\textwidth]{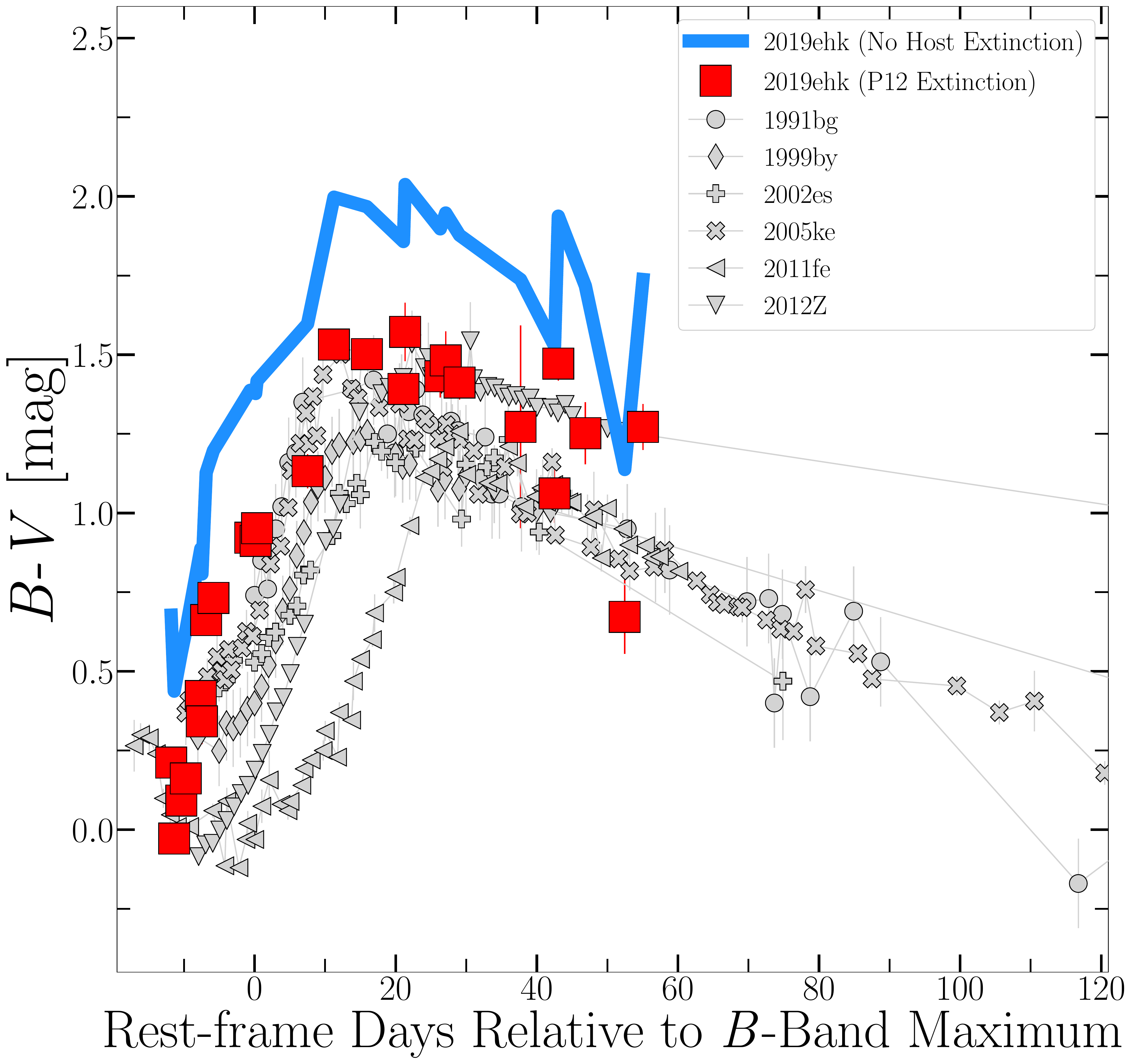}}\\
\subfigure[]{\includegraphics[width=0.49\textwidth]{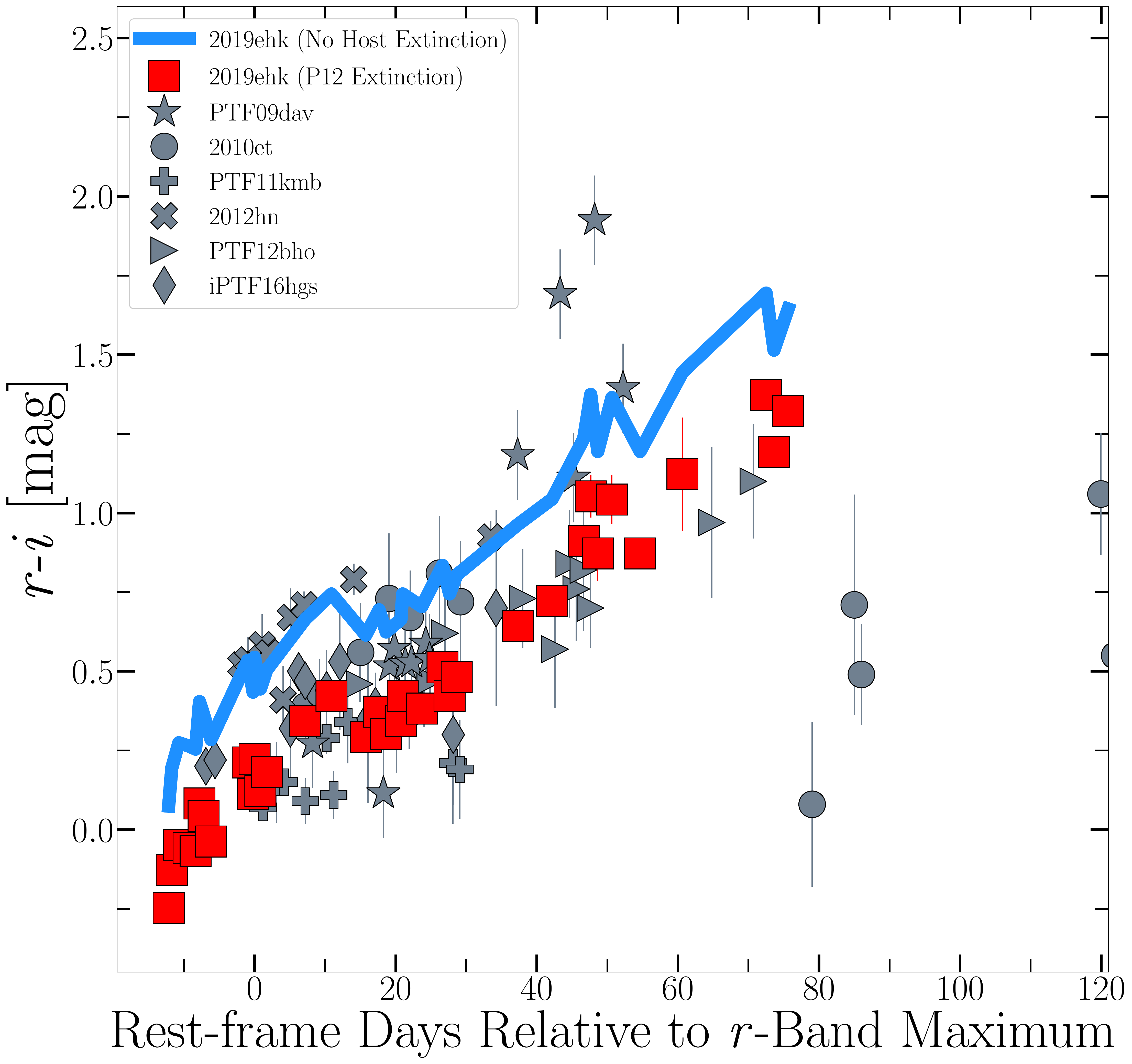}}
\subfigure[]{\includegraphics[width=0.49\textwidth]{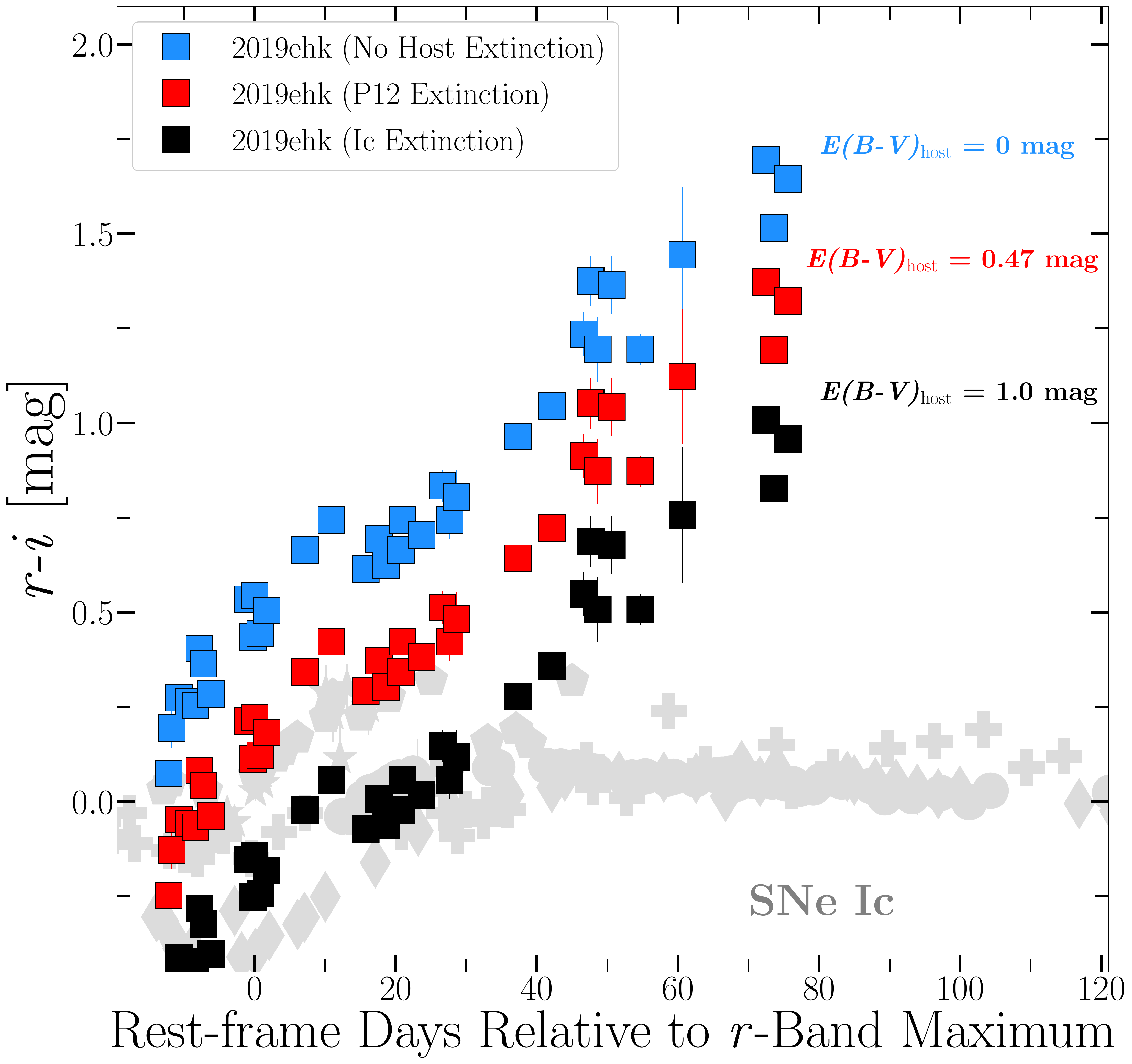}}
\caption{(a) \textit{g-r} color comparison of SN~2019ehk  and \cas. SN~2019ehk colors from photometry are presented as a blue line. The red squares represent the photometric colors that have been de-reddened according to the \cite{Poznanski12} (P12) extinction relation and host galaxy reddening $E(B-V) = 0.47$. (b) \textit{B-V} color comparison of SN~2019ehk and various types of SNe Ia. (c) \textit{r-i} color comparison of SN~2019ehk and \cas. (d) SN~2019ehk's \textit{r-i} color evolution for different levels of host extinction: 0~mag (black), 0.47~mag (red) and 1.0~mag (blue). These are compared to the \textit{r-i} colors of a sample of type Ic SNe (grey).   \label{fig:colors}}
\end{figure*}

In Figure \ref{fig:colors}, we present $g-r$, $B-V$ and $r-i$ color comparison plots of SN~2019ehk, \cas, SNe Ia/Iax and SNe Ic. Given the relative uncertainty on SN~2019ehk's host-galaxy extinction, we display color curves that have no host extinction applied (red squares) as well as colors where the adopted value of $E(B-V)=0.47$~mag is used to correct for extinction (blue line). As shown in Figures \ref{fig:colors}(a)/(c), SN~2019ehk's de-reddened colors are consistent to within 0.1~mag in $g-r$ and 0.2~mag in $r-i$ of the typical \cas; all objects exhibiting a noticeably ``red'' color evolution. Consequently, SN~2019ehk's intrinsically red colors deviate significantly from all flavors of SNe~Ia shown in Figures \ref{fig:colors}(b). SN~2019ehk is $\sim$0.3~mag redder than the reddest SN~Ia, 2005ke and SN~Iax, 2012Z. 

\begin{figure*}
\centering
\subfigure[]{\includegraphics[width=0.49\textwidth]{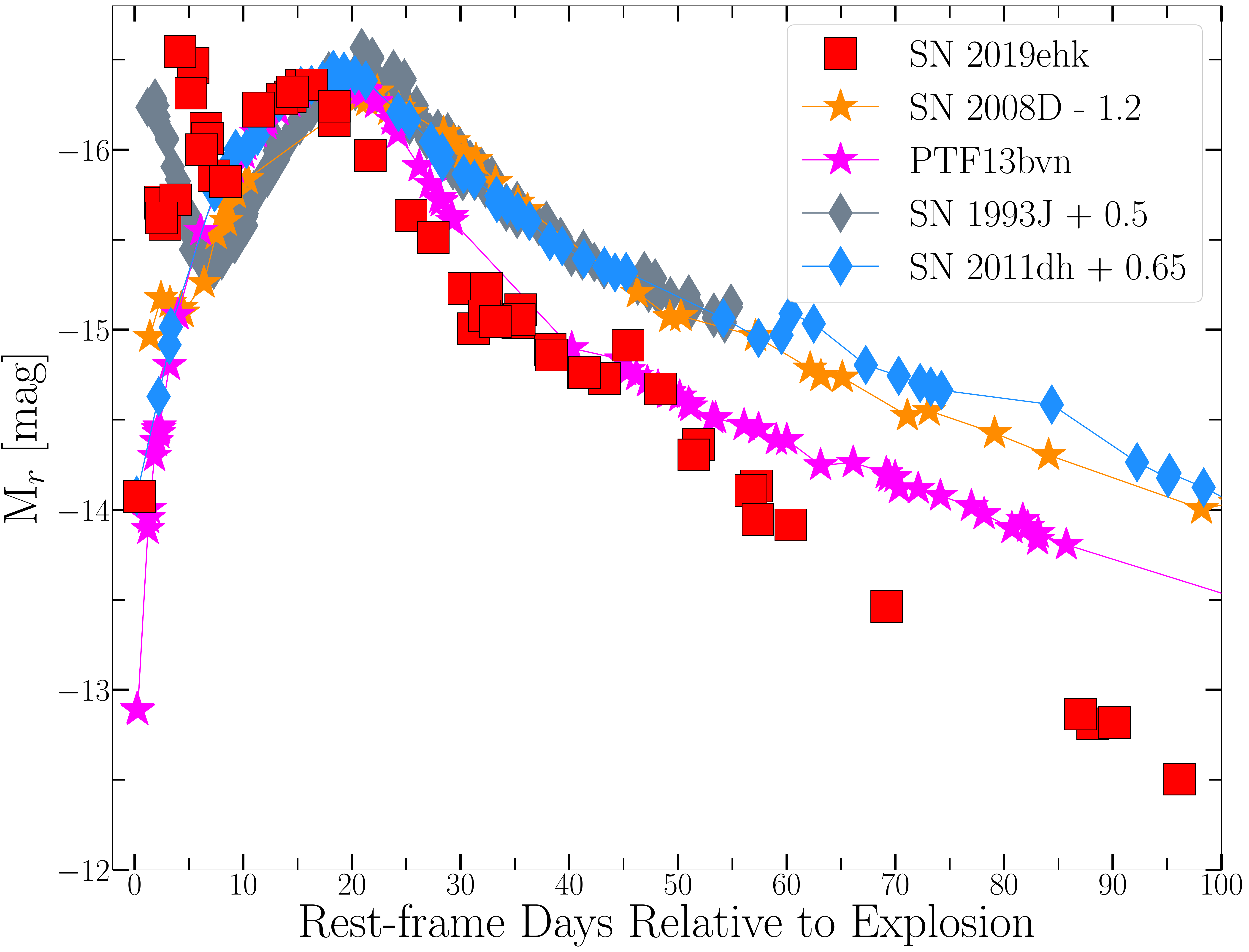}}
\subfigure[]{\includegraphics[width=0.49\textwidth]{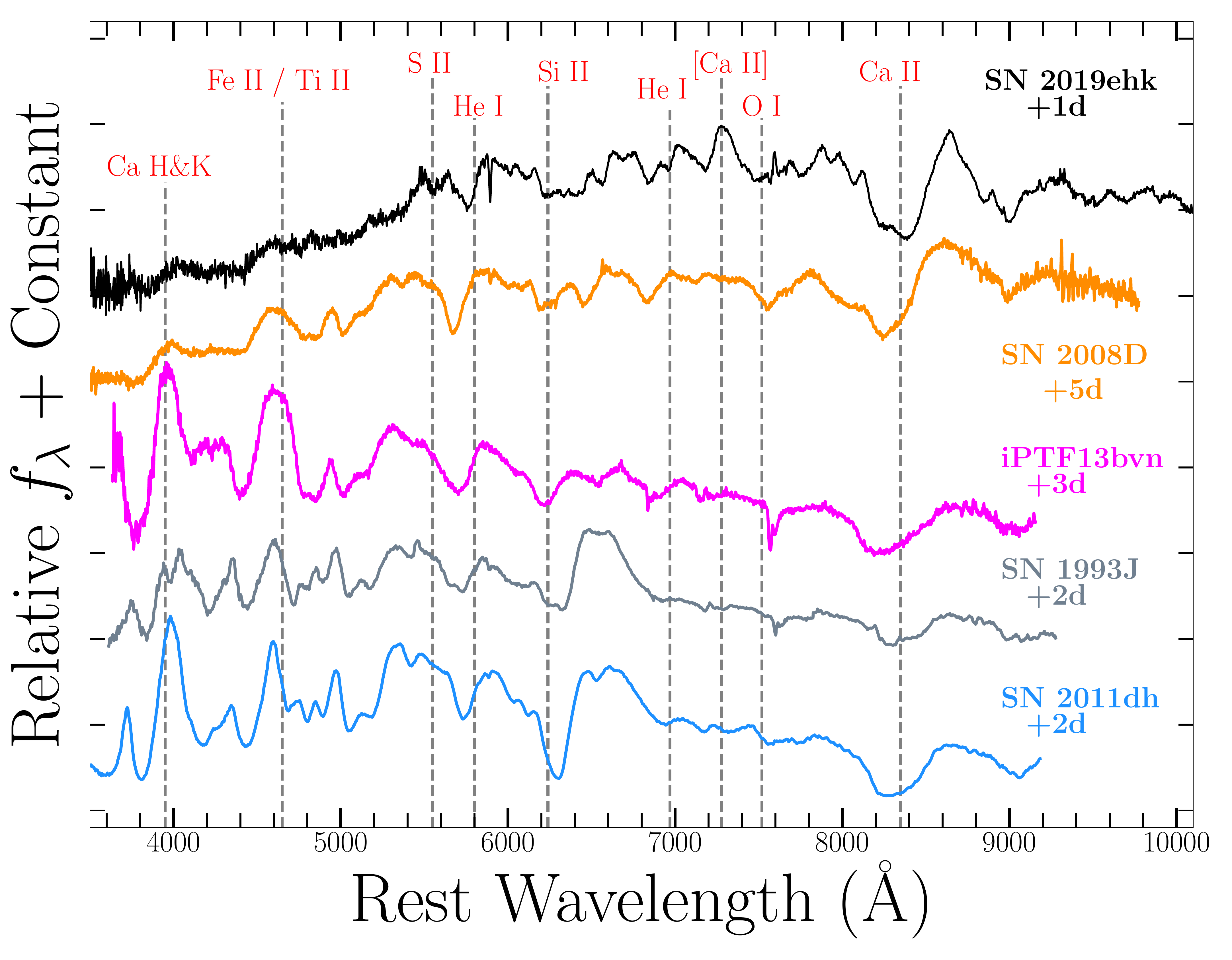}}
\caption{(a) Photometric comparison of SN~2019ehk (red squares) with respect to SNe Ib (stars; \citealt{soderberg08, Malesani09, modjaz09,fremling16}) and SNe IIb (diamonds; \citealt{wheeler93, arcavi11}). (b) Spectral comparison of SN~2019ehk (without reddening correction) and SNe Ib/IIb. While there are some individual similarities between SN~2019ehk and SNe Ib/IIb, the apparent contrast in its photometric and spectral evolution is indicative of different underlying explosion physics, which then distances this SN from a  SN~Ib/IIb classification. \label{fig:IIb_compare}}
\end{figure*}

We present $r-$band light curve comparisons of SN~2019ehk and type IIb/Ib SNe (SNe IIb/Ib) in Figure \ref{fig:IIb_compare}(a). SN~2019ehk has a similar peak magnitude to SN~Ib, iPTF13bvn and a higher peak magnitude than prototypical SN~Ib, SN~2008D. While SN~2019ehk's $r-$band evolution is quite similar to iPTF13bvn, it has a significantly shorter rise-time than any SNe~Ib. With respect to SNe IIb, SN~2019ehk is less luminous at peak and evolves faster than both SNe~1993J and 2011dh. Furthermore, the first light curve peak observed in SNe~IIb occurs on a longer timescale ($\sim$ $10-15$~days) than that observed in SN~2019ehk ($\sim$5~days). The first peak in these SNe is also typically less luminous than the secondary maximum, which is reversed in SN~2019ehk. However, the double-peaked light curve in SN~2019ehk may be physically connected to a explosion scenario wherein the SN shock ``breaks out'' into an extended envelope, which then rapidly cools. Such a mechanism has been invoked as an explanation for the primary peak in SNe~IIb and we further discuss this model in  \S\ref{sec:flare_shockcool}.

\subsection{Pseudo-Bolometric Light Curve}\label{subsec:bol_LC}

\begin{figure*}
\centering
\subfigure[]{\includegraphics[width=0.48\textwidth]{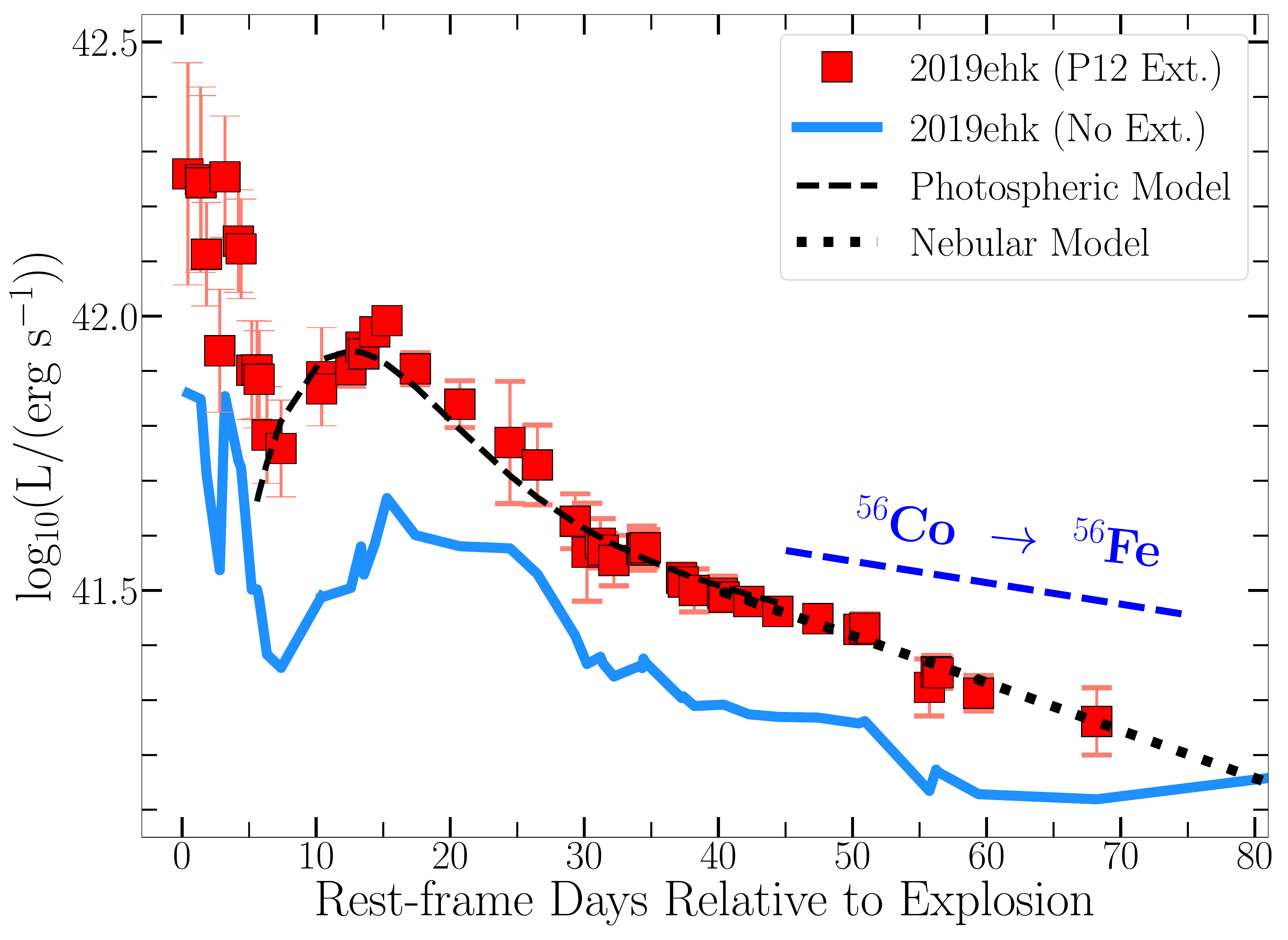}}
\subfigure[]{\includegraphics[width=0.46\textwidth]{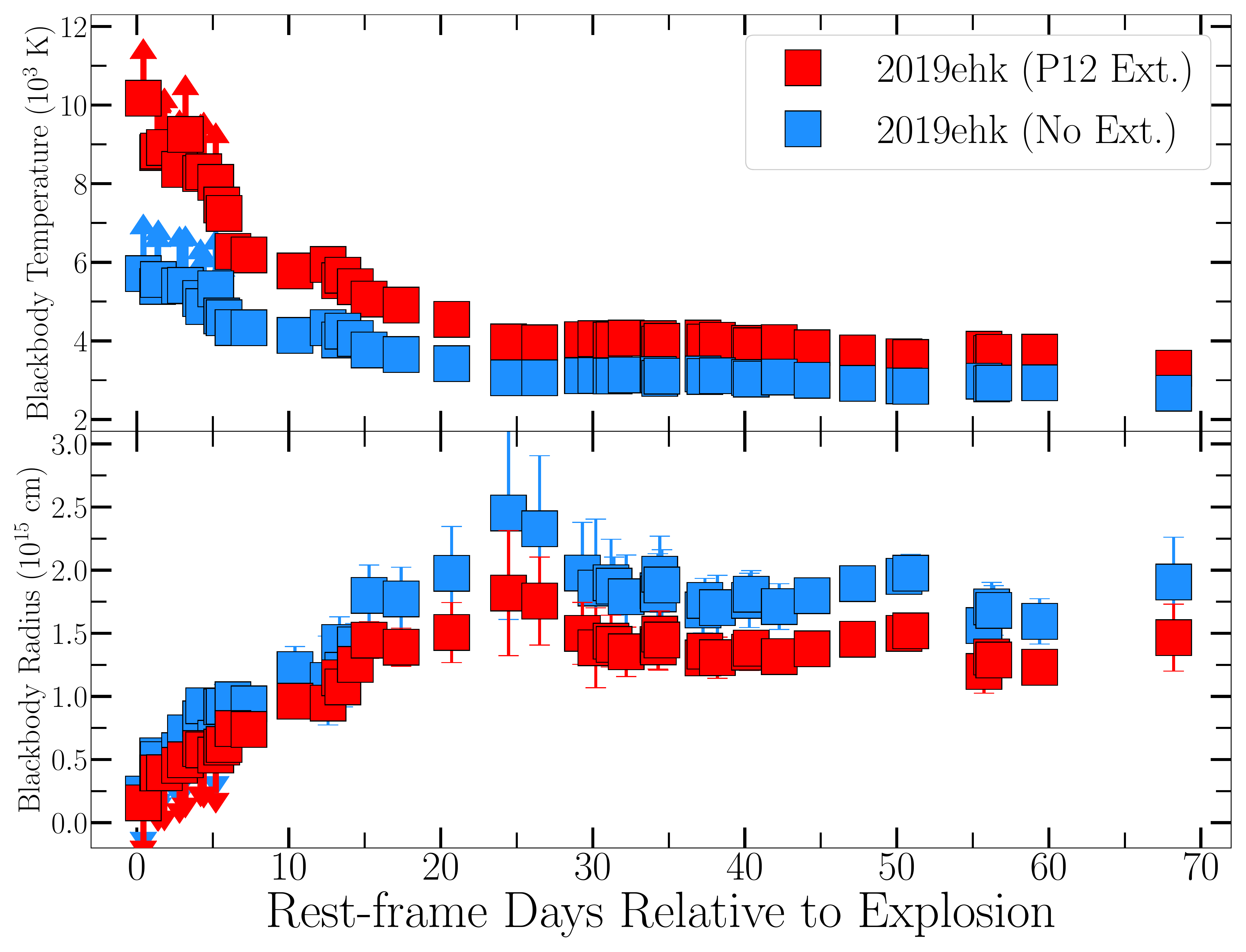}}
\caption{(a) Pseudo-bolometric light curve of SN~2019ehk for different host-galaxy reddening: $E(B-V)$ = 0.47 (red squares) and $E(B-V)$ = 0.0 (blue line). Points at $t < 6$d were calculated using a linearly increasing photosphere radius (e.g., see \S\ref{sec:flare}, Fig. \ref{fig:flare_lum}) Separate photospheric light curve models for the early-time light curve (\S\ref{subsec:bol_LC}) are plotted as dashed black line. Modeling of the nebular phase data plotted as dotted black line. (b) Blackbody radii and temperatures derived from SED modeling of all multi-color optical photometry. Red squares indicate a host extinction correction of $E(B-V)$ = 0.47 and blue squares indicate $E(B-V)$ = 0. Radii and temperatures at $t < 6$ days are displayed as upper and lower limits, respectively. \label{fig:BB_LC}}
\end{figure*}

We construct a pseudo-bolometric light curve by fitting the broad-band photometry with a blackbody model that is dependent on radius and temperature. Each spectral energy distribution (SED) was generated from the combination of multi-color optical photometry in $uBVcgoriz$ bands (3000-9000\AA). In regions without complete color information, we extrapolated between light curve data points using a low-order polynomial spline. We present SN~2019ehk's bolometric light curve in addition to its blackbody radius and temperature evolution in Figure \ref{fig:BB_LC}. We display the inferred blackbody luminosities, temperatures and radii that resulted from both host-galaxy extinction corrected photometry and non-corrected photometry. All uncertainties on blackbody radii and temperature were calculated using the co-variance matrix generated by the SED fits. It should be noted that the blackbody approximation breaks down when emission lines begin to dominate the spectrum of SN~2019ehk at  $t > 30$ days after explosion. Therefore a blackbody assumption for SN~2019ehk at late-times is most likely an over-simplification and could result in additional uncertainty on the presented bolometric luminosities and the resulting estimates on physical parameters of the SN. For the secondary, Nickel-powered light curve peak, we find a peak bolometric luminosity of $(9.81 \pm \: 0.15) \times 10^{41} \: \mathrm{erg\:s^{-1}}$.


In order to determine physical parameters of the explosion, we model the bolometric light curve with the analytic expressions presented in Appendix A of \cite{valenti08}. We exclude the first light curve peak from this analysis and model the bolometric evolution of SN~2019ehk for $t>8$ days post-explosion. These models are divided into two distinct parts: the photospheric phase ($t < 30$~days past explosion), which is based on \cite{arnett82} and the nebular phase ($t > 60$~days past explosion), which is derived from prescriptions outlined in \cite{sutherland84} and \cite{Cappellaro97} (however see \citealt{wheeler15} for corrected \citealt{arnett82} equations). Furthermore, this analytic formalism self-consistently implements the possibility of incomplete $\gamma$-ray trapping in the expanding SN ejecta throughout the modeling process. A typical opacity of $\kappa = 0.1$ cm$^2$ g$^{-1}$ is applied in each model. The free parameters of each model are kinetic energy ($E_{\rm k}$), total mass of synthesized ${}^{56}\textrm{Ni}$ ($M_{\textrm{Ni}}$), and ejecta mass ($M_{\rm ej}$). However, there is a known degeneracy within these models between kinetic energy and ejecta mass:

\begin{equation}
    M_{\rm ej} = \frac{10}{3} \frac{E_{\rm k}}{v^2}
\end{equation}

\noindent
where we follow standard practice and use $v$$\approx$$v_{ph}$, i.e. the photospheric velocity at peak. We use $v_{ph} \approx 6500 \ \kms$, which is estimated from \ion{Si}{ii} absorption at peak. Our photospheric and nebular models are presented in Figure \ref{fig:BB_LC}(a) as the dashed and dotted lines, respectively. From these models, we calculate   $M_{\textrm{Ni}}=(3.1 \pm 0.11) \times 10^{-2} \ \Msun$,  $E_{\rm k}=(1.8 \pm 0.1) \times 10^{50}$~erg and $M_{\rm ej}=(0.72 \pm 0.04) ~ \Msun$. We discuss the modeling of the first light curve peak in \S\ref{sec:flare_ni}. Furthermore, we show that the nebular phase light curve decline is slightly faster than the typical decay of ${}^{56}\textrm{Co}$ $\rightarrow$ ${}^{56}\textrm{Fe}$ that assumes complete trapping of $\gamma$-rays. 

In Figure \ref{fig:BB_LC}(b), we present the evolution of SN~2019ehk's blackbody radius and temperature for different extinction values from 0.44 to 73.2~days after explosion. For phases 0.44-6d, it should be noted that the peak of the blackbody curve is not visible in our model fits i.e., the blackbody peaks in the near-to-far UV. Thus we cannot be confident that the reported blackbody radii and temperatures during these times are completely accurate. As is further discussed in \S\ref{sec:flare_specifics}, these specific radii and temperatures are best treated as upper and lower limits, respectively. 

At the time of first detection in $g-$band, SN~2019ehk had a minimum blackbody temperature of $\gtrsim 10,200$~K and a maximum initial radius of $\leq 1.6 \times 10^{14}$~cm ($2300 \ \Rsun$). This was conservatively calculated by assuming no color evolution between first and second epochs and then fitting a blackbody model. We can thus better constrain the initial radius at $t=0.44$d by fitting a blackbody model to  the initial $g-$band detection for a range of fiducial temperatures $T=(2-4)\times10^4$ K. In this case we find photospheric radii of $7 - 4 \times 10^{13}$ cm (1000-500 $\Rsun$). Extended progenitors for SN~2019ehk are ruled out in \S\ref{subsec:HR_progenitors}. Therefore, considering a compact massive progenitor with radius of $\sim$10~$\Rsun$,  we estimate a shock velocity of $v_s \approx 1.8 \times 10^{4} \ \kms$ in order to reach a blackbody radius of $7\times 10^{13}$ cm at $t=0.44$d. This is also a reasonable estimate for shock breakout from a WD progenitor. Because the shock could be ahead of the photosphere, we consider $v_s$ to be a lower limit on the true shock velocity, which is consistent with being larger than the photospheric velocities derived from SN~2019ehk spectra.

\section{Optical/NIR Spectral Analysis}\label{sec:spectro_analysis}

\begin{figure}[h]
\centering
\includegraphics[width=0.45\textwidth]{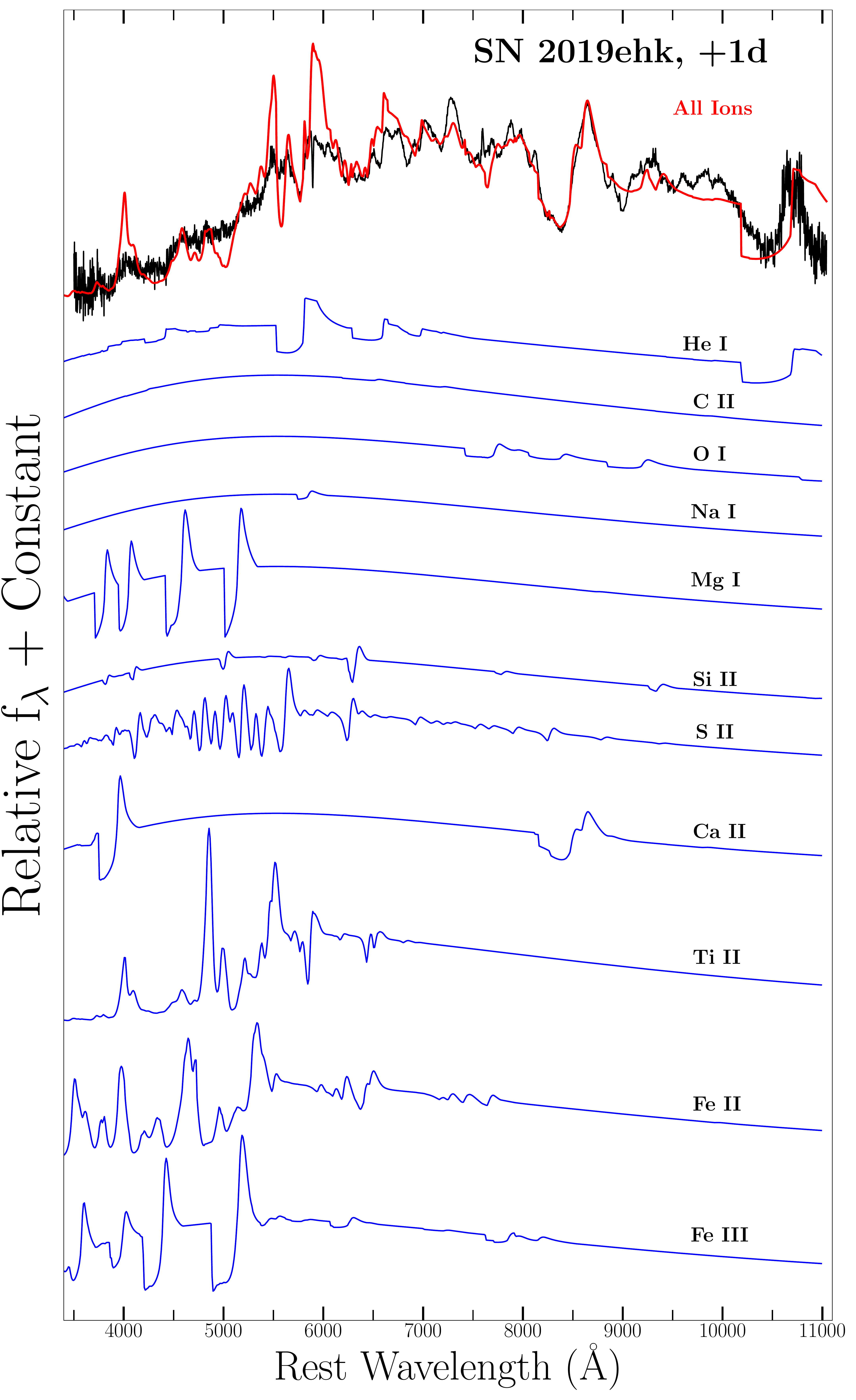} \caption{ Decomposition of active ions in \texttt{SYNAPPS} fit. Phase relative to second \textit{B}-band maximum. Total fit is shown in red, while blue lines mark each individual ion's contribution. \label{fig:synapps}}
\end{figure}

\subsection{Spectroscopic Properties}\label{subsec:spec_analysis}

We model the SN~2019ehk spectrum near peak in order to understand the chemical composition of the explosion. To do this, we utilize the spectral synthesis software \texttt{SYNAPPS} \citep{thomas11}, which is dependent on generalized assumptions about the SN such as spherical symmetry, local thermal equilibrium, and homologous expansion of ejecta. We present a \texttt{SYNAPPS} fit to the +1 day spectrum as the red line in Figure \ref{fig:synapps}. As shown in blue, we detect the following species in SN~2019ehk near peak: \ion{He}{i}, \ion{C}{ii}, \ion{O}{i}, \ion{Na}{i}, \ion{Mg}{i}, \ion{Si}{ii}, \ion{S}{ii}, \ion{Ca}{ii}, \ion{Ti}{ii}, \ion{Fe}{ii} and \ion{Fe}{iii}. While the \ion{C}{ii} absorption is weak relative to the continuum, it does appear to be contributing to the overall flux near $\lambda\lambda$6580,7234. The model also appears to be over-producing the line flux between 5500-6000\AA, which we attribute to possible deficiency in fitting species such as \ion{S}{ii}, \ion{Ti}{ii} and \ion{Fe}{ii}. However, the overall spectral profiles are matches in that region, which allows us to conclude that those ions are in fact present in the SN ejecta.  

We perform additional spectral modeling to explore the possibility that hydrogen or exotic Fe-group elements such as \ion{Cr}{ii}, \ion{Sc}{ii} and \ion{Sr}{ii} are present in SN~2019ehk. After multiple iterations of \texttt{SYNAPPS} modeling, we find no detectable H$\alpha$ or Balmer series lines in the maximum light spectrum. Furthermore, the addition of \ion{Cr}{ii}, \ion{Sc}{ii} and \ion{Sr}{ii} to our \texttt{SYNAPPS} models does not improve the overall fit, specifically bluewards, and thus we cannot claim a confident detection of these ions. All identified ions in SN~2019ehk are typical of canonical \cas \ (e.g., 2005E-like) and indicate a similar chemical composition to be expected for an object within the class.

We track the expansion velocity of different ions through modeling of P-Cygni and pure emission line profiles. We estimate the photospheric velocities of various ions from first detection of spectral line formation at -9d to the last pre-nebular spectrum taken at +59d relative to the second $B-$band peak. At -9d, the fastest moving ions in the SN ejecta is \ion{Si}{ii} at $-11700 \pm 250 \ \kms $ and \ion{Ca}{ii} at $-10400 \pm 300 \ \kms $; this is measured from the fitted minimum of the $\lambda 6355$ absorption profile. These profiles, including \ion{O}{i} and \ion{He}{i}, show similar declines in velocity as the SN expands and becomes optically thin. We also measure \ion{Ca}{ii} and [\ion{Ca}{ii}] velocities from the FWHM of the $\lambda 8542$ and $\lambda 7291$ profiles, which remain approximately constant out to nebular times at $\sim 9000~\kms$ and $\sim 6000~\kms$, respectively.

\begin{figure*}
\centering
\subfigure[]{\includegraphics[width=0.49\textwidth]{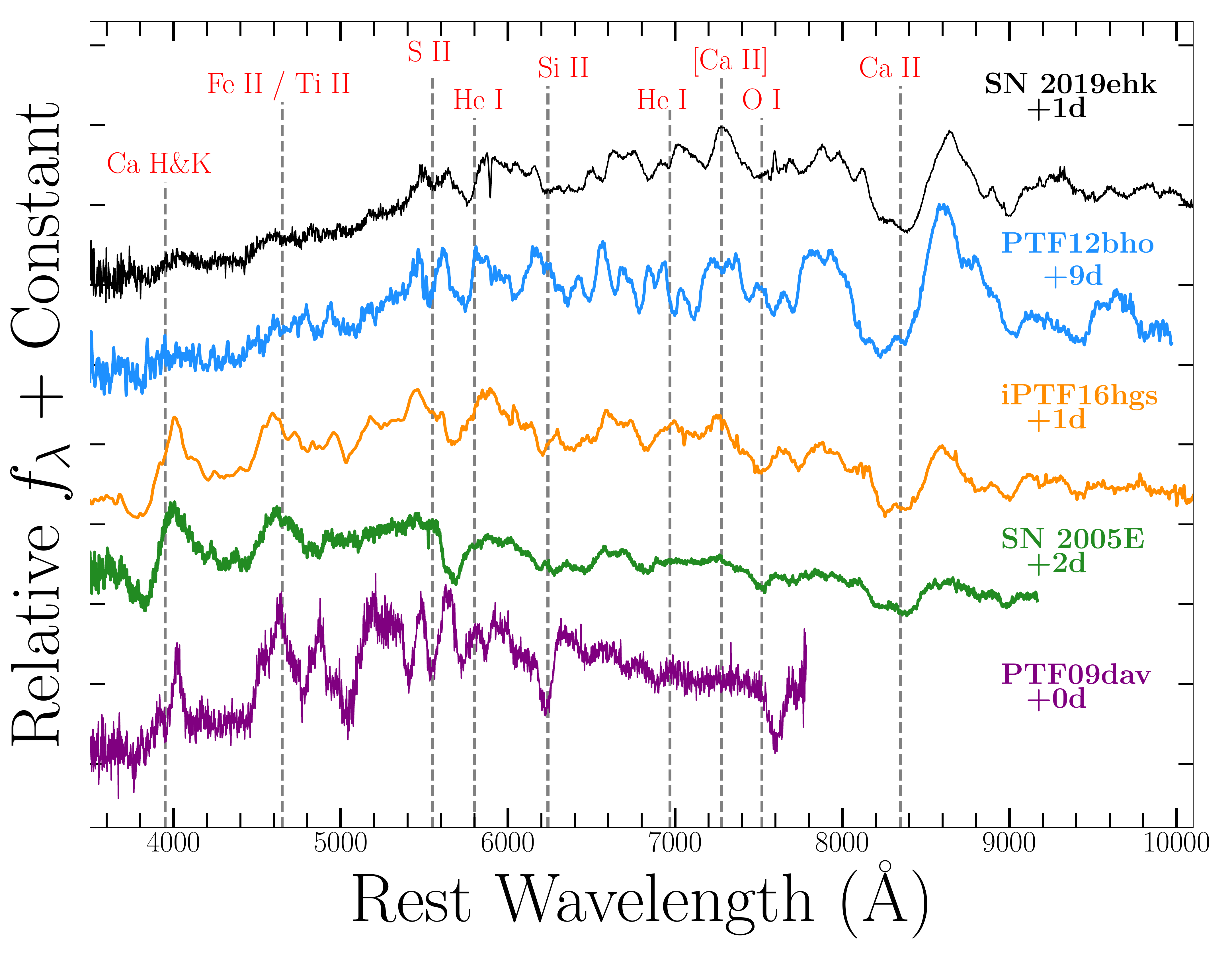}}
\subfigure[]{\includegraphics[width=0.49\textwidth]{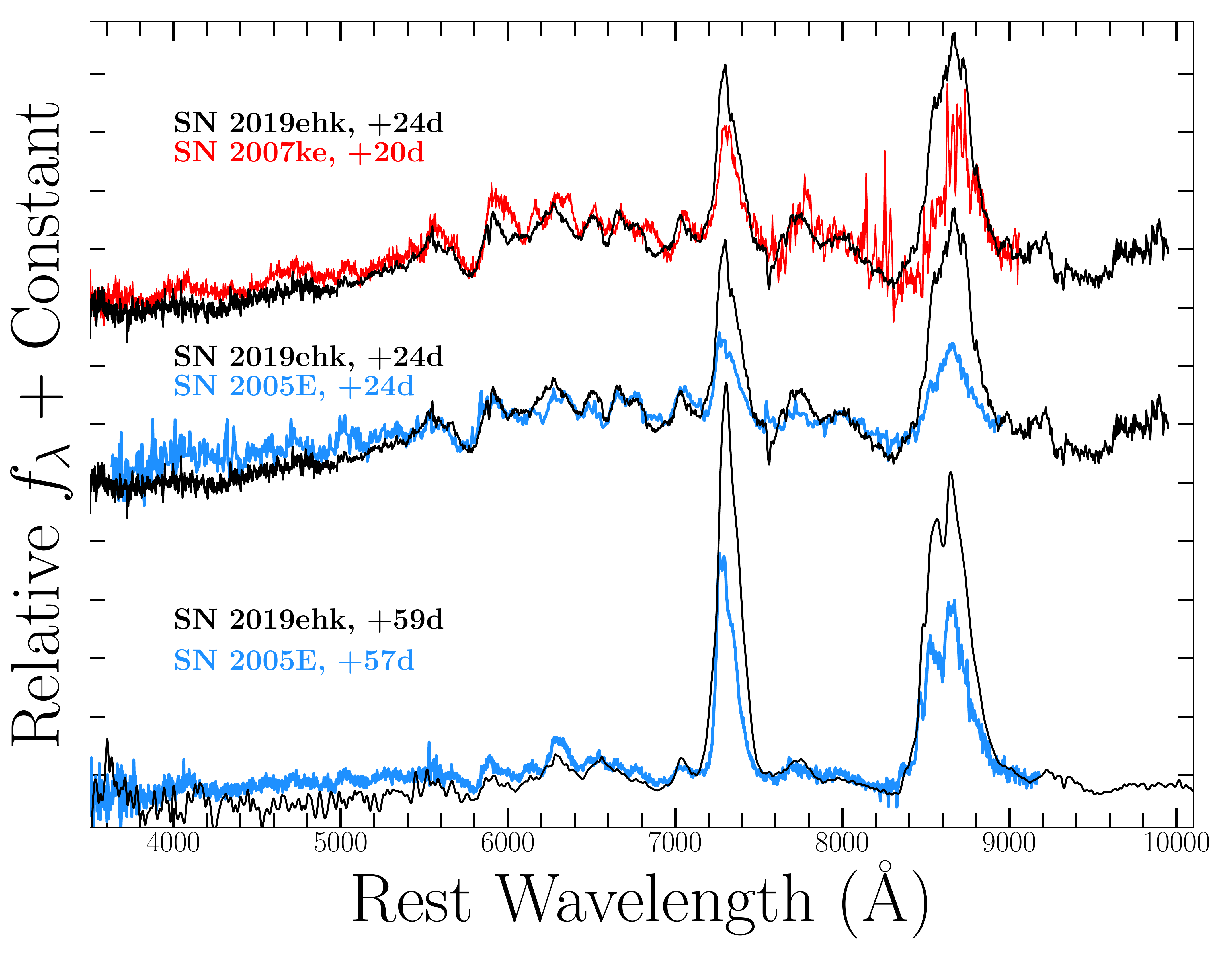}}
\caption{(a) Spectral comparison of SN~2019ehk (black) and other \cas \ at approximately the same phase \citep{perets05, sullivan11, lunnan17, de18}. Common ions are marked by grey lines.  (b) Direct spectral comparison of SN~2019ehk (black) and \cas \ SNe~2007ke and 2005E at approximately the same phase \citep{perets05, lunnan17}. Almost every line transition is matched between spectra, with SN~2019ehk showing stronger \ion{Ca}{ii} emission than both other objects. \label{fig:carich_spectra}}
\end{figure*}

In the +38d NIR/IR spectrum of SN~2019ehk (Fig. \ref{fig:IR_spectrum}), we identify similar ions to those found in our optical spectral modeling: \ion{He}{i}, \ion{C}{i}, \ion{Mg}{i} and \ion{Ca}{ii}. We present the velocity profiles of \ion{He}{i} $\lambda\lambda10850,20587$ in Figure \ref{fig:IR_spectrum-v}. Both IR \ion{He}{i} lines have identical P-Cygni line profiles, with $\lambda10850$ showing a strong emission component and faster absorption minimum. The FWHM of the $\lambda10850$ line is $7036 ~ \kms$ and the $\lambda20587$ line is $5700 ~ \kms$. 

We present early-time spectral comparisons of SN~2019ehk and other \cas \ in Figure \ref{fig:carich_spectra}. Near (second) maximum light, SN~2019ehk is most similar visually to PTF12bho \citep{lunnan17} and iPTF16hgs \citep{de18}. All three objects have strong \ion{Ca}{ii} absorption, prominent \ion{He}{i} profiles and the fast emergence of a [\ion{Ca}{ii}] profile relative to peak. SN~2019ehk and PTF12hbo both show little bluewards flux from Fe-group elements, which is unlike the prominent Fe-group transitions seen in iPTF16hgs, SN~2005E \citep{perets10} and PTF09dav \citep{sullivan11}. This may indicate either a low total Nickel mass (typical for these objects) or variation in the mixing of Fe-group elements in the outer layers of SN ejecta. This process can then result in the suppression of bluewards flux. 

\begin{figure*}
\centering
\includegraphics[width=\textwidth]{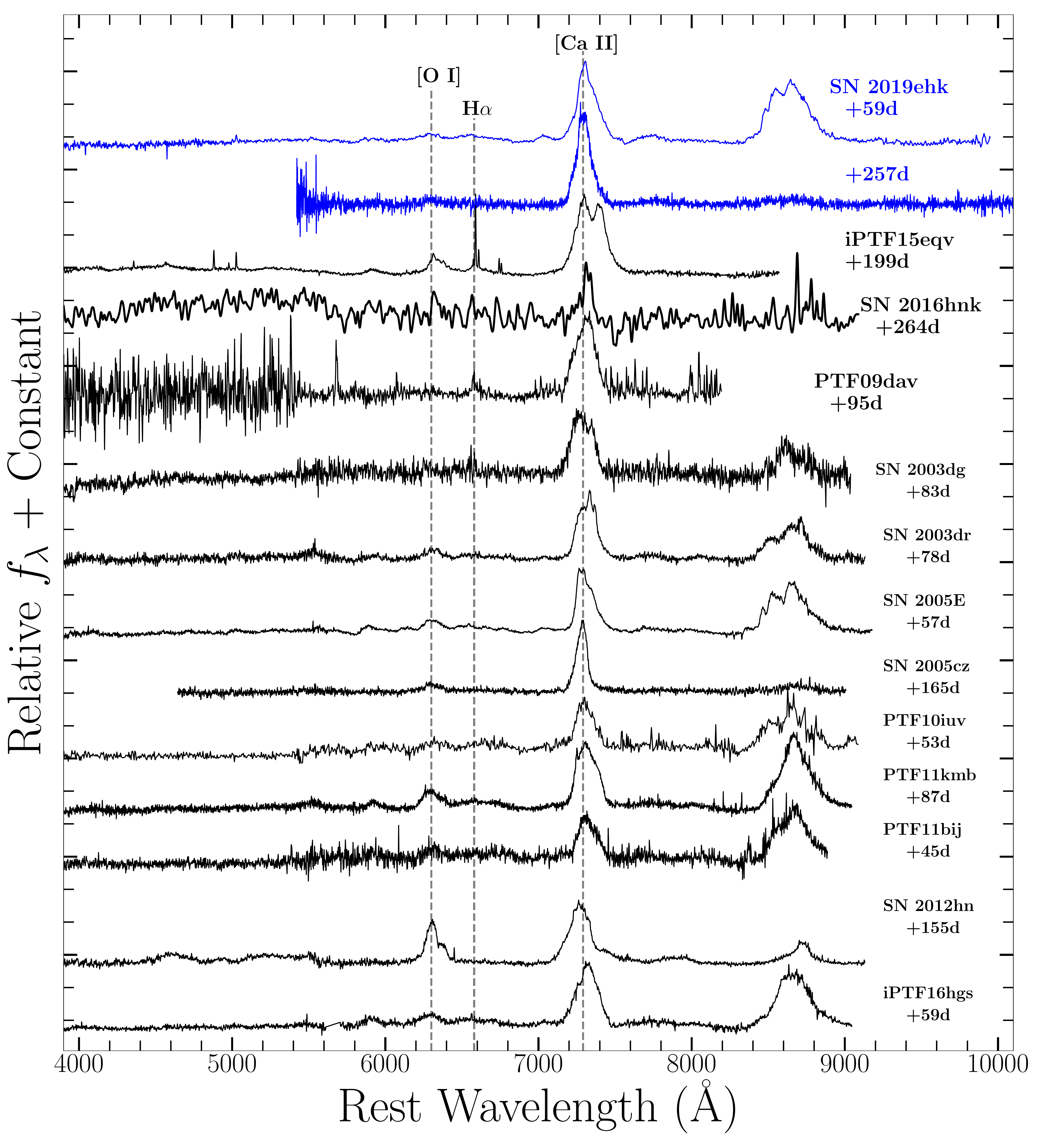}
\caption{Nebular spectra of all classified \cas. Pre-nebular (+59d) and fully nebular (+257d) spectra of SN~2019ehk shown in blue. Prominent [\ion{O}{i}] and [\ion{Ca}{ii}] lines as well as H$\alpha$ marked by dashed grey lines. \label{fig:ca-rich_nebular}}
\end{figure*}

As shown in Figure \ref{fig:carich_spectra}(b), SN~2019ehk is nearly identical to SNe~2007ke and 2005E near +24 days after second maximum light. These pre-nebular spectra are dominated by [\ion{Ca}{ii}] and \ion{Ca}{ii} emission but are not yet optically thin given the observed P-Cygni profiles of \ion{He}{i} and \ion{Ca}{ii}. Nonetheless, the prominence of [\ion{Ca}{ii}] emission at such an early phase indicates a rapid evolution towards the nebular regime and low enough ejecta densities to allow for efficient cooling through forbidden transitions. Furthermore, we compare nebular spectra of the majority of \cas \ to SN~2019ehk in Figure \ref{fig:ca-rich_nebular}. Similar to all other \cas, there is no detectable emission from Fe-group elements in the bluewards spectrum; the majority of the observed flux being in [\ion{Ca}{ii}] emission, which shows no apparent [\ion{Ni}{ii}] $\lambda 7378$ line blending.

\begin{figure}[h]
\centering
\subfigure[]{\includegraphics[width=0.42\textwidth]{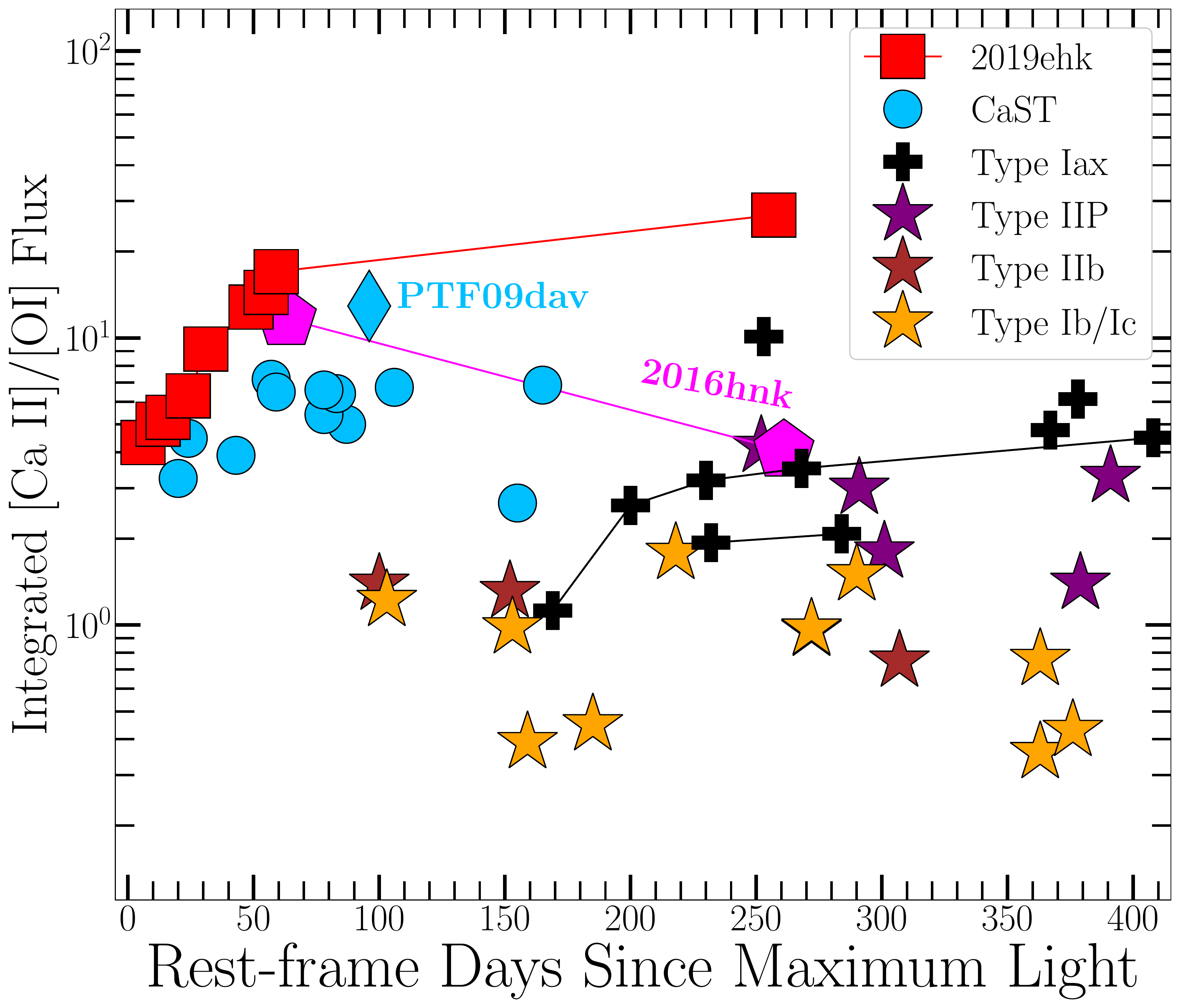}}
\subfigure[]{\includegraphics[width=0.42\textwidth]{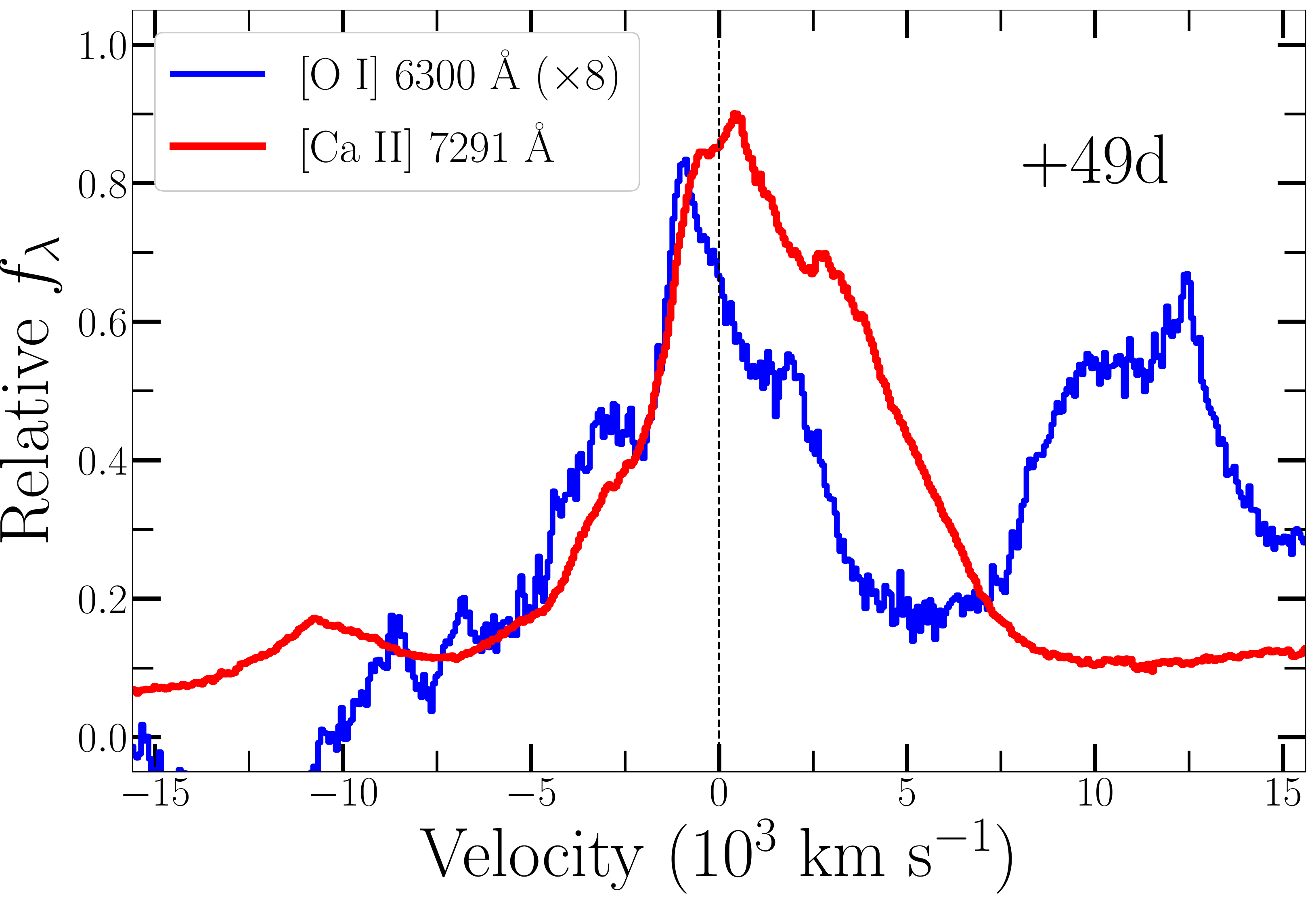}}
\subfigure[]{\includegraphics[width=0.42\textwidth]{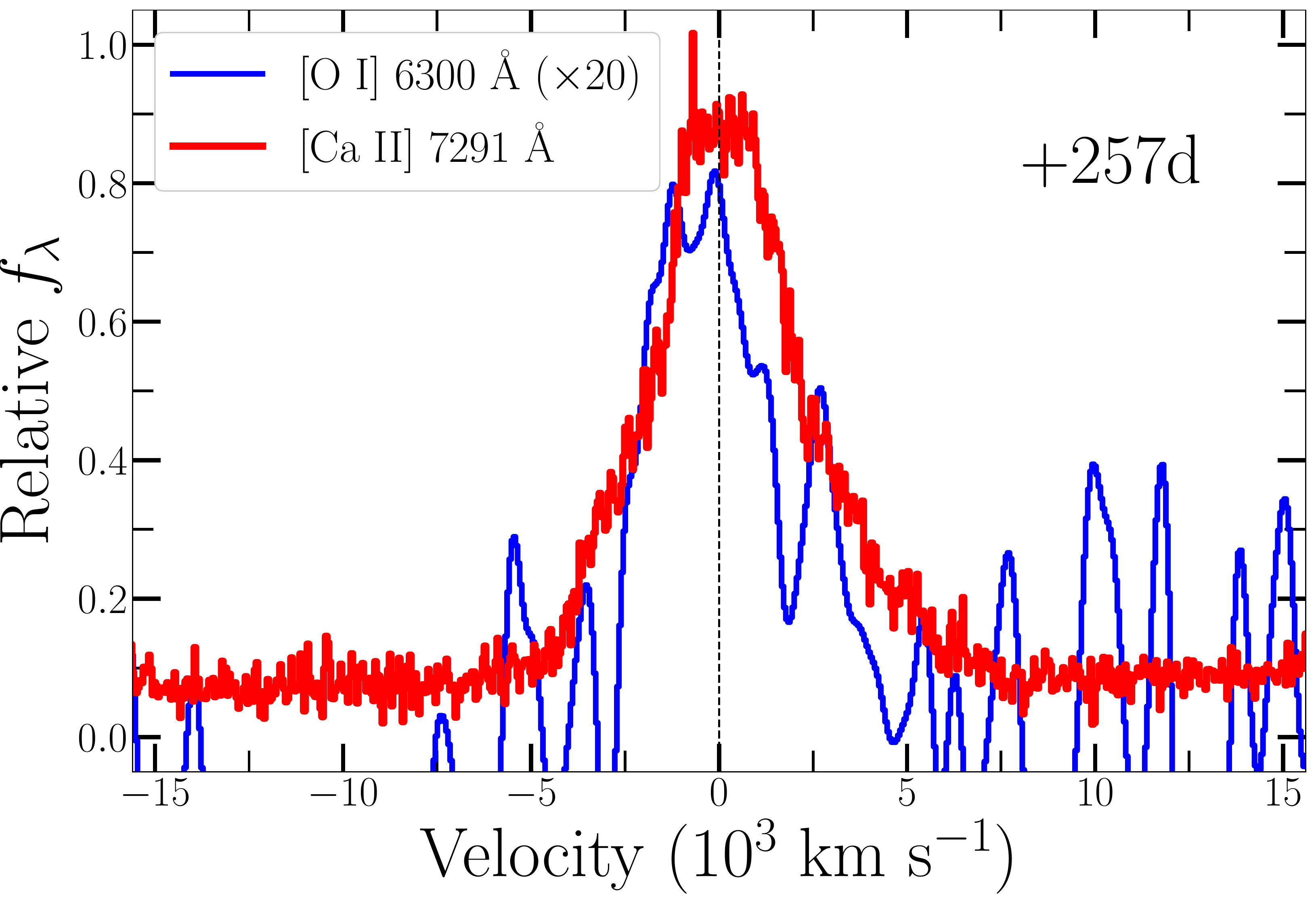}}
\caption{(a) Ratio of integrated [\ion{Ca}{ii}] and [\ion{O}{i}] flux with respect to phase for SN~2019ehk, PTF09dav, SN~2016hnk, \cas \, SNe~Iax and assorted types of core-collapse SNe. [\ion{Ca}{ii}]/[\ion{O}{i}] values for all Type II/Ibc objects from \cite{milisavljevic17}. (b)/(c) Velocity profiles of [\ion{Ca}{ii}] $\lambda\lambda$ 7291,7324 (red) and scaled [\ion{O}{i}] $\lambda\lambda$ 6300, 6364 (blue) in SN~2019ehk at +49d and +257d post second maximum light. \label{fig:caii_oi}}
\end{figure}

A common \ca \ classifier is a [\ion{Ca}{ii}]/[\ion{O}{i}] line flux ratio greater than 2. We show the evolution of this ratio, in addition to a direct comparison of [\ion{O}{i}] to [\ion{Ca}{ii}] lines in Figure \ref{fig:caii_oi}. As seen in \ref{fig:caii_oi}(a), even after reddening corrections, SN~2019ehk has the highest observed [\ion{Ca}{ii}]/[\ion{O}{i}] ratio of any known \ca \ at $t < 150$ days. This indicates that SN~2019ehk is not only more O-poor than most \cas, but it also has the fastest observed evolution to the optically thin regime. A quantitative discussion of elemental abundances in SN~2019ehk is presented in \S\ref{subsec:nebular_gas}.

While the spectral characteristics of SN~2019ehk appear to confidently place it within the \ca \ class, we explore the similarities between this SN and SNe Ib/IIb. As shown in Figure \ref{fig:IIb_compare}(b), SN~2019ehk, like other \cas, has similar spectral features to SNe~Ib 2008D near peak such as detectable \ion{He}{i} and \ion{Si}{ii} profiles and strong \ion{Ca}{ii} absorption. Compared to SNe~IIb, the most apparent difference is the lack of a P-Cygni H$\alpha$ and H$\beta$ profiles in SN~2019ehk, which only showed narrow H$\alpha$ emission within $\sim$2~days of explosion. This suggests a H-rich CSM in SN~2019ehk while the broad H$\alpha$ profiles in SNe~IIb are indicative of H attached to an expanding photosphere. Finally, the line velocities in SN~2019ehk are slower overall than the photospheric velocities observed in SNe~Ib and IIb: \ion{He}{i} velocity is $\sim$6500~$\kms$ in SN~2019ehk, $\sim$9000~$\kms$ in iPTF13bvn and $\sim$7100~$\kms$ in SN~2011dh. These spectral differences may indicate that SN~2019ehk is the result of a different explosion scenario than these core-collapse SNe, but does not necessarily rule out a massive star progenitor. 

\subsection{Inferences from ``Flash-Ionized'' H+He Spectral Lines at $t<$ 3 days}\label{subsec:FS}

\begin{figure}[h]
\centering
\includegraphics[width=0.45\textwidth]{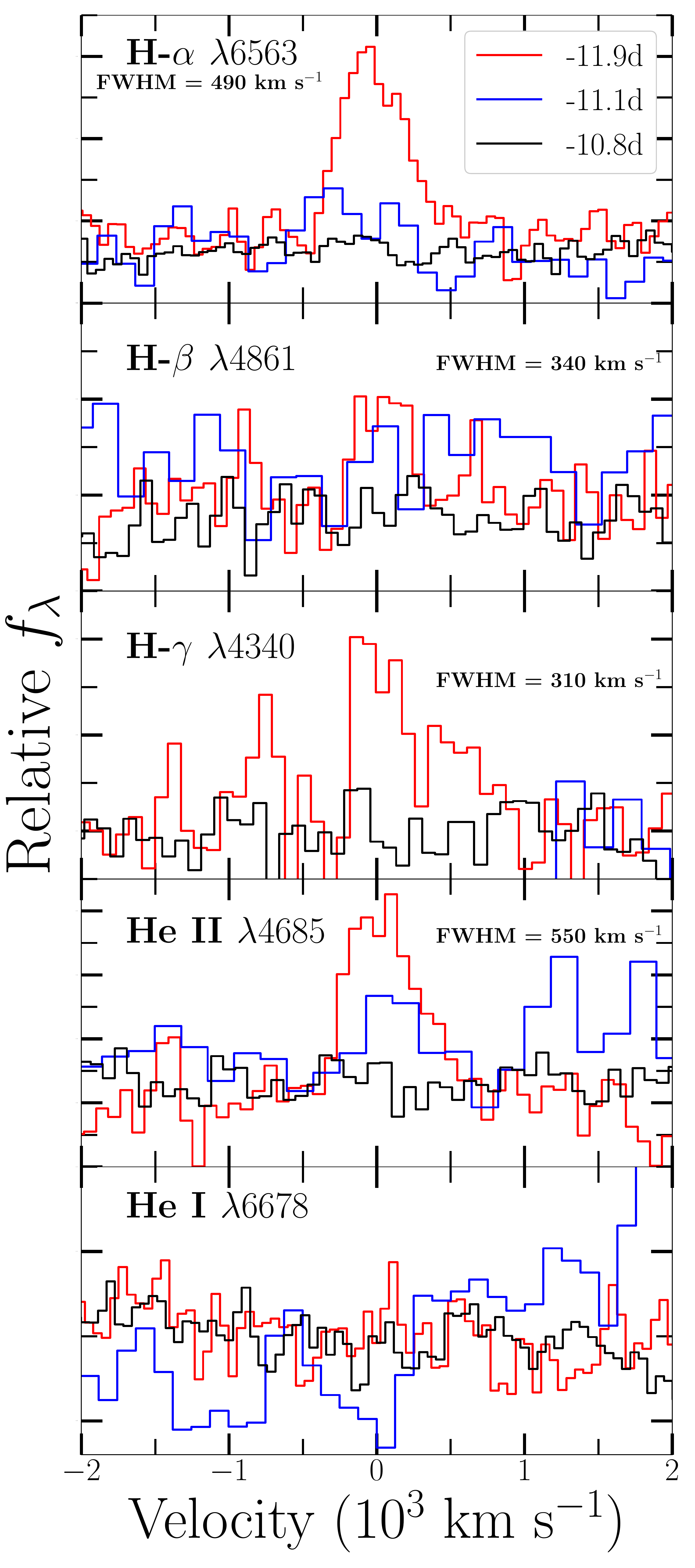} \caption{Velocity profiles of ``flash-ionized'' H Balmer series and \ion{He}{ii} lines in the first three epochs of spectroscopic observations. \ion{He}{i} $\lambda6678$ is shown in the bottom panel for reference. Phases presented are relative to the second $B$-band maximum with the red, blue and black lines at -11.9, -11.1 and -10.8 days, respectively, corresponding to 1.45, 2.33, and 2.55 days since explosion. These observations indicate the presence of pre-explosion CSM composed of H- and He-rich material moving with velocities of $\sim$400~$\kms$ and $\sim$500~$\kms$, respectively.  \label{fig:FS_vels}}
\end{figure}

The earliest spectrum obtained -11.9 days before second $B$-band maximum (1.45d since explosion) shows narrow H$\alpha$ and \ion{He}{ii} $\lambda$4686 emission lines with width of $\sim$500~$\kms$. The observed velocities are greater than the spectral resolution of the Kast spectrograph ($\lesssim100~\kms$) used to detect this spectral features. These lines are partially detected in the spectrum acquired on day -11.1, but not on day -10.8 (2.3 and 2.6 days since explosion, respectively) (Figure \ref{fig:FS_vels}). Furthermore, we visually identify potential narrow \ion{He}{i} emission near $\lambda6678$ in the earliest spectrum with a $\sim$1$\sigma$ detection confidence. We present its velocity profile for reference in Figure \ref{fig:FS_vels} and note that, if real, the species is below the 3$\sigma$ detection threshold. Accounting for the brightening of the underlying continuum we conclude that there is evidence for fading of H$\alpha$ and \ion{He}{ii} line flux by a factor $\ge$2 at 3$\sigma$ c.l. between the second and third epoch. These emission profiles are similar to those found in young core-collapse SNe and are thought to form from ``flash'' or shock-ionized CSM surrounding the progenitor star \citep{galyam14, kochanek19}.

The line width of $\sim$500~$\kms$ (Figure \ref{fig:FS_vels}), significantly lower than the velocity of material in the explosion's photosphere (\S\ref{subsec:spec_analysis}), indicates that the emission arises from CSM produced via mass-loss \emph{before} the explosion (as opposed to originating in the explosion's ejecta). The detection of H and He emission lines with these properties thus establishes the presence of H and He-rich CSM around SN~2019ehk. The time of their disappearance and relative luminosity enable inferences on the location of the CSM and its chemical composition, as we detail below.

The H$\alpha$ and \ion{He}{ii} $\lambda$4686 luminosities of  $2.0 \times 10^{38}$ and $3.1 \times 10^{38}$ $\rm erg~s^{-1}$, respectively, measured at 1.45 days since explosion imply $n_{\rm{He++}} / n_{\rm{H+}} = 0.44$ assuming Case B recombination \citep{hummer87}. The luminosity limit of the \ion{He}{i} $\lambda$7065 line $<4.0 \times 10^{37}$ $\rm erg~s^{-1}$ can be used to infer an  upper limit on the amount of He$^+$ using
recombination rates from \citet{benjamin99}, so that we find
\begin{equation}\label{eq:nhe_nh}
    0.44 < n_{\rm{He}} / n_{\rm{H}} < 0.88
\end{equation}
\noindent
implying partial burning of hydrogen. 

The SN shock break out radiation cannot be responsible for the ionization of the CSM at $t\ge1.4$ days, as the recombination timescale for H+ and He++ is $t_{\rm rec}$$\propto$$1/n_{e}$ and $t_{\rm rec}\le$ a few hours for gas temperatures $\sim$~$10^{5}-10^{6}$ K and free electron densities $n_e\ge10^{8}\,\rm{cm^{-3}}$ (e.g. \citealt{lundqvist96}). The source of ionizing radiation can be provided by the luminous X-ray emission (Fig. \ref{fig:xray_radio_LC}) that resulted from the SN shock interaction with the CSM (\S\ref{SubSec:Xraydensity}). In this scenario the fading of the H and He recombination lines is related to the time when the SN shock overtakes the CSM shell. We infer an outer CSM shell radius  $r\leq10^{15}$ cm, for the SN shock to reach it in $\sim$2 days (for a typical shock velocity $\sim$0.1c), and an emission measure $EM\approx 4 \times 10^{63}~\rm cm^{-3}$ to account for the observed recombination line luminosities at 1.45 days after explosion. From these inferences we derive a CSM density\footnote{Note that this is the density of the unshocked CSM gas  illuminated and ionized by the X-ray emission from the SN shock.} $n\approx10^9~\rm cm^{-3}$ and a CSM shell mass of  $M_{\rm CSM} \approx2\times10^{-3} \ \Msun$ assuming a spherical shell ($R_{\rm CSM} = 10^{15}$cm) and unity filling factor.  Note that the filling factor cannot be less than about 0.3 without reducing the ionization parameter to less than 30 and producing too much \ion{He}{i} emission. Based on the abundance by number shown in Eqn. \ref{eq:nhe_nh}, we estimate a  CSM H mass in the range $(1.2 - 4.8) \times 10^{-4} \ \Msun$ and a He mass of $(1.5 - 1.9)\times 10^{-3} \ \Msun$. 

We end by noting that for these physical parameters, the resulting ionization parameter  $\xi=L_{\rm ion}/nr^2$ (where $L_{\rm ion}$ is supplied by the X-ray luminosity) has values intermediate between those needed to doubly ionize helium (as observed), but lower than those necessary to produce high ionization lines such as [\ion{Fe}{x}], which are  not seen in the spectra of SN~2019ehk (but detected in other SNe with CSM interaction, e.g., SN2014C, \citealt{Milisavljevic15}). 

\subsection{Inferences from Nebular Phase Spectroscopy at $t\ge$ 30 days}\label{subsec:nebular_gas}

Table \ref{tbl:line_lums} lists the emission line luminosities measured from spectra acquired on 31, 38 and 59 days since second $B$-band maximum.  Recombination lines of 
\ion{He}{i}, \ion{C}{i}, \ion{O}{i} and \ion{Mg}{i} are detected, along with forbidden lines of [\ion{O}{i}] and [\ion{Ca}{ii}] and permitted lines of \ion{Ca}{ii}, while we consider the possible H$\alpha$ feature to be an upper limit.  Uncertainties in the underlying continuum and the wavelength ranges of some of the lines cause up to factor of $\sim$2 errors in the inferred luminosities, especially for lines that show prominent P-Cygni profiles. With this caveat in mind, we find that the ratios of \ion{He}{i} line fluxes approximately agree with those predicted with the atomic rates of \citet{benjamin99} for densities up to $\sim 10^{10}~\rm cm^{-3}$ 
at a temperature of $\sim$~$10^4$ K.

The inferred blackbody radius at $\sim$59~days after second $B$-band maximum ($\sim$72~days since explosion) is $\sim$~$1.5\times 10^{15}$~cm (Figure \ref{fig:BB_LC}). The maximum velocity shift of the [\ion{Ca}{ii}] emission feature is $v_{\rm{[\ion{Ca}{ii}]}}\approx$~5000~$\kms$, corresponding to a radius of $v_{\rm{[\ion{Ca}{ii}]}}\times t\approx 3.2\times 10^{15}$~cm.  The nebular emission is produced between those radii, so we take the volume to be about $1.1 \times 10^{47}~\rm cm^{3}$. The observed \ion{He}{i} line luminosities and inferred volume  require $n_e~n_{\rm{\ion{He}{ii}}}\approx 10^{16}~\rm cm^{-6}$ at 59~days since second B-maximum.  We note that the \ion{He}{i} $\lambda$7065 line is stronger than expected, probably because of repeated scatterings that convert \ion{He}{i} $\lambda$3889 photons into emission at $\lambda$7065. This scenario is supported by the  prominent P-Cygni profiles of the \ion{He}{i} NIR lines, which indicate large optical depths and a substantial population of the 1s2s $^3$S metastable level.  

The relative luminosities and recombination rates from \citet{hummer87} and \citet{julienne74} imply number density ratios $n_{\rm{H II}} / n_{\rm{\ion{He}{ii}}}$ $ \leq1.3$ and  $n_{\rm{\ion{O}{ii}}} / n_{\rm{\ion{He}{ii}}}$ = 1.5.  If no other elements contribute significant numbers of free electrons, the densities of electrons, He$^+$, H$^+$ and O$^+$ are 3.8, 1.0, $<$1.3 and $1.5\times 10^8~\rm cm^{-3}$, respectively. If carbon contributes free electrons, $n_e$ will be correspondingly higher and the densities of the ions correspondingly lower.  The limit on the ratio of He$^+$ to H$^+$ is similar to the He/H ratio derived for the CSM (\S\ref{subsec:FS}), so H:He:O ratio may be similar to the values above.

At densities above $10^7~\rm cm^{-3}$, the [\ion{O}{i}] and [\ion{Ca}{ii}] lines are in their high density limits, and their luminosities are given by the populations of the excited states multiplied by the Einstein A values:
\begin{equation}
L_{\rm{[\ion{O}{i}]}} = n_{\rm{\ion{O}{i}}} ~ A_{\rm{[\ion{O}{i}]}} ~ h\nu_{\rm{[\ion{O}{i}]}} ~(5/14)~e^{-22000/T} 
\label{Eq:LOI}
\end{equation}
\begin{equation}
L_{\rm{[\ion{Ca}{ii}]}} = n_{\rm{\ion{Ca}{ii}}} ~ A_{\rm{[\ion{Ca}{ii}]}} ~ h\nu_{\rm{[\ion{Ca}{ii}]}} ~(10/11)~e^{-19700/T}
\label{Eq:LCaII}
\end{equation} 

\noindent
where h$\nu$ is the photon energy, the exponentials are the Boltzmann factors ($T$ is in K) and the numerical factors are statistical weights. The observed luminosities of the [\ion{Ca}{ii}] lines are much higher than the [\ion{O}{i}] luminosities ( $L_{\rm{[\ion{Ca}{ii}]}}/L_{\rm{[\ion{O}{i}]}}\approx25$ at 257 days since second $B$-band maximum, Figure \ref{fig:caii_oi}).  From Eqn. \ref{Eq:LOI}-\ref{Eq:LCaII}:
\begin{equation}
\frac{L_{\rm{[\ion{Ca}{ii}]}}}{L_{\rm{[\ion{O}{i}]}}}=\frac{28}{11}\frac{n_{\rm{\ion{Ca}{ii}}} }{n_{\rm{\ion{O}{i}}}}\frac{A_{\rm{[\ion{Ca}{ii}]}}}{A_{\rm{[\ion{O}{i}]}}}\frac{\nu_{\rm{[\ion{Ca}{ii}]}}}{\nu_{\rm{[\ion{O}{i}]}}} e^{2300/T} \approx\,1100 \frac{n_{\rm{\ion{Ca}{ii}}} }{n_{\rm{\ion{O}{i}}}}
\end{equation} 
\noindent
where we used $T=10^4$ K, $A_{\rm{[\ion{Ca}{ii}]}}=2.6\,\rm{s^{-1}}$ and  $A_{\rm{[\ion{Ca}{ii}]}}\approx390 A_{\rm{[\ion{O}{i}]}}$. We thus infer $\frac{n_{\rm{\ion{O}{i}}} }{n_{\rm{\ion{Ca}{ii}}}}\approx 30$. Since there is a strong overlap of the temperature and ionization parameter ranges where \ion{O}{i} and \ion{Ca}{ii} exist, we expect $\frac{n_{\rm{OI}} }{n_{\rm{CaII}}}\approx\frac{n_{\rm{O}} }{n_{\rm{Ca}}}$, implying that, as in iPTF15eqv \citep{milisavljevic17}, the prominent Ca lines result from the density and ionization state of the ejecta rather than an overabundance of Ca with respect to O. We note that the [\ion{O}{i}] and [\ion{Ca}{ii}] lines are likely to arise from a region of lower electron density than the recombination lines, because the forbidden lines are strongly suppressed at densities above $10^8~\rm cm^{-3}$, leading to higher ratios of the $\lambda$8600 calcium triplet to the [\ion{Ca}{ii}] lines than are observed.  The \ion{Ca}{ii} feature at $\lambda$11873 is much stronger than expected for optically thin emission, even if the 4s-5p lines from the ground state are converted to $\lambda$11873 through multiple scatterings.  It is possible that the \ion{He}{ii} $\lambda$1640 line pumps the 4s-5p transition, since the separation is about 1700$\,\kms$.  If so, the \ion{He}{ii} line is formed by recombination, and this would  be the only indication of doubly ionized helium in the nebular gas.

Assuming temperatures of $\sim$10$^4$ K for the recombination lines and 5000 K for the forbidden lines, the densities and volume yield rough mass estimates from the day +59 spectrum (from second $B$-band max) of 0.008, 0.037, 0.10 and 0.004 $\Msun$ for He$^+$, O$^+$,
O$^0$ and Ca$^+$, respectively. It should be noted that at these phases the SN is not fully nebular and therefore the derived masses may be lower than the true elemental masses in the explosion.

\begin{deluxetable}{ccccc}
\tablecaption{Nebular emission line luminosities for three epochs of spectroscopy  at 31, 38 and 59 days after second $B$-band maximum light.\label{tbl:line_lums}}
\tablecolumns{5}
\tablewidth{0.45\textwidth}
\tablehead{
\colhead{Wavelength} & \colhead{Line ID} & \colhead{Day +31} & \colhead{Day +38} & \colhead{Day +59}\\
\colhead{(\AA)} & \colhead{} & \colhead{($10^{38}~\rm erg~s^{-1}$)} & \colhead{($10^{38}~\rm erg~s^{-1}$)} & \colhead{($10^{38}~\rm erg~s^{-1}$)}
}
\startdata
5876       & \ion{He}{i}  &  17 & -- & 2.3 \\
6303,6363  & [\ion{O}{i}] &  19 & -- & 7.3 \\
7065       & \ion{He}{i}  &    14  & -- & 4.5 \\
7291,7324  & [\ion{Ca}{ii}]&  220  & -- & 130 \\
7774       & \ion{O}{i}  &   18   & -- & 4.9 \\
8579       & \ion{Ca}{ii} &  290   & -- & 150 \\
9224       & \ion{O}{i}  & 21  & -- & 5.0 \\
10830      & \ion{He}{i}  &  94 &  162 & 29. \\
11873      & \ion{Ca}{ii} & -- &   72 &  \\
14878      & \ion{Mg}{i}  & -- &   14 &  \\
15900      & \ion{C}{i}   & -- &   16 &  \\
20589      & \ion{He}{i}  & -- &   16 &  \\
\enddata
\tablecomments{In our latest spectrum at +257d, [\ion{Ca}{ii}] and [\ion{O}{i}] luminosities are $3.1\times 10^{38}$~erg~s$^{-1}$ and $1.2\times 10^{37}$~erg~s$^{-1}$, respectively. }
\end{deluxetable}

\begin{figure*}
\centering
\subfigure[]{\includegraphics[width=0.32\textwidth]{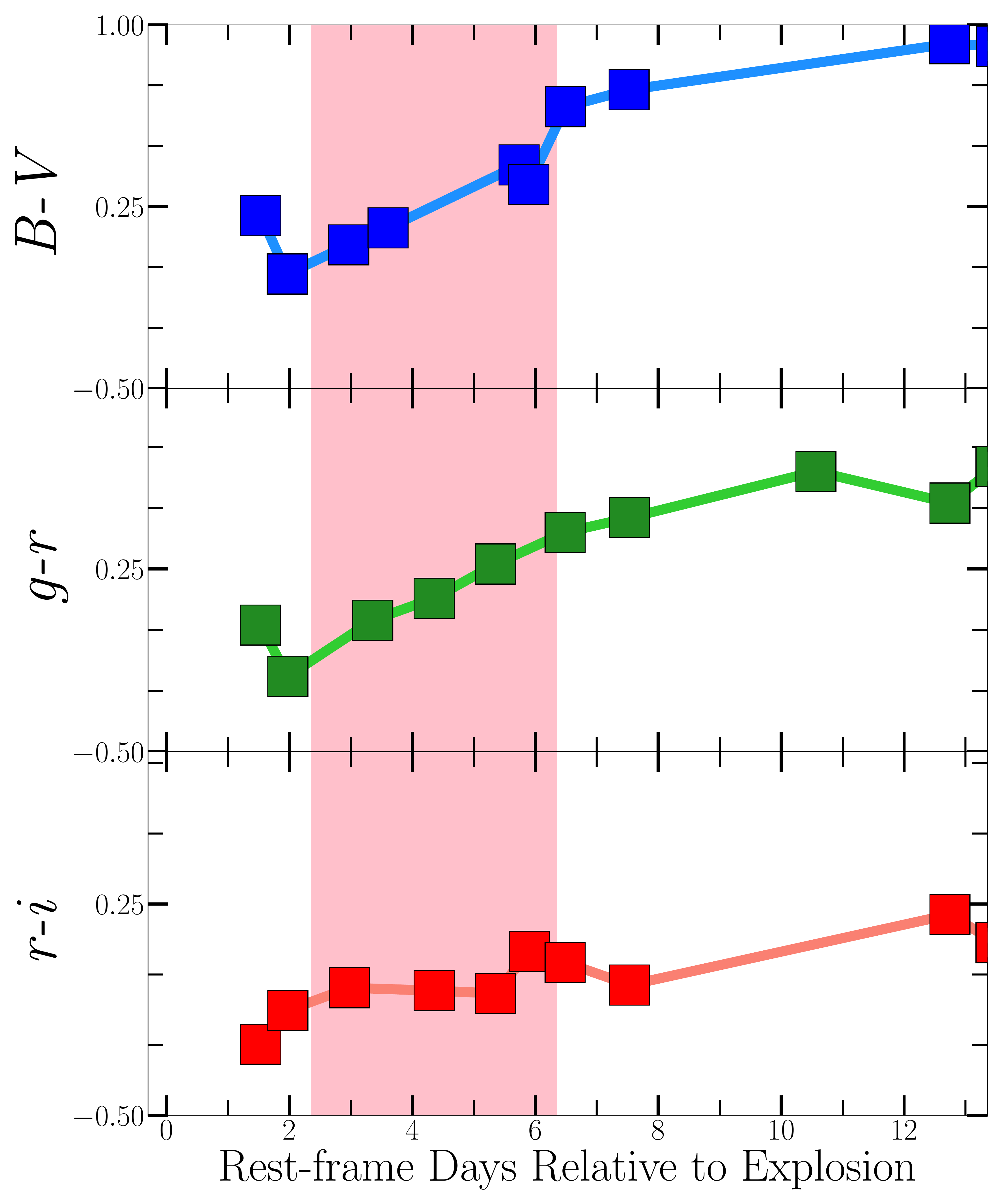}}
\subfigure[]{\includegraphics[width=0.32\textwidth]{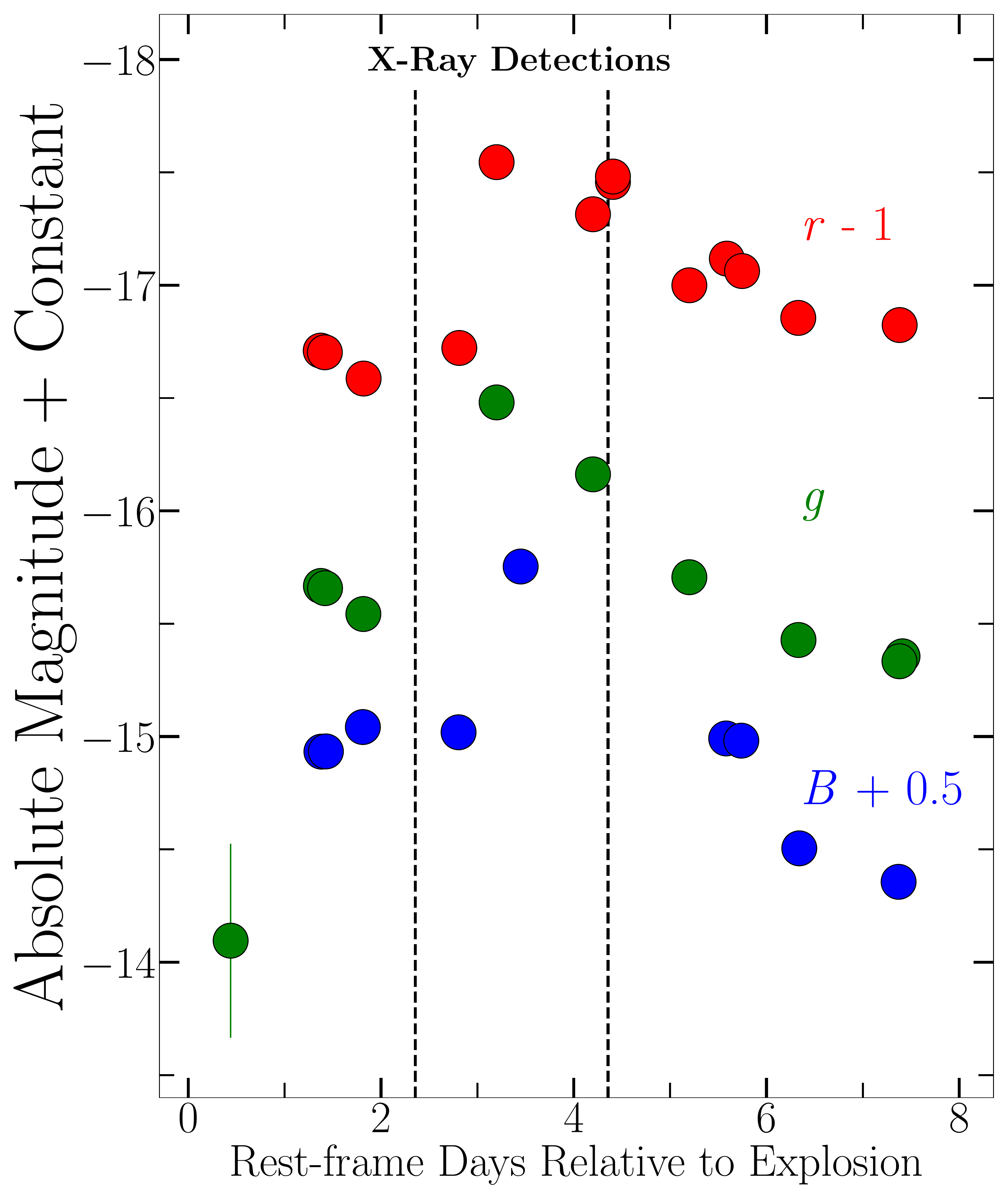}}
\subfigure[]{\includegraphics[width=0.32\textwidth]{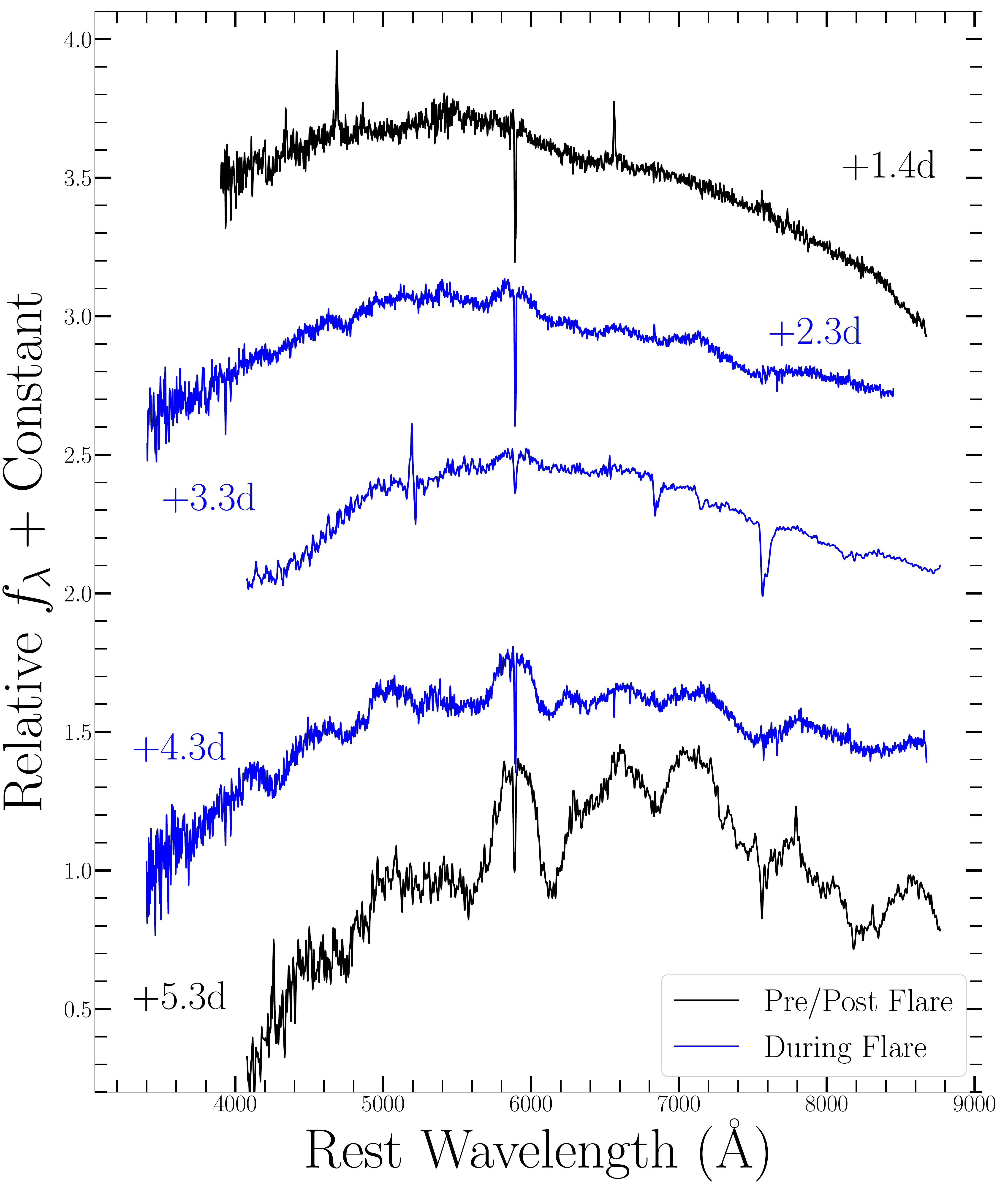}}
\caption{ (a) Highlighted by the pink shaded region are extinction corrected colors in SN~2019ehk during the optical ``flare.'' This color evolution indicates that the flare was quite blue as the colors do not become redder until after the first light curve peak. (b) $Bgr$-band, extinction corrected photometry during the flare with the times of the X-ray detections from \textit{Swift}-XRT shown as vertical black dashed lines. (c) Spectral evolution during the flare shown in blue, with observations before and after presented in black. The peak of the flare occurs at +3.3 days with respect to explosion, which has an observed increase in optical flux as shown in the spectrum.  Following the flare, regions of line formation in the photosphere emerge in the spectra and known ions can be more easily identified.
\label{fig:flare_optical}}
\end{figure*}

\section{The optical ``Flare''}\label{sec:flare}
Here we describe the observational properties of the first light curve peak and present physical models that can explain this initial increase in total flux. In an effort to be succinct, we hereafter refer to this primary light curve evolution as the ``flare.'' In this section, all times are referred with respect to the explosion.
\subsection{Observational Properties}\label{sec:flare_specifics}
The flare is observed across all UV, optical and NIR photometric bands from the first $g-$band detection at 0.44~days until $\sim$7~days after explosion. We present SN~2019ehk's color, photometric and spectral evolution during the flare in Figure \ref{fig:flare_optical}. We observe an initial rise in $g-$band flux from 0.44-1.38~days and then seemingly constant flux between 1.38 and 2.81~days. However, in some photometric bands (e.g., $gVri$) the flux in this phase range appears to be decreasing. This indicates that there could be 2 separate peaks within the flare or possibly separate emission mechanisms at these early-times. Then, as shown in Figure \ref{fig:flare_optical}(b), the most dramatic flux increase occurs in $<1$~day and peaks at $t_{p} = 3.2 \pm 0.1$~days. This is reflected by a $\sim$1~mag flux increase in all photometric bands. During the early rise, the flare spectrum is blue and mostly featureless, with transient H and He recombination lines that soon subside (\S\ref{subsec:FS}). Clear photospheric spectral features (e.g., \ion{Si}{ii}, \ion{O}{i}, \ion{Ca}{ii}) first appear after the flare's peak at $t \approx 3$~days after explosion (Figure \ref{fig:flare_optical}c).

\begin{figure}
\centering
\includegraphics[width=0.49\textwidth]{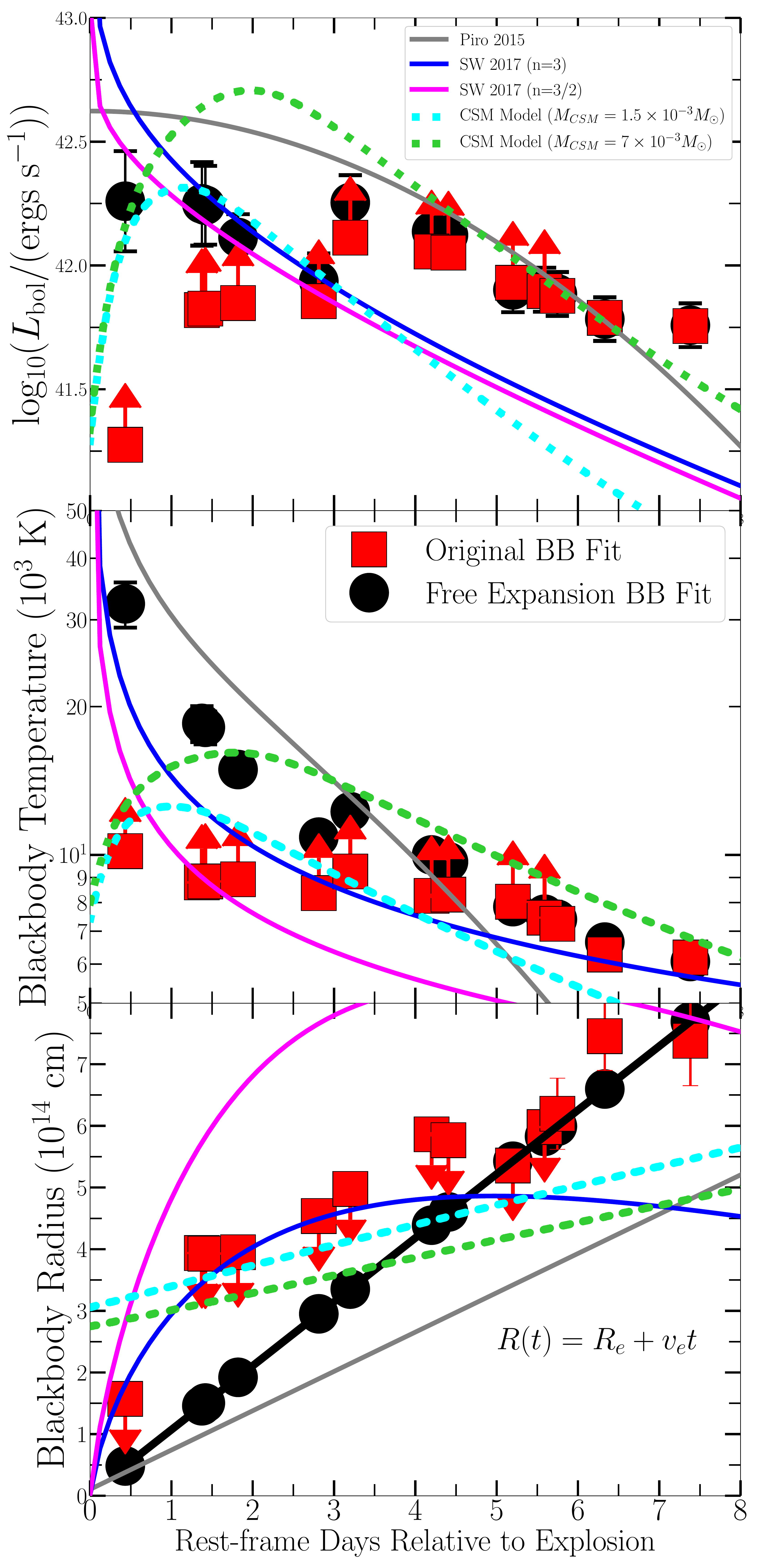} \caption{\textit{Top:} Inferred bolometric luminosity during the flare presented as red squares and black dots (fixed blackbody radius). Shock interaction models for different CSM masses are plotted in green and cyan dashed lines (see \S\ref{sec:flare_csm}). Shock cooling models are plotted as solid lines: \cite{piro15} in grey, \cite{sapir17} n = 3/2[3] in pink[blue]. \textit{Middle:} Lower limits and more realistic estimates of the blackbody temperature during the flare. For the interaction model  we show the effective blackbody temperature.  \textit{Bottom:} Upper limits and more realistic estimate of the blackbody radius assuming a linear increase of the photospheric radius with time ($v_e \approx 12,000$~$\kms$). The shock interaction model presents the radius of the emitting region. \label{fig:flare_lum} }
\end{figure}

We present SN~2019ehk's blackbody radius $R(t)$, temperature $T(t)$ and resulting bolometric luminosity evolution during the flare in Figure \ref{fig:flare_lum} (shown as squares). As discussed in \S\ref{subsec:bol_LC}, at $t\lesssim5$~days the blackbody SED peak lies in the mid-UV, outside the range covered by our complete photometric dataset. At these times the data provide lower limits on the blackbody temperature and upper limits on the radius, which results in a lower limit on the true bolometric luminosity (arrows in Figure \ref{fig:flare_lum}). The bolometric light curve at $t< 3$~days was likely dominated by UV radiation and decreased rapidly from a peak luminosity potentially larger than $L_{\rm bol} (t_p) \approx 10^{42}$~erg~s$^{-1}$  shown in Figure \ref{fig:flare_lum} with red squares.

A reasonable assumption for stellar explosions at early times is that of a photosphere expanding homologously in time (e.g., \citealt{Liu18}). Here we make the simplistic assumption of a linear evolution of the photospheric radius with time, $R(t) = R_e + v_e*t\approx  v_e*t$, where we take $v_e\approx 12000 \ \kms$, similar to the velocities observed in the first photospheric spectra and $R_e$ is the initial envelope radius (black dotted line in Fig. \ref{fig:flare_lum}, lower panel). Interestingly, the resulting $R(t)$ matches the photospheric radius at $t\ge 5$ days. Freezing the blackbody radius to the values implied by the linear evolution with time in our blackbody fits leads to larger inferred temperatures, as expected (Fig. \ref{fig:flare_lum}, middle panel). The resulting bolometric luminosity is also consequently larger (Fig. \ref{fig:flare_lum}, upper panel). While we consider these estimates to lead to a more realistic bolometric output at early times, we caution that the assumption of a linearly increasing photospheric radius is likely an over-simplification and that accelerated expansion could have a significant influence on the very early-time SN evolution.

\subsection{Nickel Powered Model}\label{sec:flare_ni}
A possible power source for the flare emission is the radioactive decay of an amount of $^{56}\rm{}Ni$ that was heavily mixed into the outer layers of ejecta. This $^{56}\rm{}Ni$ mass is distinct from the centrally located $^{56}\rm{}Ni$ that is responsible for the main SN optical peak. As discussed in \cite{de18} for iPTF16hgs, this distribution of $^{56}\rm{}Ni$ could result in two distinct light curve peaks, each powered by its own supply of $^{56}$Ni (e.g., see also \citealt{drout16}). We test the validity of this model for SN~2019ehk by applying the same analytic model for a radioactively powered light curve as that presented in \S\ref{subsec:bol_LC}. We find $E_{\rm{k}} \approx 10^{47}$ ergs and $M_{\rm{Ni}} \approx 3 \times 10^{-2}~\Msun$. A total ejecta mass of $M_{\rm{ej}} \approx 10^{-4}~\Msun$ is estimated using $v_{\rm{ph}} \approx 12000$~$\kms$, which is derived from \ion{Si}{ii} absorption near the peak of the flare.

This model both produces a poor fit to the flare's bolometric luminosity as well as results in a $M_{\rm Ni} / M_{\rm ej}$ ratio greater than 1, which is clearly unphysical. Furthermore, this model is disfavored because it does not naturally explain the presence of early-time X-ray emission. If an exterior plume of $^{56}$Ni is the power-source behind the flare, an additional, independent ingredient would need to be invoked to explain the X-rays, which would have occurred coincidentally at the same time as the optical flare, but would otherwise have no physical connection to the flare. More natural scenarios are those where the optical flare and the X-ray emission are different manifestations in the electromagnetic spectrum of the same physical process (\S\ref{sec:flare_shockcool}, \S\ref{sec:flare_csm}).

\begin{figure*}
\centering
\includegraphics[width=\textwidth]{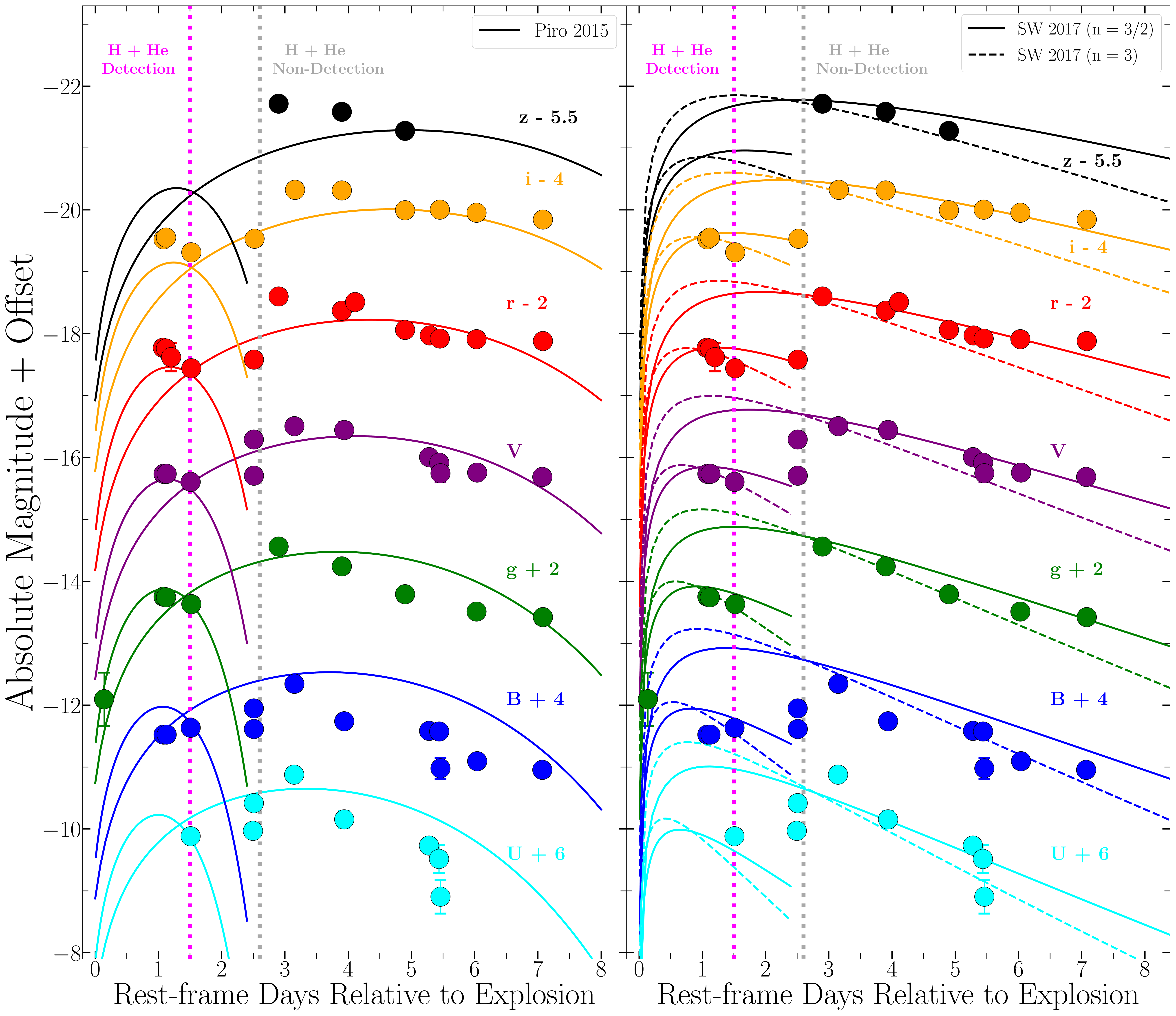} \caption{Multi-color shock cooling model fits to the flare assuming a blackbody SED. \textit{Left:} \cite{piro15} models are presented as solid lines with the phases of ``flash-ionized'' H and He detection and non-detection presented as dotted magenta and grey lines, respectively. \textit{Right:} \cite{sapir17} models shown as dashed (n=3) and solid (n=3/2) lines. We model the flare in two components due to temporal variability at $t<2$~days. Model specifics are discussed in \S\ref{sec:flare_shockcool} and physical parameters are presented in Table \ref{tbl:shocktable}.   \label{fig:shock_cool} }
\end{figure*}

\subsection{Shock Breakout and Envelope Cooling Model}\label{sec:flare_shockcool}
It is now understood that shock breakout through an extended distribution of material (e.g., stellar envelope) can increase the SN flux above the typical radioactively powered continuum emission. The resulting observational signature is a double-peaked light curve where the first peak originates from the expansion and cooling of the shocked envelope, followed by the standard SN peak of emission. This is typically observed in SNe IIb (e.g., SNe~1993J, 2011dh and 2016gkg; \citealt{wheeler93, arcavi11, arcavi17, piro17}) and numerous models have been put forward to explain this observational signature with breakout and cooling emission into an expanding envelope  \citep{nakar14, piro15, sapir17}. 

As discussed in \S\ref{sec:flare_specifics}, the light curve exhibits nearly constant flux at $t < 2.5$~days before the dramatic rise and decline in magnitude from $3 < t < 6$~days. Furthermore, as illustrated by the magenta and grey dotted lines in Figure \ref{fig:shock_cool}, H and He emission lines persist in SN~2019ehk spectra until $t \approx 2.5$~days  and fade in visibility when the primary peak of the flare occurs at $t \approx 3$~days. These observational signatures suggest separate emission components within the flare: one that allows for H + He spectral emission in addition to unremitting flux ($t < 2$~days), and one that induces a substantial rise in flux without ``flash-ionized'' spectral lines ($2 < t < 6$~days). Consequently, we choose to model each of the observationally distinct regimes within the flare separately. 

In the following sections, we describe and apply three models for a shock cooling emission mechanism to explain the entire evolution of the optical flare in SN~2019ehk. At the time of explosion, each model produces constraints on the envelope mass, $M_e$, envelope radius, $R_e$, the velocity of the shock or envelope, $v_e$ and the time offset from explosion $t_o$ (consistent with our explosion time estimate). In this analysis, we use \texttt{emcee}, a Python-based application of an affine invariant MCMC with an ensemble sampler \citep{foreman-mackey13}. We compile the best fit parameter estimates from each model in Table \ref{tbl:shocktable}. While in \S\ref{sec:flare_shockcool} we model the flare emission with two cooling-envelope components, we note that the presence of H and He emission in the first flare component requires a persistent source of ionizing radiation that might not be provided by pure cooling-envelope models, which motivates our investigation of models that also include ongoing CSM interaction in \S\ref{sec:flare_csm} 

\subsubsection{\cite{nakar14} Model}\label{sec:N14_model}
\cite{nakar14} present scaling relations for ``non-standard'' core-collapse SN progenitors with compact cores surrounded by extended envelopes. By showing that the peak of the optical flux will occur when the mass depth (i.e., photon diffusion distance within mass) is equal to the envelope mass ($M_e$), they construct the following analytic expression for $M_e$:

\begin{equation}
    M_e \approx 5\times 10^{-3}\kappa_{0.34}^{-1}\Bigg(\frac{v_e}{10^9 \ \rm{cm \ s}^{-1}}\Bigg) \Bigg(\frac{t_p}{1 \ \rm{day}}\Bigg)^2 \Msun
\end{equation}

\noindent 
where $v_e$ is the expansion velocity of the extended envelope, $t_p$ is the time to first light curve peak and the opacity is $\kappa_{0.34} = \kappa/0.34$~cm$^{2}$~g$^{-1}$. As discussed in \S\ref{sec:flare_shockcool}, it is likely that the flare is the product of separate emission mechanisms, each occurring on different timescales. As a result, we apply the \cite{nakar14} model to each ``peak'' within the flare at times $t_{p1} = 0.44 \pm 0.10$ and $t_{p2} = 3.2 \pm 0.10$~days. We estimate an envelope velocity of $\sim 1.2 \times 10^9 $~cm~s$^{-1}$ from the absorption minimum of the \ion{He}{i} $\lambda5976$ transition, which is the first detectable spectral feature to appear at $2.2 \pm 0.10$~days after explosion.

Furthermore, from \cite{nakar14}, the envelope radius can be expressed as: 

\begin{equation}
    R_e = 2\times10^{13}\kappa_{0.34} L_{43} \Bigg(\frac{v_e}{10^9 \ \rm{cm \ s}^{-1}}\Bigg)^{-2} \rm{cm}
\end{equation}

\noindent 
where $L_{43} = L_{bol}(t_p)/10^{43}$~erg~s$^{-1}$. At $t_{p1} = 0.44 \pm 0.10$~days and $t_{p2} = 3.2 \pm 0.10$~days, we calculate peak bolometric luminosities of $L_{bol}(t_{p1}) = 1.8 \pm 0.9 \times 10^{42} $ and $L_{bol}(t_{p2}) = 1.8 \pm 0.10 \times 10^{42} $~erg~s$^{-1}$, respectively. They also predict the observed temperature at $t_p$ as:

\begin{equation}
    T_{\rm{obs}}(t_p) \approx 3\times10^{4} \kappa_{0.34}^{-0.25} \Bigg(\frac{R_e}{10^{13} \ \rm{cm}}\Bigg)^{0.25} \Bigg(\frac{t_p}{1 \ \rm{day}}\Bigg)^{-0.5} \ \rm{K}
\end{equation}

Using this expression for $t_{p1}$ and $t_{p2}$, we calculate observed flare temperatures of $2.5 \pm 0.30 \times 10^4$ and $1.7 \pm 0.90 \times 10^4$~K, respectively; both of which are consistent with derived blackbody temperatures at the same phase as shown in Figure \ref{fig:flare_lum}. Overall, we caution against the accuracy  of these model outputs due to uncertainties surrounding the bolometric luminosities at $t<6$~days. As discussed in \S\ref{subsec:bol_LC}, we can only place solid constraints on upper and lower limits on the blackbody radii and temperature during the flare, which then affect the bolometric luminosity at those times. Thus the $M_e$ and $R_e$ values derived from the \cite{nakar14} models should be treated as lower limits given the uncertainty on each peak luminosity. For the main peak of the flare and opacity $\kappa = 0.2$~cm$^{2}$~g$^{-1}$, we estimate an envelope mass of $M_e \approx 0.1 \ \Msun$ and radius of $R_e \approx$ 100~$\Rsun$.

\subsubsection{\cite{piro15} Model}\label{sec:P15_model}
Starting from the scaling relations at $t_p$ from \cite{nakar14}, \cite{piro15} presents a generalized analytic model that allows a direct, detailed comparison to the observed flux evolution with time. The SN shock is assumed to propagate into extended material of mass $M_e$ of unknown chemical composition surrounding the progenitor star core with mass $M_c$. This is a one-zone model that does not include a prescription for the density profile, gradient or chemical composition of the extended material.  Following \cite{piro15} the expansion velocity  $v_e$ and the energy $E_e$  passed into the extended material read:

\begin{equation}\label{eq:P15_ve}
    v_e \approx (2\times10^9)E_{51}^{0.5}\Big(\frac{M_c}{\Msun}\Big)^{-0.35}\Big(\frac{M_e}{0.01\Msun}\Big)^{-0.15} \textrm{cm s}^{-1}
\end{equation}

\begin{equation}\label{eq:P15_Ee}
    E_e \approx (4\times 10^{49})E_{51}\Big(\frac{M_c}{\Msun}\Big)^{-0.7}\Big(\frac{M_e}{0.01\Msun}\Big)^{0.7} \textrm{erg}
\end{equation}

\noindent
where $E_{51} = E_{SN}/10^{51}$erg. \cite{piro15} show that the shocked extended material will expand (with characteristic radius $R(t)=R_e+v_e\,t$) and cool, with an observed peak of emission occurring at time $t_p$:
    
\begin{equation}\label{eq:P15_tp}
    t_p \approx 0.9 \kappa_{0.34}^{0.5}E_{51}^{-0.25}\Big(\frac{M_c}{\Msun}\Big)^{0.17}\Big(\frac{M_e}{0.01\Msun}\Big)^{0.57} \textrm{day}
\end{equation}

\noindent

In their Eqn. 15, \cite{piro15} present a predicted bolometric luminosity from shock cooling as:

\begin{equation}\label{P15_Lt}
    L(t) = \frac{t_eE_e}{t_p^2} \textrm{exp}\Big[\frac{-t(t+2t_e)}{2t_p^2}\Big]
\end{equation}
\noindent
where $t_e = R_e / v_e$. 

Following \cite{arcavi17} and \cite{piro15}, we model the emission from the extended mass as a blackbody spectrum with radius $R(t) = R_e + v_et$ and temperature:

\begin{equation}\label{P15_T}
    T(t) = \Big[\frac{L(t)}{4\pi \sigma_{\rm{SB}} R^2(t)}\Big]^{1/4}
\end{equation}

We calculate the expected apparent magnitudes for individual photometric bands from this model using the \texttt{pysynphot} Python package and we fit these models to the (extinction-corrected) apparent magnitudes of SN~2019ehk in $uBVgriz$ bands at $t<6$~days. As before, we fit the data at $t < 2$d and $t < 6$~days as two separate components. For all models we adopt $E_{SN} = 1.8 \times 10^{50}$~erg, $\kappa = 0.2$~cm$^2$~g$^{-1}$ and $M_c = 1~\Msun$ (\S\ref{subsec:bol_LC}). It should be noted that the chosen core mass $M_c$ has little impact on the final inferred parameters. We present all multi-color light curve fits using these models as the solid lines in Figure \ref{fig:shock_cool}. As shown in the plot, this simplified model provides a reasonable match to the data for both components of the flare. The best fitting values for both components are reported in Table \ref{tbl:shocktable}. 

\subsubsection{\cite{sapir17} Model}\label{sec:SW17_model}
\cite{sapir17} present an updated version of the model by \cite{rabinak11}, which applies to the immediate post-shock breakout evolution at $t\approx$ few days, when the emission is dominated  by radiation from the external envelope layers, and extends the solutions by \cite{rabinak11} to later times, when the observed emission originates from the inner envelope layers and depends on the progenitor density profile. \cite{sapir17} adopt a progenitor structure with a polytropic hydrogen-dominated envelope, which they demonstrate numerically can power an initial light curve peak through shock cooling.  

Below we present the analytic expression for the envelope's bolometric luminosity that was derived by \cite{arcavi17} starting from \cite{sapir17}:

\begin{equation}\label{SW16_Lt}
    \begin{aligned}
    L(t) = 1.88[1.66]\times10^{42}\\
    \times \ \Big(\frac{v_{s,8.5}^2R_{13}}{\kappa_{0.34}}\Big)\Big(\frac{v_{s,8.5}t^2}{f_pM\kappa_{0.34}}\Big)^{-0.086[-0.175]}\\
    \times \ \textrm{exp}\Big\{-\Big[\frac{1.67[4.57]t}{(19.5\kappa_{0.34}M_ev_{s,8.5}^{-1})^{0.5}}\Big]^{0.8[0.73]}\Big\} \textrm{erg~s}^{-1}
    \end{aligned}
\end{equation}

\noindent
where $R_{13} \equiv R_e / 10^{13}$cm, $v_{s,8.5} \equiv v_s / 10^{8.5}$ cm s$^{-1}$, $M = M_e + M_c$ and $t$ is in days. This model is for a polytropic index of $n = 3/2[3]$, which encompasses both stars with convective envelopes as well as radiative envelopes e.g., RSGs[BSGs], respectively. Same as for the \cite{piro15} models, we adopt $M_c =1\, \Msun$. The dimensionless scaling factor $f_p$ from \cite{sapir17} is:
\begin{equation}
    f_p \approx 
    \begin{cases}
    (M_e/M_c)^{0.5},& n=3/2\\
    0.08(M_e/M_c),& n=3
    \end{cases}
\end{equation}

\noindent
Finally, \cite{arcavi17} present an envelope temperature derived by \cite{sapir17} to be:

\begin{equation}\label{eq:SW16_Tt}
    \begin{aligned}
    T(t) \approx 2.05[1.96] \times 10^4\\
    \times \ \Bigg( \frac{v_{s,8.5}^2 t^2}{f_pM\kappa_{0.34}}\Bigg)^{0.027[0.016]}\Bigg(\frac{R_{13}^{0.25}}{\kappa_{0.34}^{0.25}}\Bigg) t^{-0.5} \ \textrm{K}
    \end{aligned}
\end{equation}

We assume a blackbody spectrum and perform the same analysis as in \S\ref{sec:P15_model} to extract apparent magnitudes from the predicted luminosity and temperature. We model the flare by the same methods and present light curve fits for an $n=3/2$ and $n=3$ polytropes as solid and dashed lines, respectively, in Figure \ref{fig:shock_cool}. We find that the first flare component at $t < 2$d can be fit accurately with our MCMC model. For the first peak within the flare, we estimate envelope radii and masses of $R_e \approx 40[30] \ \Rsun$ and $M_e \approx 0.8[0.2] \ \Msun$ for n = 3[3/2] polytropes. The MCMC routine, however, does not formally converge when we attempt to fit the entire data set at $t < 6$d. In Figure \ref{fig:shock_cool} we show a representative model, with parameter values indicated in Table \ref{tbl:shocktable}. These values should be treated as order of magnitude estimates.  

We end by noting that the model by \cite{sapir17} is valid for times:
\begin{equation}\label{eq:SW16_t1}
    t > 0.2 \frac{R_{13}}{v_{s,8.5}} \textrm{max}\Bigg[0.5, \frac{R_{13}^{0.4}}{(f_p\kappa_{0.34}M)^{0.2}v_{s,8.5}^{0.7}}\Bigg] \textrm{days}
\end{equation}

\begin{equation}\label{eq:SW16_t2}
    t < 7.4 \Big(\frac{R_{13}}{\kappa_{0.34}}\Big)^{0.55} \textrm{days}
\end{equation}
\noindent
We test the validity of our derived model parameters (Table \ref{tbl:shocktable}) with Equations \ref{eq:SW16_t1} and \ref{eq:SW16_t2} and we find that our model parameters satisfy the relations above. For the first peak in the flare we find: $\sim 0 < t < 4.09$ days (n = 3/2) and $\sim 0 < t < 8.94$ days (n = 3). For the second peak we find: $\sim 0 < t < 4.31$ days (n = 3/2) and $\sim 0 < t < 9.74$ days (n = 3). All derived timescales are valid for the duration of the flare. 

In the previous three subsections we have investigated a shock cooling model as a power source for the flare. Because of its temporal structure, we have modeled the flare in two components ($t<2$ and $t < 6$ days) in order to derive physical parameters (e.g., radius, mass, velocity) of a shock heated envelope needed to match optical the optical light curve. Figure \ref{fig:shock_cool} demonstrates that modeling the entire flare with one shock cooling model cannot reproduce the observations but the corresponding radii and masses for each model represent upper limits on the total amount of shocked material capable of powering the flare.

\subsection{CSM Interaction Model}\label{sec:flare_csm}

Another potential source of energy to power the optical flare emission is via ongoing SN shock interaction with the medium. This scenario has physical similarities to that discussed in \S\ref{sec:flare_shockcool} with the key difference being that rather than powering this rapid light curve peak via post-breakout cooling emission, the CSM interaction model allows for continuous energy injection due to the ongoing conversion of shock kinetic energy into radiation. The presence of CSM around the SN~2019ehk progenitor is evident given the detection of flash-ionized H and He features in the first optical spectrum at 1.45 days since explosion. The estimated blackbody radius at the time of the first spectrum is $\leq 4\times10^{14}$cm (\S\ref{subsec:bol_LC}) and the velocities of H- and He-rich material are $\sim 400$ and $500\kms$, respectively (\S\ref{subsec:FS}). The flash-ionized CSM lies in front of the photosphere at radii $>4\times10^{14}$cm. Therefore, this H+He rich material was lost by the stellar progenitor to the environment $\gtrsim$ 3 months prior to explosion. 

We quantitatively test the scenario of a SN shock interacting with a shell of CSM through 1D numerical radiation hydrodynamics simulations with the \texttt{CASTRO} code \citep{almgren10}. Equations for radiation hydrodynamics are solved using a gray flux-limited non-equilibrium diffusion approximation. The models are similar to those applied to the SN~Ic-BL, 2018gep \citep{ho19} and the fast-evolving luminous transient KSN~2015K \citep{rest18}, but have been adapted to the observables in SN~2019ehk. 

Our simulations assume spherical symmetry wherein the SN ejecta expands homologously and is characterized by a broken power-law density profile ($\rho_{\rm{ej}} \propto r^{-n}$, with $n=3$), ejecta mass $M_{\rm{ej}}$, energy $E_{ej}$, initial outer radius $R_{\rm{ej}}$, outer velocity $v_{\rm{ej}}$ and ejecta temperature $T_{\rm{ej}} = 10^4$ K. The CSM shell is assumed to have constant density and is initialized with temperature $T_{\rm{csm}} = 10^3$ K. We adopt a static CSM (i.e. $v_{\rm{csm}} = 0\, \kms$) whose velocity has no affect on the model results so long as $v_{\rm{csm}} << v_{\rm{ej}}$. The shell is described physically by its mass $M_{\rm{csm}}$, radius $R_{\rm{csm}}$ and thickness $\delta R_{\rm{csm}}$. Once the ejecta have reached homology we use the radiative transfer code \texttt{Sedona} \citep{kasen06} to generate synthetic bolometric light curves as well as the temporal evolution of the effective blackbody temperature and radius in each model. Unlike other CSM interaction codes (e.g., MOSFIT, \citealt{guillochon18}; TigerFit, \citealt{Chatzopoulos16}) that use the semi-analytic Arnett approximation with a parameterized heating term, our simulations self-consistently solve for the time-dependent light curves by evolving the coupled radiation hydrodynamics equations with \texttt{CASTRO}.

From a grid of shock interaction simulations, we find that the first component of the flare is best fit by shock breakout emission into a CSM characterized by the following parameters: mass $M_{\rm{csm}} = 1.5\times10^{-3} \ \Msun$, radius $R_{\rm{csm}} = 2\times 10^{14}$ cm, thickness $\delta R_{\rm{csm}} = 4\times 10^{13}$ cm and opacity $\kappa = 0.4$ cm$^{2}$ g$^{-1}$. This model was initialized for a SN with $M_{\rm{ej}} \approx 1 \ \Msun$, which is based on observations as constrained by our modeling of  \S\ref{subsec:bol_LC}. This model is presented with respect to SN~2019ehk's bolometric luminosity, temperature and radius evolution during the flare in Figure \ref{fig:flare_lum}. We also show a CSM interaction model that is able to power the entire flare ($t < 7$d) with $M_{\rm{csm}} = 7\times10^{-3} \ \Msun$ and the same physical parameters as above. These CSM properties are consistent with the masses independently inferred from the optical spectral modeling of \S\ref{subsec:FS} and X-ray modeling of \S\ref{SubSec:Xraydensity}.

\section{Radio/X-ray data Modeling} \label{Sec:Radio_Xray_Modeling}

\subsection{Inferences on the explosion's local environment from X-ray observations}
\label{SubSec:Xraydensity}
The luminous ($L_x$$\approx $$10^{41}\,\rm{erg\,s^{-1}}$), rapidly-decaying X-ray emission ($L_x\propto t^{-3}$) with a hard spectrum is consistent with thermal bremsstrahlung from shocked CSM gas in adiabatic expansion. In this scenario the X-ray luminosity scales as the emission measure $EM=\int n_e n_I dV$, and $EM \propto r^{-3}$$ \propto $$ t^{-3}$ once the shock has swept up most of the CSM gas.  For $n_e\approx n_I$, the $EM$ measured from the first epoch of X-ray observations at $\sim2.8$ d indicates a particle density $n\approx  10^{9}R_{\rm{csm,15}}^{-1}\delta R_{\rm{csm,15}}^{-0.5}f^{-0.5}\,\rm{cm^{-3}}$, where $R_{\rm{csm,15}}$ and $\delta R_{\rm{csm,15}}$ are the radius and thickness of the shocked shell of gas in units of $10^{15}\,\rm{cm}$, respectively, and $f$ is a volume filling factor.  This density estimate is remarkably similar to the density of the pre-shocked CSM gas that we have inferred from the H and He recombination lines (\S\ref{subsec:FS}). The inferred mass of the shocked gas is $M_{\rm{csm}}\approx0.01 R_{\rm{csm,15}}^{-1}\delta R_{\rm{csm,15}}^{-0.5}f^{-0.5}\,\rm{M_{\odot}}$.

For a typical SN shock velocity of $\sim0.1c$, the forward shock radius at $2.8$ d is $r\approx 7\times 10^{14}\,\rm{cm}$. The disappearance of the H and He recombination lines by $2.4$ d post explosion and the rapid fading of the X-ray luminosity detected at 2.8 d indicate that the shock has overtaken the shell of CSM by this time. Using $R_{\rm{csm}}\approx 7\times 10^{14}\,\rm{cm}$ and assuming $\delta R_{\rm{csm}}\approx R_{\rm{csm}}$ we infer a particle density of $n\approx  10^{9}\,\rm{cm^{-3}}$ and a total CSM shell mass of $M_{\rm{csm}}\approx7\times 10^{-3}\,\rm{M_{\odot}}$ (for $f=1$). This result is consistent with the mass of pre-shocked CSM gas $\sim2\times 10^{-3}\,\rm{M_{\odot}}$ that was in front of the shock at $t=1.4$ d since explosion derived in \S\ref{subsec:FS}. Together with the modeling of the flare optical continuum of \S\ref{sec:flare}, these  results strengthen the scenario where the detected X-rays and continuum optical emission originate from pre-existing H/He rich CSM shocked by the SN blastwave, while the H and He recombination lines result from pre-shocked CSM gas lying in front of the SN shock and ionized by its X-ray emission. If the chemical composition of the entire shell is similar to that constrained by the H+He emission lines of \S\ref{subsec:FS}, and under the assumption of $f\approx 1$, the total CSM H mass is in the range $(4.- 17.) \times 10^{-4} \ \Msun$ and the total CSM He mass is constrained within the range $(5.3 - 6.7)\times 10^{-3} \ \Msun$.

\subsection{Inferences on the explosion's environment at $R\ge10^{16}\,\rm{cm}$ from radio observations}
\label{SubSec:radiomassloss}

\begin{figure}[h]
\centering
\includegraphics[width=0.49\textwidth]{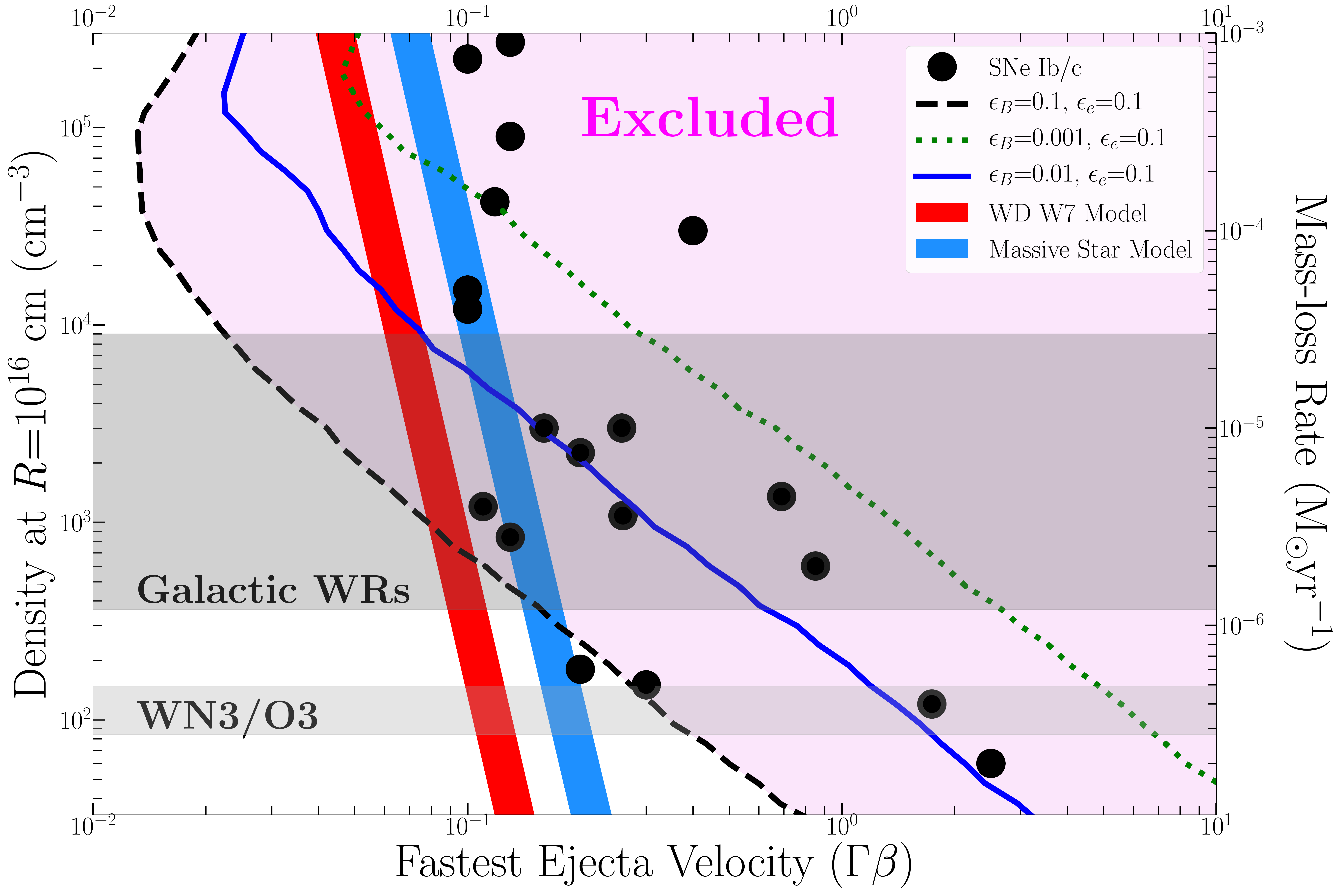} \caption{Environment density $\rho_{\rm{CSM}}\propto r^{-2}$ vs. shock velocity parameter space.  Radio non-detections of SN\,2019ehk rule out the vast majority of the parameter space of Ib/c SNe (black dots, \citealt{drout16}), for which $\epsilon_B=0.1$ and $\epsilon_e=0.1$ are typically assumed (black dashed line). The parameter space to the right of the thick blue line and green dotted line is ruled out for a different choice of microphysical parameters ($\epsilon_B=0.01$ and  $\epsilon_B=0.001$, respectively).   Red (blue) band: range of SN~2019ehk shock velocities during our radio monitoring ($\delta t=30-220 $ d) for an explosion with $E_{\rm k}=1.8\times 10^{50}\,\rm{erg}$ and $M_{\rm{ej}}=0.7\,\rm{M_{\odot}}$ (\S\ref{subsec:bol_LC}) and a massive star (blue) or WD (red) outer ejecta density profile. Grey shaded regions: range of mass-loss rates $\dot M$ for Galactic WRs \citep{crowther07, massey15} for a wind velocity $v_w=1000\,\rm{km\,s^{-1}}$.  \label{fig:mdot}}
\end{figure}

We interpret the radio upper limits of \S\ref{SubSec:VLA} in the context of synchrotron emission from electrons accelerated to relativistic speeds at the explosion's forward shock, as the SN shock expands into the medium. We adopt the synchrotron self-absorption (SSA) formalism by \cite{Chevalier98} and we self-consistently account for free-free absorption (FFA) following \cite{Weiler02}. For the calculation of the free-free optical depth $\tau_{\rm ff}(\nu)$ we adopt a wind-like density profile $\rho_{\rm{csm}}\propto r^{-2}$ in front of the shock, and we conservatively assume a gas temperature $T=10^4\,\rm{K}$ (higher gas temperatures  would lead to tighter density constraints). The resulting SSA+FFA synchrotron spectral energy distribution depends on the radius of the emitting region, the magnetic field, the environment density and on the shock microphysical parameters $\epsilon_B$ and $\epsilon_e$ (i.e.~the fraction of post-shock energy density in magnetic fields and relativistic electrons, respectively). 

\begin{figure*}
\centering
\includegraphics[width=0.95\textwidth]{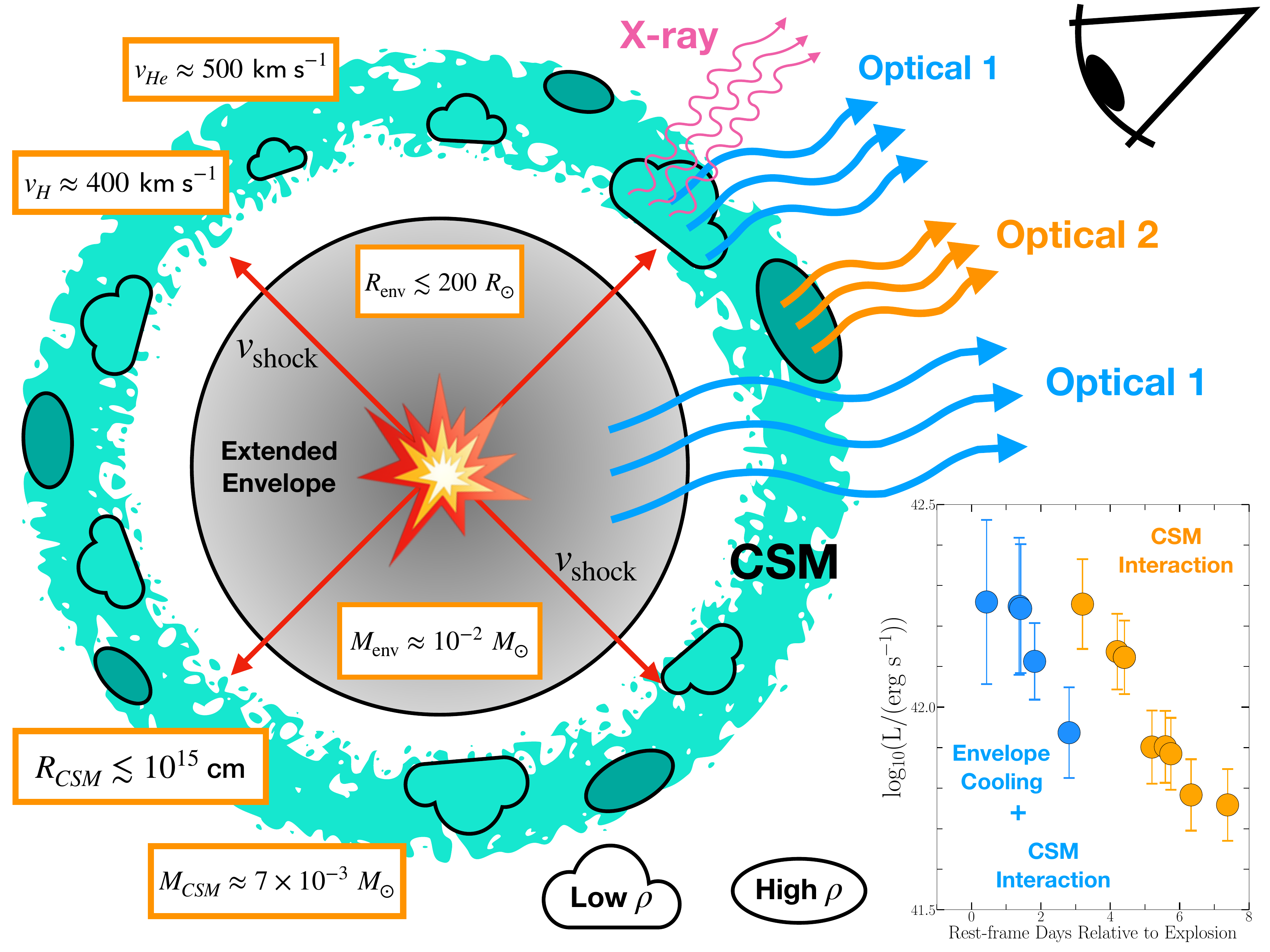} \caption{Visual representation  of SN~2019ehk's progenitor environment at the time of explosion (\S\ref{subsec:model}). Here, the SN shock breaks out from an extended envelope and collides with lower density, outer CSM, inducing X-ray emission and flash-ionized spectral lines. A combination of envelope cooling and shock interaction produces the first part of the flare (blue light curve points), while high density or ``clumpy'' CSM causes delayed optical emission at $t>2$d (orange light curve points). CSM velocities and abundances are derived from flash-ionized spectral lines, while the total mass is calculated from X-ray detections. The physical scale and mass of the inner extended material are estimated from shock cooling models. The bolometric light curve during the flare is presented in lower right for reference. \label{fig:model}}
\end{figure*}

\begin{figure*}
\centering
\includegraphics[width=0.95\textwidth]{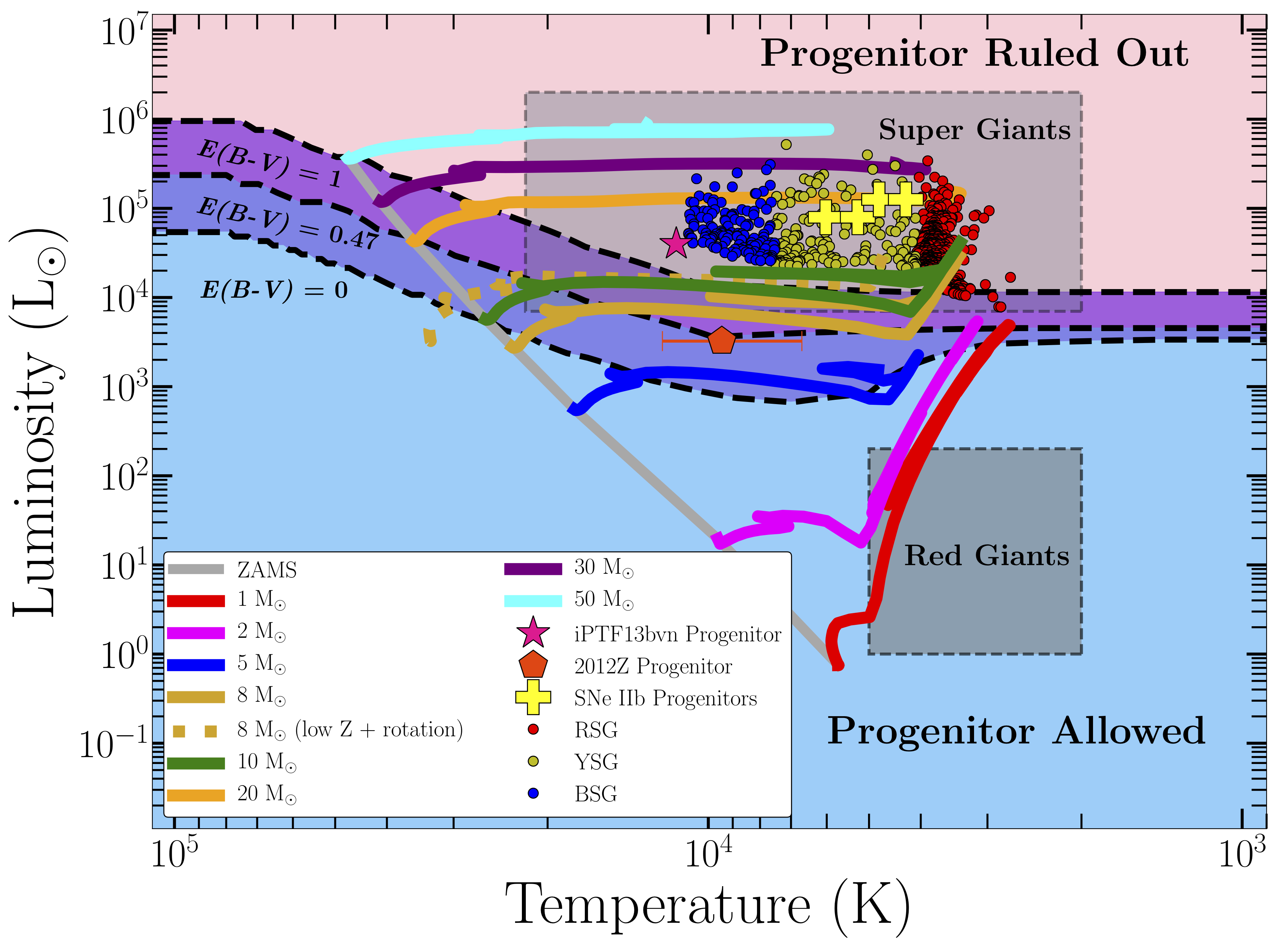} \caption{
H-R diagram showing the inferred limits on the stellar progenitor of SN~2019ehk from pre-explosion HST imaging. Permitted regions for different local extinctions are shown as dashed lines in the shaded violet-to-blue region and the excluded region is presented in pink. We plot \texttt{MESA} stellar evolutionary tracks from 1-50$\,\Msun$ single stars with no rotation, and solar metallicity as solid lines. For completeness, an $8 \ \Msun$ single star track with rotation and sub-solar metallicity is plotted as a dashed line. The progenitors of SN~Ib iPTF13bvn and SN~Iax, 2012Z are displayed as a pink star and a red diamond, while progenitors of SNe~IIb are shown as yellow plus signs \citep{maund11, cao13, mccully14}. Dashed grey squares represent the range of Supergiants (top) and Red Giants (bottom). A representative sample of Red, Yellow and Blue supergiants in the LMC are plotted as circles \citep{neugent12}. With the most conservative choice of local extinction ($E(B-V)=1$ mag) the HST limits rule out all single massive stars capable of exploding, while a realistic choice of extinction correction ($E(B-V)=0.47$ mag) extends the masses of single stars progenitors that are ruled out to $\gtrsim 8 \ \Msun$.  \label{fig:HR}}
\end{figure*}

Figure \ref{fig:mdot} shows the part of the density vs. shock velocity parameter space that is ruled out by the upper limits on the radio emission from SN\,2019ehk for three choices of microphysical parameters. Specifically, we show the results for $\epsilon_B\approx0.1$ and $\epsilon_e\approx0.1$ (which have been widely used in the SN literature) to allow a direct comparison with other SNe (black dots in Figure \ref{fig:mdot}). We find that SN\,2019ehk shows a combination of lower environment density and lower shock velocity  when compared to core-collapse SNe with radio detections. As a final step, we self-consistently solve for the shock dynamics in a wind medium adopting the explosion's parameters inferred in \S\ref{subsec:bol_LC} (kinetic energy $E_{\rm k}\approx1.8\times 10^{50}\,\rm{erg}$ and ejecta mass $M_{\rm{ej}}\approx0.7\,\rm{M_{\odot}}$). We show the resulting shock velocity $\Gamma\beta$ as a function of the environment density for an outer density profile of the ejecta of the exploding star typical of compact massive stars ($\rho_{\rm{ej}}\propto v^{-n}$ with $n\approx10$, \citealt{Matzner99}) or relativistic WDs (e.g., \citealt{Chomiuk12} and references therein). The SN shock decelerates with time as  it plows through the medium. Figure \ref{fig:mdot} illustrates the range of shock velocities during the time of our radio observations at $\delta t\approx 30-220$ d for the two choices of stellar progenitors. For more realistic choices of microphysical parameters  ($\epsilon_B=0.01$, $\epsilon_e=0.1$), our results imply a mass-loss rate limit $\dot M<10^{-5}\,\rm{M_{\odot}yr^{-1}}$ for an assumed wind velocity $v_w=500\,\rm{km\,s^{-1}}$ similar to the observed velocities of H and He-rich material (Figure \ref{fig:FS_vels}). This limit applies to distances $r\approx10^{16}-10^{17}\,\rm{cm}$ from the explosion site, and it is shown in Figure \ref{fig:density} in the context of  predictions from WD merger models. These merger simulations are discussed in greater detail in \S\ref{SubSec:wd_models}.

\begin{figure}[h]
\centering
\includegraphics[width=0.45\textwidth]{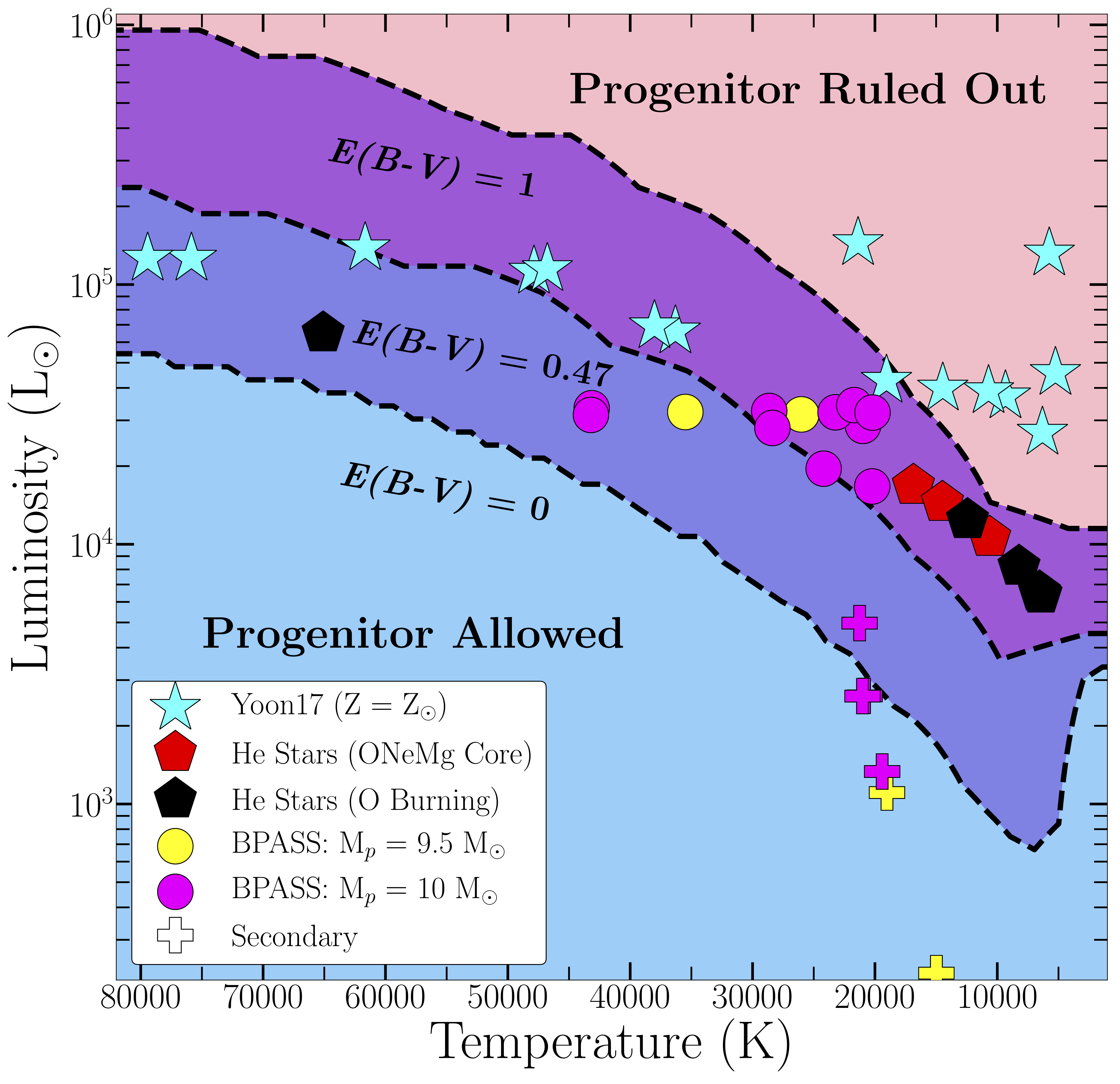} \caption{\texttt{BPASS} models consistent with pre-explosion limits and SN properties. Primary and secondary stars are shown as circles and plus signs, respectively. Cyan stars represent the primary star in binary models by \cite{yoon17} that result in SNe~IIb/Ib. Helium star models where different amounts of the envelope is removed are shown as polygons. Final states of these models are either an ONeMg core (red) or O-burning (black) in the core. The same color coding as Fig. \ref{fig:HR} is used to indicate allowed regions of the parameter space for different intrinsic $E(B-V)$. \label{fig:binary}}
\end{figure}

\section{Discussion} \label{Sec:discussion}

\subsection{A Physical Progenitor Model} \label{subsec:model}

Panchromatic observations have provided an unprecedented picture of this \ca \ both before and after explosion. In Figure \ref{fig:model}, we attempt to combine inferences made from observation and modeling to create a visualization of the explosion and surrounding environment. Our model is a snapshot of the SN at explosion and contains physical scales and parameters such as distance, velocity and composition estimates. 

It is most likely the case that the flare is powered by shock interaction or cooling emission in an extended mass of material, regardless of the type of progenitor that exploded. The progenitor could have accrued an extended envelope located at $<200~\Rsun$ (light grey circle; Fig. \ref{fig:model}), while mass-loss in the progenitor's final months may have placed H- and He-rich material in the circumstellar environment (shown in sea foam green; Fig. \ref{fig:model}) with velocities of $\sim \ 400-500 \ \kms$ and at distances $\lesssim 10^{15}$ cm. The detection of early-time X-ray emission and flash-ionized H and He spectral lines is clear evidence for a SN shock colliding with removed CSM. The observed CSM velocities might be difficult to explain given typical WD escape velocities of $\gtrsim 1000 \ \kms$ needed for mass ejection from a WD surface. However, material might be ejected at low velocities during mass-transfer in WD binaries prior to the merger (see \S\ref{SubSec:wd_models}).

Based on our modeling of the flare in \S \ref{sec:flare_shockcool} and \S\ref{sec:flare_csm}, we propose a physical scenario that could have produced this first optical light curve peak. In the picture, the flare is powered by two physically distinct emission components: shock interaction with more distant CSM in addition to the cooling of hot, shocked material a smaller radii. Following shock breakout, the inner extended envelope will cool, producing some of the emission on timescales $t<2$ days (blue light curve points; Fig. \ref{fig:model}). Once the shock collides with more distant H- and He-rich CSM it will induce ``flash-ionized'' spectral lines that are powered until 1.5 days via X-ray emission from the shock propagating through the CSM shell. The same low density region of the CSM responsible for X-rays and narrow emission lines can also power the early-time light curve ($t < 2$d). Our analysis has indicated that this shell had a mass of $\sim 7 \times 10^{-3} \ \Msun$ and is located between $4\times 10^{13}$ - $10^{15}$ cm from the progenitor. 

At $t > 2$d, the flare's power source and the complete explosion picture becomes more ambiguous. By the start of the main peak of the flare (orange light curve points; Fig. \ref{fig:model}), the narrow emission lines are no longer detectable and the X-ray emission from the initial shock is rapidly decaying, suggesting that the shock has overtaken the entire CSM shell. Here we propose two plausible expanations for the rapid increase in flux at $t \approx 2$d. (i) Delayed optical emission from the high density, optically thick regions of the CSM shell begin to cool and radiate in the optical bands following shock interaction. (ii) The shock encounters additional CSM material at $r > 10^{15}$cm which induces optical emission from shock interaction. 

While this physical progenitor model does account for most of the observables, there are many caveats and unknowns about such a system. First, this model assumes spherically symmetric distributions of mass, both in the inner extended envelope and the outer CSM. Alternatively, this material could have formed a torus where more mass is located in the equatorial regions rather than at the poles. Secondly, neither the shock cooling (\S\ref{sec:flare_shockcool}) nor the shock interaction models (\S\ref{sec:flare_csm}) takes into account the chemical composition of the shocked material that then causes the flare. It is likely that the extended masses have significant density gradients, which could lead to variations on how the radiation is able escape the material. Such a scenario would be best tested through numerical modeling (e.g., \citealt{piro17}) in which the density gradients and composition are taken into account, but is ultimately beyond the scope of this paper. Nonetheless, our observations have allowed for the most complete picture of \ca \ explosion mechanisms and their circumstellar environments. In the following sections we discuss the stellar systems capable of producing the SN~2019ehk observables. 

\subsection{Pre-Explosion Constraints on a Massive star Progenitor} \label{subsec:HR_progenitors}
Figure \ref{fig:HR} shows the constraints on the progenitor system of SN~2019ehk in the H-R diagram, as derived from pre-explosion HST multi-band imaging. In the context of single stars, only compact objects (e.g., WD, NS, BH) and massive stars ($8-10~\Msun$) are consistent with observations. Specifically, we plot the \texttt{MESA} evolutionary tracks \citep{choi16} of non-rotating single massive stars with the same metallicity as the host galaxy ($Z=Z_{\sun}$). We find that only stars with mass $\sim$8-10 $\Msun$ satisfy the limits for the most extreme choice of intrinsic $E(B-V)\approx 1$ mag. This is also true for low metallicity stellar tracks with rotation included e.g., dashed goldenrod line of 8~$\Msun$ progenitor. However, a more realistic choice of intrinsic $E(B-V) = 0.47$~mag would effectively rule out the vast majority of parameter space corresponding to various types of single massive stars ($\gtrsim 8~\Msun$). Furthermore, we explore the potential of a single He star progenitor (Table \ref{tbl:he_star_table}) that would be responsible for a core-collapse SN~Ib-like explosion. As shown in Figure \ref{fig:binary}, this model is only consistent with the most highly reddened pre-explosion limits and requires a mechanism to remove its outer H-rich envelope. Overall, we conclude that single massive stars are unlikely progenitors of SN~2019ehk.

We then explore the possibility of a binary progenitor system. To this aim, we employ the large grid of Binary Population and Spectral Synthesis (\texttt{BPASS}) models  by \cite{eldridge17} to find binary systems that fit the observational parameters of SN~2019ehk. Firstly, we exclude binary models whose final luminosity and temperature do not reside within the ``allowed'' parameter region of Figure \ref{fig:HR} (blue shaded regions). This includes the final luminosity and temperature of both the primary and secondary stars; neither of which should be detected in archival \textit{HST} imaging. Additionally, we only include systems whose final helium mass is $>0.1$~$\Msun$ and final hydrogen mass is $<0.01$~$\Msun$ (e.g., \S\ref{SubSec:Spec}). To meet the \texttt{BPASS} condition for a resulting SN, we only include systems where the primary's CO core mass is >1.35$\Msun$ and total mass is >1.5 $\Msun$. Following these conditions, we look for systems whose ejecta mass is $< 1.0\, \Msun$ for a weak SN explosion ($E_{\rm k}\approx10^{50}$erg), both of which are inferred from observations (\S\ref{subsec:bol_LC}). When this cut is made on predicted ejecta mass, we recover \emph{no} consistent binary systems within the SN~2019ehk parameter space. However, because parameters associated with a predicted SN in \texttt{BPASS} are uncertain, we choose to include systems that have a predicted ejecta mass $M_{\rm{ej}} < 2 \ \Msun$ for completeness. We plot the final luminosities and temperatures of 13 potential binary systems in Figure \ref{fig:binary} and display significant BPASS parameters of each model in Table \ref{tbl:bp_table1}. Overall, these binary configurations have primary stars with masses of $9.5-10$~$\Msun$ and radii $<15$~$\Rsun$. 

We further test the possibility that SN~2019ehk is the result of a more exotic binary system through He-star modeling in \texttt{MESA}. We initialize $2.7-3.0$~$\Msun$ He-stars with C/O cores and track their luminosity and temperature evolution until the exhaustion of He-burning and the onset of O core burning or the formation of an ONeMg core. We test the following mass-loss scenarios: no mass-loss, standard Wolf-Rayet (WR) winds, artificial envelope removal and binary interaction with NS companion (with varying orbital periods). We present the specifics of each model in Table \ref{tbl:he_star_table} and plot each final luminosity/temperature as red and black polygons in Figure \ref{fig:binary}. These are compared to binary models in \cite{yoon17} that result in normal SNe~Ib/IIb (plotted as cyan stars).  

Overall, our presented He-star models are consistent with the pre-explosion parameter space for host extinctions of $E(B-V)$ = 0.5 - 1 mag. We can rule out some of these systems based on the final mass if we assume that the total ejecta mass will be this mass minus $\sim$1.4~$\Msun$. The estimated ejecta mass in SN~2019ehk is $\sim$0.7~$\Msun$, which is consistent with an artificial envelope removal (models \#2, 4) and a He-star + NS binary (models \#7,8), both ending in O core burning. However, these models do not naturally reconcile the presence of H-rich CSM in the SN~2019ehk progenitor environment. 

We can further constrain the presence of a dusty progenitor for SN~2019ehk by utilizing the \emph{Spitzer} pre-explosion limits (Table \ref{tbl:spitzer_table}). We use the most constraining limit of $>23.87$~mag from Channel 2 and assume that the majority of the flux is emitted about an effective wavelength of $\lambda_{\rm eff} = 4.493~\mu$m. We then apply the spherically symmetric dust shell model shown in Equation 1 of \cite{kilpatrick18a}. As in their study, we also assume that the dust shell emits isotropically in the optically thin limit \citep{fox10} and have a flux density that goes as $F_{\nu} \approx M_d B_{\nu}(T) \kappa_{\nu} / d^2$, where $M_d$ is the shell mass, $d$ is the distance to SN~2019ehk and $B_{\nu}(T)$ is the Planck function. Applying this simple approximation, we derive dust shell masses limits of $< 6.6\times10^{-8} - 5.3\times 10^{-6}~\Msun$ for  shell temperatures $T_{\rm s} = 1500 - 500$~K, respectively. 

Our inferred dust mass is a factor $\sim4$ smaller than that derived by \cite{kilpatrick18a} for LBV outburst Gaia16cfr and an order of magnitude lower than typical dust masses observed around type IIn SNe \citep{fox11}. Furthermore, the total dust luminosity of the Gaia16cfr progenitor was $2.4 \times 10^5$~L$_{\odot}$, which is more than an order of magnitude larger than the NIR F160W \emph{HST} pre-explosion limits (e.g., Fig. \ref{fig:HR}). Since our derived dust shell mass is similar to Gaia16cfr, a massive star progenitor with a small dust shell would have been detected in pre-explosion images of the SN explosion site. While this analysis is highly simplified, our findings make a dusty progenitor for SN~2019ehk highly unlikely given the observations. 

Finally, it should be noted that the luminosity limit derived from \textit{Chandra} pre-explosion imaging does not constrain the existence of a luminous supersoft X-ray source (SSS) at the location of SN~2019ehk. Such a system has been invoked as a precursor to SNe~Ia wherein a nuclear-burning WD accretes mass from a non-degenerate companion. This process in turn produces X-ray luminosities of order $10^{38}$~erg~s$^{-1}$. However, it has been demonstrated that there are not enough observed SSSs that retain luminous X-ray emission on the same timescale as is needed for quasi-steady burning on the WD surface \citep{stefano10}. Therefore a single-degenerate scenario, or related event, cannot be constrained with our current \textit{Chandra} X-ray limits. 

From this analysis, we can rule out all single massive stars $> 8\ \Msun$ as progenitors of SN~2019ehk. With regards to binary systems, the pre-explosion parameter space allows for only the lowest mass massive star binaries ($9.5-10 \ \Msun$) or He stars whose envelopes are removed through a mass-loss mechanism. However, while our pre-explosion limits greatly constrain the massive star parameter space, progenitor systems involving a WD cannot be excluded based on detection limits. 

\subsection{White Dwarf Explosion Models}
\label{SubSec:wd_models}
 
Given the pre-explosion limits, every progenitor system involving a WD is permitted in the progenitor parameter space of SN~2019ehk. Nevertheless, we can exclude some of these scenarios based on observed properties of the explosion. As shown in Figure \ref{fig:HR}, the progenitor of SN~Iax, 2012Z is not ruled out and has been proposed to be a He star + WD binary \citep{mccully14}. However, this progenitor channel cannot account for the H-rich material observed in SN~2019ehk's circumstellar environment nor the photospheric He in its spectra without significant buildup of unburned He on the WD surface at the time of explosion. Furthermore, explosion models for this configuration generally produce SN~Ia, or Iax-like events from failed detonation/deflagration \citep{Jor+12,Kro+13} that do not match the observed photometric or spectroscopic evolution of \cas. The same reasoning rules out a main sequence (MS) companion model typical of SN~Ia models. Because common single degenerate progenitor channels appear unlikely for SN~2019ehk, we explore double degenerate explosion scenarios capable of reproducing \ca \ observables such as those from SN~2019ehk.

Recently, \cite{perets19} suggested a double WD (DWD) merger scenario for the origin of SNe Ia, where a CO-WD merges with a hybrid HeCO WD (\citealt{Zenati2019} and references therein). In this hybrid + CO DWD (HybCO) model, the disruption of a hybrid WD by a more massive (>0.75 $M_\odot$) CO WD can give rise to normal SNe Ia \citep{perets19}, through a detonation of a He-mixed material on the CO WD surface, followed by a detonation of a CO core due to its compression by the first He-detonation. In cases where the primary WD was of a low mass ($\lesssim 0.65$ M$_\odot$), only the first He-detonation occurs while the CO core is left intact leaving a remnant WD behind. In such cases, and in particular when the progenitor is a hybrid-WD disrupting a lower-mass CO WD (or another hybrid WD), Zenati et al. 2020 (in prep.) find that the He-detonation gives rise to a faint transient, potentially consistent with \cas. 

In this specific double-degenerate channel, mass that is lost from the secondary WD prior to its disruption can give rise to CSM, possibly consistent with the observations of SN~2019ehk, as we describe below (a more detailed discussion will be provided in Bobrick et al. 2020, in prep.). This scenario has been explored in the context of SNe~Ia wherein the merger is preceded by the ejection of mass as ``tidal tails" and placed at distances $r \approx 10^{15}$~cm \citep{raskin13}. Further in, material around the primary WD can ``settle down'' to form an extended envelope ($r \approx 10^{11}$~cm); this process can occur on timescales of $< 1000$~yrs before merger \citep{shen12, schwab16}.

Before the actual merger, DWDs spiral-in due to gravitational wave emission. As the binary components gradually come into contact and the donor starts losing mass, the mass transfer rate in the system gradually grows, starting from small values below $10^{-12} \ \Msun {\rm yr}^{-1}$ and continually increasing, which leads to the eventual disruption over several years' time. Mass-loss during this phase leads to material ejected at typical velocities of likely a few hundreds up to thousand $\kms$, which expands to characteristic radii of $10^{15}-10^{16}\,{\rm cm}$  by the time the actual merger happens, while some material could be ejected shortly before the final merger of the WDs. Here we focus on the mass transfer and ejection prior to the merger/disruption of the WDs, which can contribute to the CSM far from the WD and may explain the observations. \cite{Lev+17} discussed the possibility of very high velocity CSM from material ejected after the disruption of a WD, and just shortly before the merger; this is however unlikely to explain, or be consistent with, the observations shown here.     

We compute the density distribution in the ejecta by solving the equations of secular evolution of the mass transfer rate and binary orbital properties (masses and separation, $M_1, M_2, a$) in DWD binaries driven by gravitational wave emission starting from early phases of mass transfer (see e.g., \citealt{Marsh2004,Gokhale2007,Bobrick2017}). We represent the WD donor by one-dimensional, corotating and perfectly-degenerate models following the Helmholtz equation of state \citet{Timmes2000} and calculate the mass transfer rate following \citet{Kolb1990}. The binaries are evolved from the moment mass transfer rate reaches $10^{-12} \ \Msun {\rm yr}^{-1}$ until the mass transfer rate reaches $10^{-2} \ \Msun {\rm yr}^{-1}$, shortly before the merger. We assume that a fixed fraction of mass is ejected from the systems at some characteristic velocity during the process of mass transfer, both parameters being free parameters of the model.
    
\begin{table}
\centering
\begin{tabular}{c c c c c} 
 \hline
 Model Name & $M_{\rm donor} + M_{\rm acc}$ & $f_{\rm ej}$ & $v_{\rm ej}$ & Abund.\\
  & ($\Msun$) & & ($\kms$) & (donor)\\
 \hline
 Fiducial & $0.5+0.6$ & $0.99$ & $300$ & CO \\ 
 Reduced mass loss & $0.5+0.6$ & $0.1$ & $500$ & CO \\
 Fast ejecta & $0.5+0.6$ & $0.99$ & $1000$ & CO \\
 Heavy accretor & $0.5+0.9$ & $0.99$ & $500$ & CO \\
 Hybrid donor & $0.53+0.6$ & $0.99$ & $500$ & HeCO \\ 
 Super-Chandra & $0.75+0.95$ & $0.99$ & $500$ & CO \\
 \hline
\end{tabular}
\caption{WD explosion models presented in \S\ref{SubSec:wd_models}. The columns show the model name, the masses of the primary and the secondary in solar masses, the fraction of the transferred material which is ejected from the system, the velocity of the ejecta and the chemical composition of the donor. The accretor has a C/O composition.}
\label{tab:DWDModels}
\end{table}

\begin{figure}[h]
\centering
\includegraphics[width=0.49\textwidth]{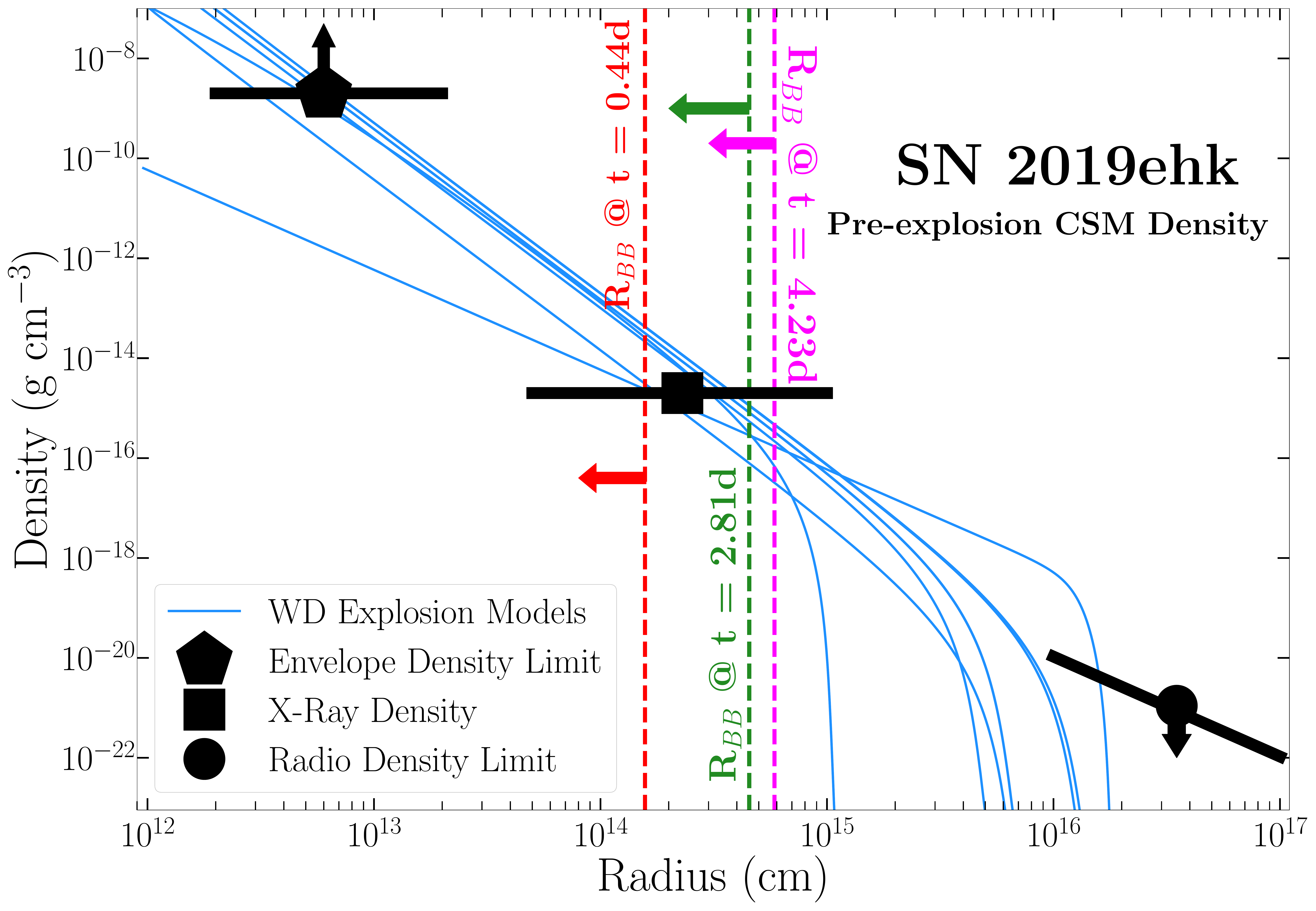} \caption{Density profile of the SN~2019ehk explosion environment. Shown as black squares are density limits derived from X-ray detections and presented at radii derived from blackbody modeling. The black circle is the density limit derived from modeling of the radio non-detections. Blue lines are CSM models for WD mergers at the time of explosion (see \S\ref{SubSec:wd_models}). \label{fig:density}}
\end{figure}

We explored several physically-motivated cases which cover most of the parameter space of possible assumptions in the model, as summarised in Table~\ref{tab:DWDModels}. As the fiducial model, we chose a $0.5+0.6\,M_\odot$ DWD binary, which represents potential progenitors of CaSTs in the hybCO scenario. In the fiducial model, we assume that $99$\% of mass is lost due to direct-impact accretion expected in these binaries and we assigned ejecta velocities of $500 \ \kms$, comparable to the orbital velocity in the binary. In the exploratory models, we considered the cases where only $10$\% of mass is lost, where the ejecta is launched at $1000 \ \kms$, where the accretor is a $0.9\,M_\odot$ CO WD and where the donor is a hybrid HeCO WD $0.53\,M_\odot$, based on the detailed model from \citet{Zenati2019}. Additionally, we simulated a super-Chandrasekhar binary with 0.75+0.95~$\Msun$ CO WDs, which is expected to produce brighter SNe Ia instead. As may be seen from Figure~\ref{fig:density},
the density distributions from the models agree well with the density limits derived from the X-ray detections, flash-ionized spectral lines and radio non-detections. The agreement is also robust to the assumptions in the model, apart from the model with the $0.9\,M_\odot$ CO accretor, for which the ejecta density at late times (small radii) disagrees with the X-ray limits. Indeed, this latter case is not expected to give rise to a CaST SN in the HybCO model. 

Throughout the evolution, mass transfer gradually peels the donor starting from the outermost layers, and therefore the ejected mass inherits the composition profiles of the donor WD. We use  \texttt{MESA} models of WDs stripped during binary evolution and find that $0.53\,M_\odot$ CO WDs contain about $3\times10^{-3}\,M_\odot$ of H, while hybrid WDs contain less. For example, a $0.53 \ \Msun$ HeCO WD model contains only $2\times 10^{-5}\,M_\odot$ of H and is based on the models from \citet{Zenati2019}. In contrast, models of single WDs predict $\sim$ $10^{-4}~\Msun$ of surface H \citep{Lawlor2006} for low-mass WDs ($\lesssim0.6$ M$_\odot$) and orders of magnitude lower H abundances on higher mass WDs. Since H is initially in the outermost layer of the donor, it ends up in the outermost parts of the CSM, being replaced-by/mixed-with He at typical separations $10^{14}$--$10^{15}\,{\rm cm}$, assuming H layers between $10^{-3}\,M_\odot$ and $10^{-4}\,M_\odot$. Depending on the mass of the He layer, He is replaced by CO at separations between $10^{12}$ and $10^{14}\,{\rm cm}$, assuming He fraction between $10^{-2}\,M_\odot$ and $10^{-3}\,M_\odot$. For hybrid-WDs containing $>0.03$ M$_\odot$ of He, no CO is stripped until the final disruption of the hybrid-WD. 

The inferred composition of the CSM around SN~2019ehk is broadly consistent with a $\sim 0.53$ M$_\odot$ CO or a $\sim 0.48$ M$_\odot$ hybrid-WD donor model; both of which formed during binary evolution, and not as isolated single WDs. In particular, these are consistent with the expectations of the HybCO model for CaST SNe progenitors. In the HybCO model interpretation, future observations of CaSTs may potentially be used to put strong constraints on the progenitor systems, and even the surface composition of WDs.

The exact velocity and the geometry of the material lost to the surroundings are the main uncertainties in the pre-merger stripping model. While this material is expected to have velocities comparable to the orbital velocities, the exact detailed hydrodynamical picture of the secular mass-loss in direct-impact DWD binaries is uncertain. In particular, the material may be ejected in an outflow from the disc, a more tightly-collimated jet from near the accretor, or as a more isotropic cloud-like structure powered by the feedback from accretion. When it comes to the fraction of the mass lost from the binary, even within a wide range of assumed efficiencies of mass-loss (range of $5-100\%$ ejection efficiency), the CSM ejecta profiles agree well with the observations. They explain (i) the cut-off at large separations, due to the time when the secondary WD gradually overfills its Roche lobe and before which no significant stripping initiates; (ii) the overall density profile of the CSM; (iii) the overall composition and the transition between the outer and inner regions due to the compositional structure on the stripped WD surface and (iv) the observed low CSM velocities derived from early-time H$\alpha$ and \ion{He}{ii} lines. It should also be noted that similar observables could be obtained from the disruption of HybCO (hybrid He/C/O) WD by a NS (e.g., see \citealt{fernandez19}) although the rates associated with such binary systems are not consistent with \cas.

\subsection{Tidal Disruption by an Intermediate-Mass Black Hole} \label{subsec:imbh}
A proposed model for \cas\, is the tidal disruption of a low mass WD by an intermediate mass black hole (IMBH) \citep{rosswog08, metzger12, macleod14, sell15, tanikawa17}. One prominent signature of this accretion process would be the presence of X-ray emission above the Eddington luminosity. Since we observed luminous X-ray emission from the \cas\, SN~2019ehk for the first time, we briefly discuss this scenario here. \cite{sell15} first explore this scenario for the \ca~2012hn to constrain the potential masses of the IMBH and of the disrupted WD via X-ray upper limits at 533 days after explosion. \cite{milisavljevic17} employ a similar method for iPTF15eqv, for which these authors infer an IMBH  mass $\lessapprox 100 \Msun$ on an accretion timescale of <164 days. We apply the same method to SN~2019ehk here. 

The X-ray luminosity of SN~2019ehk $L_x\approx10^{41}\,\rm{erg\,s^{-1}}$ at $\sim3$ days since explosion (Figure \ref{fig:xray_radio_LC}) is consistent with the Eddington luminosity of a $\sim10^3\,\rm{M_{\sun}}$ BH, for which the timescale of fallback accretion is (e.g. \citealt{milisavljevic17}):

\begin{equation}
    t_{\textrm{Edd}} = \Big(\frac{M_{\textrm{BH}}}{10^3 \ \Msun}\Big)^{-2/5} \Big(\frac{M_{\textrm{WD}}}{0.6 \ \Msun}\Big)^{1/5}
    \Big(\frac{R_{\textrm{WD}}}{5 \times 10^{-2} \ \Rsun}\Big)^{3/5} \textrm{yr},
\end{equation}

\noindent
which indicates that for fiducial values of $M_{\rm{WD}}$ and $R_{\rm{WD}}$, such a transient would have an accretion luminosity above the Eddington limit for $t_{\rm{Edd}} \approx$ 1 yr. This timescale is not consistent with observations of SN~2019ehk as its X-ray emission fades quickly on timescales of days as $L_x\propto t^{-3}$ after the first detection. The IMBH scenario can be further constrained by using the deepest X-ray luminosity limit of $< 3.3 \times 10^{38}$~erg~s$^{-1}$ obtained with \textit{Chandra} at $292$~days since explosion. Using $t_{\rm{Edd}} = 292.2$~d (the phase of observation), we calculate a limit on the BH mass of $\lesssim 2000 ~ \Msun$, assuming fiducial WD parameters. From the X-ray luminosity limit, and assuming an accretion efficiency of $10\%$, we calculate a BH mass limit of $\lesssim 33 ~ \Msun$. Furthermore, as discussed in \S\ref{Sec:Radio_Xray_Modeling}, we find no evidence for an off-axis jet in our modeling of the radio emission, which is assumed to be associated with an accretion event such as the tidal disruption of a WD by an IMBH. Lastly, a IMBH progenitor is expected to be associated with a cluster, yet we find no sources near the SN location in the pre-explosion images. Based on these inferences, we conclude that the tidal disruption of a WD by an IMBH is an extremely unlikely physical scenario for SN~2019ehk.

\subsection{SN~2019ehk in the ``Calcium-strong'' Class} \label{subsec:ca-rich_19ehk}
SN~2019ehk is currently the CaST with the most extended and detailed observational data set across the electromagnetic spectrum. A key question is how representative SN~2019ehk is of the entire ``Calcium-strong'' class of transients? As discussed in \S\ref{subsec:phot_properties}, the optical light curve ($M_{B}^{\mathrm{peak}} = -15.10 \pm 0.0210$ mag, 
$\Delta$m$_{15} = 1.71 \pm 0.0140$~mag) and color evolution of SN~2019ehk are consistent with  the class of \cas\, (e.g., Figures \ref{fig:ca_rich_compare} \& \ref{fig:colors}). However, the main photometric difference is its prominent double (triple?) peaked light curve with the initial ``flare'' only matching one other object in the class, iPTF16hgs \citep{de19}. iPTF16hgs was not discovered as early but does show consistency spectroscopically (Figure \ref{fig:carich_spectra}(a)) to SN~2019ehk. Both objects were found in star forming host galaxy environments, in contrast with the majority of the sample (e.g. \citealt{shen19}). Furthermore, \cite{de18} also find that shock breakout emission can reproduce the increase in flux prior to the Ni-powered light peak. These combined similarities suggest a shared progenitor scenario amongst these two objects (and potentially other \cas).

Spectroscopically, SN~2019ehk shows near remarkable consistency with CaSTs SNe~2005E and 2007ke (Figure \ref{fig:carich_spectra}b). This level of similarity is intriguing given that the large-scale environments of these two CaSTs relative to SN~2019ehk are quite different (SNe~2005E and 2007ke are located on the outskirts of early-type galaxies while SN~2019ehk is embedded in a late-type spiral galaxy). SN~2005E was modeled via a helium shell detonation of a sub-Chandra WD \citep{perets05, waldman11}, and SN~2007ke is thought to arise from a compact object progenitor given the lack of star formation at its explosion site \citep{lunnan17}. Generally, SN~2019ehk shares clear spectroscopic similarities with the rest of the class: Type I spectrum, visible He I, weak Fe-group element and \ion{O}{i} transitions, and dominant \ion{Ca}{ii} emission at late-times. SN~2019ehk has the largest [\ion{Ca}{ii}]/[\ion{O}{i}] ratio yet observed amongst CaSTs (and known transients as a whole) and has the earliest visible detection of [\ion{Ca}{ii}] (-5 days). Out to nebular times, SN~2019ehk shows persistent [\ion{Ca}{ii}] emission that is similar to other CaSTs. Therefore, SN~2019ehk's [\ion{Ca}{ii}]/[\ion{O}{i}] ratio is consistent with the overall classification of \cas\, and it is the ``richest'' known object in Ca emission. 

SN~2019ehk is located in a star-forming region of a barred spiral host-galaxy. SN~2019ehk thus adds to the increasing evidence for a wide distribution of both early and late type host galaxies for CaSTs. The SN is also embedded in its host galaxy (offset $\sim$ 2~kpc), which suggests that \ca \ class cannot be completely defined by large galactic offsets. Overall, a large fraction of the current \cas \ sample are located at large offsets from early-type galaxies and/or with limited to no visible star formation \citep{perets10, perets11, kasliwal12, Lym+13,lyman14, lunnan17, de20}. However, multiple confirmed CaSTs and candidate objects have deviated from this trend. iPTF15eqv, iPTF16hgs and SN~2016hnk are all located in spiral host-galaxies and analysis of the explosion sites indicate the presence of star formation \citep{milisavljevic17, de18, galbany19, wjg19}. Similarly, \cas\, PTF09dav, SN~2001co, SN~2003H, SN~2003dr and 2003dg appear to have exploded in or offset from disk-galaxies \citep{sullivan11, kasliwal12, Perets2014, foley15}.  

The older stellar populations where a large fraction of \cas\, are found makes it difficult to reconcile a massive star progenitor for the entire class. In the context of WD progenitors, the increased discovery of CaSTs in late-type galaxies with a young stellar population component is still compatible with an older progenitor given the frequency of WDs in a variety of host environments. A larger sample of stellar ages near \ca \ explosion sites will confirm whether a broad(er) delay time distribution is needed to explain the presence of some \cas \ in younger stellar populations. Nevertheless, the existence of star-forming host galaxies does potentially still allow for a massive stellar progenitor channel (and hence a core-collapse origin) as an explanation for some CaSTs, as suggested by \cite{milisavljevic17}. SN~2019ehk has greatly constrained the massive star progenitor parameter space by illustrating that only the lowest mass stars ($\sim$8-10~$\Msun$) in binary systems are permitted progenitors of a \ca. Increasing the sample size of CaSTs with detailed observational coverage across the spectrum will help to reveal whether this class truly has multiple associated progenitor scenarios. 

Finally, the detection of luminous X-ray emission in SN~2019ehk represents a newly discovered observational signature of \cas. Based on the observational coverage at X-ray wavelengths, it has become apparent that \cas\, may only exhibit X-ray emission at very early-times. No other \cas\, has X-ray observations before +25d after explosion yet we now know that X-ray emission in SN~2019ehk only lasted until +4 days. This indicates two possibilities: the explosion and environment of SN 2019ehk are unique \textit{or} CaSTs do show X-ray emission directly after explosion that has been missed observationally until now. If the latter is true, then extremely early observations of CaSTs is imperative to understand the progenitor environments of these objects. 

\section{Summary and Conclusions} \label{Sec:conclusion}
In this paper we have presented pre- and post-explosion (0.4-292 days) panchromatic observations of the nearby \ca\, SN~2019ehk located in a region of high star formation near the core of the SAB(rs)c galaxy M100 at $d\sim16.2$ Mpc. Our observations cover the electromagnetic spectrum from the X-rays to the radio band, before and after the explosion. Below we summarize the primary observational findings that make SN~2019ehk the CaST with the richest data set to date: 
\begin{itemize}
\item SN~2019ehk was detected $\sim$0.44 days after explosion and its UV/optical/NIR photometric evolution shows a double-peaked light curve in all multi-color bands, similar to \ca \ iPTF16hgs \citep{de18}. However, different from iPTF16hgs, these very early observations of SN~2019ehk were also able to capture the rapid rise to the first light curve peak. With respect to its second broader light curve peak, SN~2019ehk has a rise-time $t_r = 13.4 \pm 0.210$~days, a peak $B$-band magnitude $M_B = -15.1 \pm 0.0210$~mag and \cite{phillips93} decline parameter of $\Delta$m$_{15}$(B) = 1.71 $\pm 0.0140$ mag.
\item Within 24~hrs of discovery, three optical spectra were acquired starting at $t\approx1.4$ days since explosion, and revealed the rapid disappearance of ``flash-ionized'' H Balmer series and \ion{He}{ii} emission lines with velocities of $\sim 400$ and $\sim 500$~$\kms$, respectively. These spectral features were detected at the time of the first light curve peak, and provide first evidence for H+He-rich CSM in the immediate vicinity of a \ca.
\item SN~2019ehk showed luminous, rapidly-decaying X-ray emission ($L_x \approx 10^{41}$~erg~s$^{-1}$ with $L_x \propto t^{-3}$). The luminous X-ray emission detected with \textit{Swift}-XRT at +3 and +4d after explosion constitutes a newly discovered observational signature of \cas\, and results from the exploration of a pristine portion of the X-ray parameter space within this class.
The X-ray emission is temporally coincident with the first optical light curve peak (``the flare''). At later times (+292d) \textit{Chandra} observations provided the deepest constraints on a \ca\, to date ($L_x<3.3\times 10^{38}$~erg~s$^{-1}$).
\item Our deep radio monitoring with the VLA provided the tightest constraints on the radio luminosity from a \ca\, at phases >30 days after explosion $L_{\nu}<10^{25}\,\rm{erg~s^{-1}~Hz^{-1}}$.

\item SN~2019ehk has the latest spectroscopic follow-up of any \ca \ at +257d after explosion. The spectrum revealed the largest [\ion{Ca}{ii}]/[\ion{O}{i}] line flux ratio yet reported ($\sim25$). 

\item The explosion site of SN~2019ehk has extremely deep pre-explosion imaging with \textit{Chandra}, \textit{Spitzer} and \textit{HST}. No source is detected in any archival image with an astrometric uncertainty of $\sigma_{\alpha} =  4.05\times 10^{-4}\arcsec$ and $\sigma_{\delta} = 2.71\times 10^{-4}\arcsec$.

\end{itemize}

By modeling these observations we place tight constraints on the SN progenitor, its environment and the explosion mechanism: 

\begin{itemize}
    
\item Bolometric light curve models show that the explosion synthesized $(3.1\pm0.11) \times 10^{-2} \, \Msun$ of ${}^{56}\textrm{Ni}$, produced $0.72 \pm 0.04\, \Msun$ of ejecta and had a kinetic energy of $(1.8\pm 0.1) \times 10^{50}$~erg. 

\item The H+He-rich material is part of the CSM and preceded the SN explosion. ``Flash-ionized'' emission lines indicate the presence of pre-shock CSM gas with mass $M_{\rm{csm}} \approx 2\times 10^{-3} \ \Msun$ and composition in the range $0.44 < n_{\rm{He}} / n_{\rm{H}} < 0.88$ by number. The total CSM mass as inferred from X-ray observations  is $M_{\rm{csm}} \approx 7\times 10^{-3} \Msun$, comprised of $(4-17) \times 10^{-4}$ and $(5.3-6.7) \times 10^{-3} \ \Msun$ of H- and He-rich material, respectively. Both observations combined revealed a CSM density of $\rho_{\rm csm} = 2\times10^{-15}$ g cm$^{-3}$ at $R_{\rm{csm}} = (0.1 - 1) \times 10^{15}$cm.

\item For realistic microphysical parameters ($\epsilon_B = 0.01$ and $\epsilon_e = 0.1$), radio non-detections suggest a mass-loss rate of $\dot{M}<10^{-5}\,\rm{M_{\odot}yr^{-1}}$ for a wind velocity $v_w = 500 \ \kms$ at distances $r \approx 10^{16} - 10^{17}$ cm from the explosion site.

\item We model the early-time optical emission with two models: (i) shock interaction with CSM and (ii) shock cooling following breakout into extended material. Given an observed SN ejecta mass $M_{\rm{ej}} \approx 1\, \Msun$, the former yields a CSM mass of $M_{\rm{csm}} = 1.5\times 10^{-3} \ \Msun$ and radius of $R_{\rm{csm}} = 4\times 10^{13}$ cm. This model can adequately power the persistent SN optical emission at $t < 6$d and is consistent with the duration of visible H+He emission lines. The latter model provides a potential physical mechanism for the increased optical emission at $t < 2$d and indicates extended material of mass $M_e \approx 7\times10^{-2} \, \Msun$ and radius $R_e \approx 200\, \Rsun$. These values are broadly consistent with our inferences from the H+He spectral lines and the modeling of the X-ray emission, suggesting that the presence of an extended distribution of (H+He rich) material with which the SN shock interacted can reasonably account for three key observational findings in SN~2019ehk (e.g., the X-ray emission, the optical flare and the transient H+He lines).
    
\end{itemize}

Pre-explosion imaging at the location of SN~2019ehk rules out a vast portion of the parameter space associated with both massive stars and WD explosions. Specifically, we find that pre-explosion limits rule out all \emph{single} massive stars with mass $\gtrsim 8 \ \Msun$ as the progenitor of \ca \ SN~2019ehk for a host reddening of $E(B-V) = 0.47$. We explore the available binary system parameter space and find that our limits only allow for systems with a $9.5-10 \ \Msun$ primary star or a low-mass He star whose envelope was removed through mass-loss and/or binary interaction. Furthermore, the observed explosion properties make it unlikely that SN~2019ehk was produced by the explosion of a C/O WD with a He or main sequence star companion. However, we find that a model for the disruption of a low-mass C/O WD or a hybrid HeCO WD ($\sim 0.5-0.6 \ \Msun$) by another, likely low-mass hybrid WD is consistent with the CSM densities, abundances and dynamics inferred for SN~2019ehk, and would possibly be able to account for the increasingly large fraction of \cas\, embedded in young stellar populations by allowing for a broader time delay distribution. Complete multi-wavelength observations of future \cas \ will be instrumental in differentiating between these two possible progenitor scenarios. 

\section{Acknowledgements} \label{Sec:ack}

We thank David Matthews, Enrico Ramirez-Ruiz, Brian Metzger and Nathan Smith for stimulating conversations and constructive comments in regards to this work.

W.J-G is supported by the National Science Foundation Graduate Research Fellowship Program under Grant No.~DGE-1842165 and the Data Science Initiative Fellowship from Northwestern University.
R.M. is grateful to KITP for hospitality during the completion of this paper. This research was supported in part by the National Science Foundation under Grant No. NSF PHY-1748958. R.M. acknowledges support by the National Science Foundation under Award No. AST-1909796. Raffaella Margutti is a CIFAR Azrieli Global Scholar in the Gravity \& the Extreme Universe Program, 2019. The Margutti's team at Northwestern is partially funded by the Heising-Simons Foundation under grant \# 2018-0911 (PI: Margutti). 

The National Radio Astronomy Observatory is a facility of the National Science Foundation operated under cooperative agreement by Associated Universities, Inc. GMRT is run by the National Centre for Radio Astrophysics of the Tata Institute of Fundamental Research.

The scientific results reported in this article are based in part on observations made by the Chandra X-ray Observatory. This research has made use of software provided by the Chandra X-ray Center (CXC) in the application packages CIAO. Partial support for this work was provided by the National Aeronautics and Space Administration through Chandra Award Number DD0-21114X issued by the Chandra X-ray Center, which is operated by the Smithsonian Astrophysical Observatory for and on behalf of the National Aeronautics Space Administration under contract NAS8-03060.

W. M. Keck Observatory [and/or MMT Observatory] access was supported by Northwestern University and the Center for Interdisciplinary Exploration and Research in Astrophysics (CIERA).


The UCSC transient team is supported in part by NSF grant AST-1518052, NASA/{\it Swift} grant 80NSSC19K1386, the Gordon \& Betty Moore Foundation, the Heising-Simons Foundation, and by a fellowship from the David and Lucile Packard Foundation to R.J.F. Research at Lick Observatory is partially supported by a generous gift from Google.

A.H. is partially supported by the Future Investigators in NASA Earth and Space Science and Technology (FINESST) Award No. 80NSSC19K1422. HBP acknowledges support from the Kingsely distinguished-visitor program at Caltech, and the KITP visitor program. KJS is supported by NASA through the Astrophysics Theory Program (NNX17AG28G). PJB is supported by a NASA Swift Guest Investigator grant 80NSSC19K0316. TMB was funded by the CONICYT PFCHA / DOCTORADOBECAS CHILE/2017-72180113. SCY is supported by the National Research Foundation of Korea (NRF) grant (NRF-2019R1A2C2010885). MG is supported by the Polish NCN MAESTRO grant 2014/14/A/ST9/00121. DAH, JB, DH, and CP are supported by NSF AST-1911225 and NASA grant 80NSSC19K1639. JD and EB are supported in part by NASA Grant 80NSSC20K0456. SJS acknowledges funding from STFC Grant ST/P000312/1. M.R.S.\ is supported by the National Science Foundation Graduate Research Fellowship Program under Grant No.\ 1842400.

Research by DJS is supported by NSF grants AST-1821967, 1821987, 1813708, 1813466, and 1908972. This research is based on observations made with the NASA/ESA Hubble Space Telescope obtained from the Space Telescope Science Institute, which is operated by the Association of Universities for Research in Astronomy, Inc., under NASA contract NAS 5-26555. These observations are associated with program 15645.

X. Wang is supported by the National Natural Science Foundation of China (NSFC grants 11325313, 11633002, and 11761141001), and the National Program on Key Research and Development Project (grant no. 2016YFA0400803). We acknowledge the support of the staff of the Xinglong 2.16-m telescope. This work was partially supported by the Open Project Program of the Key Laboratory of Optical Astronomy, National Astronomical Observatories, Chinese Academy of Sciences."  Funding for the LJT has been provided by Chinese Academy of Sciences and the People's Government of Yunnan Province. The LJT is jointly operated and administrated by Yunnan Observatories and Center for Astronomical Mega-Science, CAS.

Based on observations obtained at the Southern Astrophysical Research (SOAR) telescope, which is a joint project of the Minist\'{e}rio da Ci\^{e}ncia, Tecnologia, Inova\c{c}\~{o}es e Comunica\c{c}\~{o}es (MCTIC) do Brasil, the U.S. National Optical Astronomy Observatory (NOAO), the University of North Carolina at Chapel Hill (UNC), and Michigan State University (MSU).

Observations reported here were obtained at the MMT Observatory, a joint facility of the University of Arizona and the Smithsonian Institution.

Some of the data presented herein were obtained at the W.\ M.\ Keck Observatory, which is operated as a scientific partnership among the California Institute of Technology, the University of California and the National Aeronautics and Space Administration. The Observatory was made possible by the generous financial support of the W.\ M.\ Keck Foundation.  The authors wish to recognize and acknowledge the very significant cultural role and reverence that the summit of Maunakea has always had within the indigenous Hawaiian community. We are most fortunate to have the opportunity to conduct observations from this mountain. We wish to recognize the destructive history of colonialism endured by native Hawaiians as we strive to hear the voice of those whose sacred land we continue to utilize for scientific gain.

This work includes data obtained with the Swope Telescope at Las Campanas Observatory, Chile, as part of the Swope Time Domain Key Project (PI: Piro, Co-Is: Drout, Phillips, Holoien, French, Cowperthwaite, Burns, Madore, Foley, Kilpatrick, Rojas-Bravo, Dimitriadis, Hsiao). We wish to thank Swope Telescope observers Jorge Anais Vilchez, Abdo Campillay, Nahir Munoz Elgueta and Natalie Ulloa for collecting data presented in this paper.

This research has made use of the XRT Data Analysis Software (XRTDAS) developed under the responsibility of the ASI Science Data Center (ASDC), Italy.
The SN group at Konkoly Observatory is supported by the project ``Transient Astrophysical Objects" GINOP 2.3.2-15-2016-00033 of the National Research, Development and Innovation Office (NKFIH), Hungary, funded by the European Union. 
Based on observations collected at the European Southern Observatory under ESO programme 1103.D-0328.

This work has made use of data from the Asteroid Terrestrial-impact
Last Alert System (ATLAS) project. ATLAS is primarily funded to search
for near earth asteroids through NASA grants NN12AR55G, 80NSSC18K0284,
and 80NSSC18K1575; byproducts of the NEO search include images and
catalogs from the survey area.  The ATLAS science products have been
made possible through the contributions of the University of Hawaii
Institute for Astronomy, the Queen's University Belfast, and the Space
Telescope Science Institute.

\facilities{\emph{Hubble Space Telescope}, \emph{Spitzer Space Telescope}, \emph{Chandra} X-ray Observatory, Neil Gehrels \emph{Swift} Observatory, VLA, Las Campanas Observatory, Zwicky Transient Facility, Konkoly Observatory, ATLAS, Thacher Observatory, Swope:1m, Multi Unit Spectroscopic Explorer (MUSE),  Shane (Kast), Xinglong Observatory (BFOSC), MMT (Binospec), Bok (B\&C), Faulkes North (FLOYDS), NTT (EFOSC2), LJT (YFOSC), SOAR (Goodman, Triple-Spec), Keck I (LRIS)}

\software{emcee \citep{foreman-mackey13}, SNID \citep{Blondin07}, Superfit \citep{Howell05}, IRAF (Tody 1986, Tody 1993), AstroDrizzle \citep{astrodrizzle}, photpipe \citep{Rest+05}, DoPhot \citep{Schechter+93}, HOTPANTS \citep{becker15}, Sedona \citep{kasen06}, SYN++/SYNAPPS \citep{thomas11}, Castro \citep{almgren10}, BPASS \citep{eldridge17}, MESA \citep{choi16}, sextractor \citep{sextractor}, HEAsoft (v6.22; HEASARC 2014) }


\bibliographystyle{aasjournal} 
\bibliography{references} 

\clearpage
\appendix

\renewcommand\thetable{A\arabic{table}} 
\setcounter{table}{0}

\begin{figure}
\centering
\includegraphics[width=0.95\textwidth]{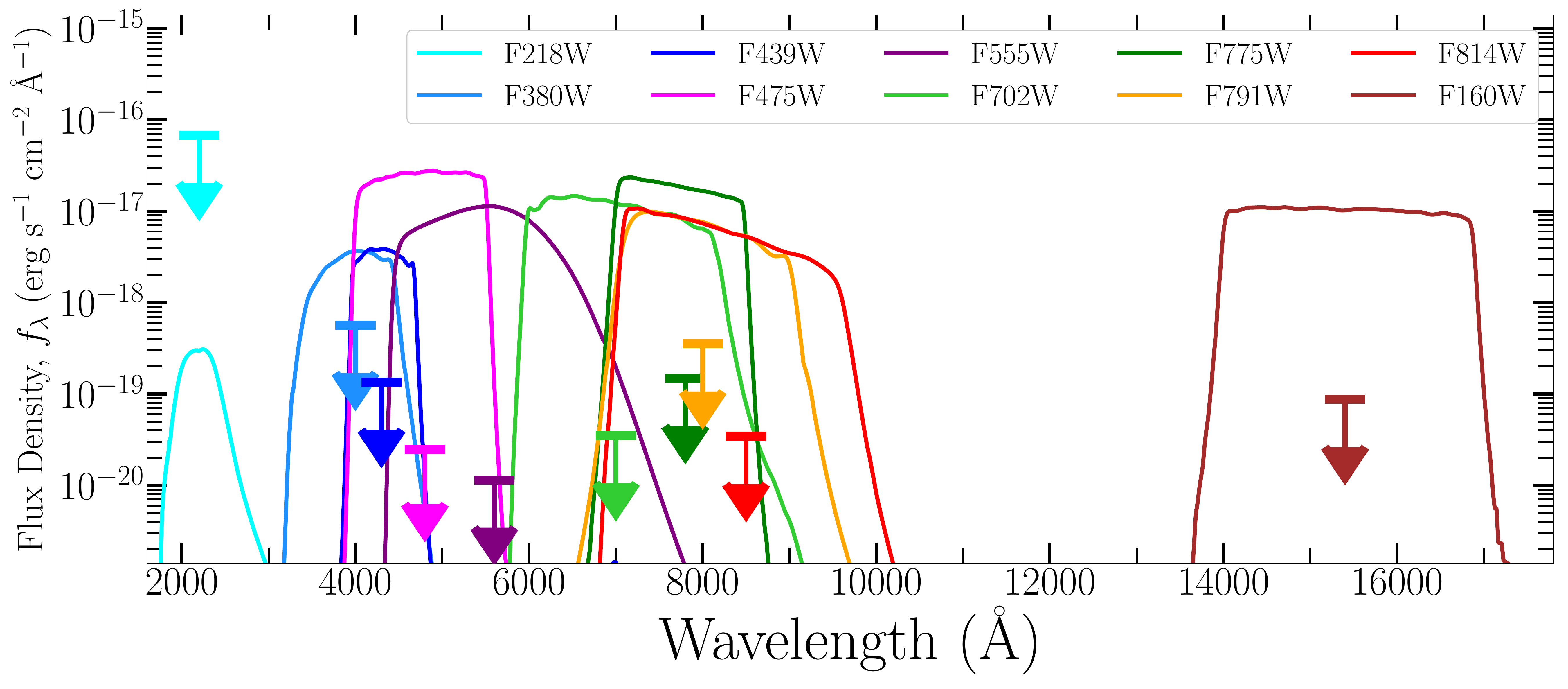} \caption{\emph{HST} pre-explosion limits with respect to filter functions. \label{fig:hst_sed}}
\end{figure}

\newpage 

\begin{figure}
\centering
\includegraphics[width=0.95\textwidth]{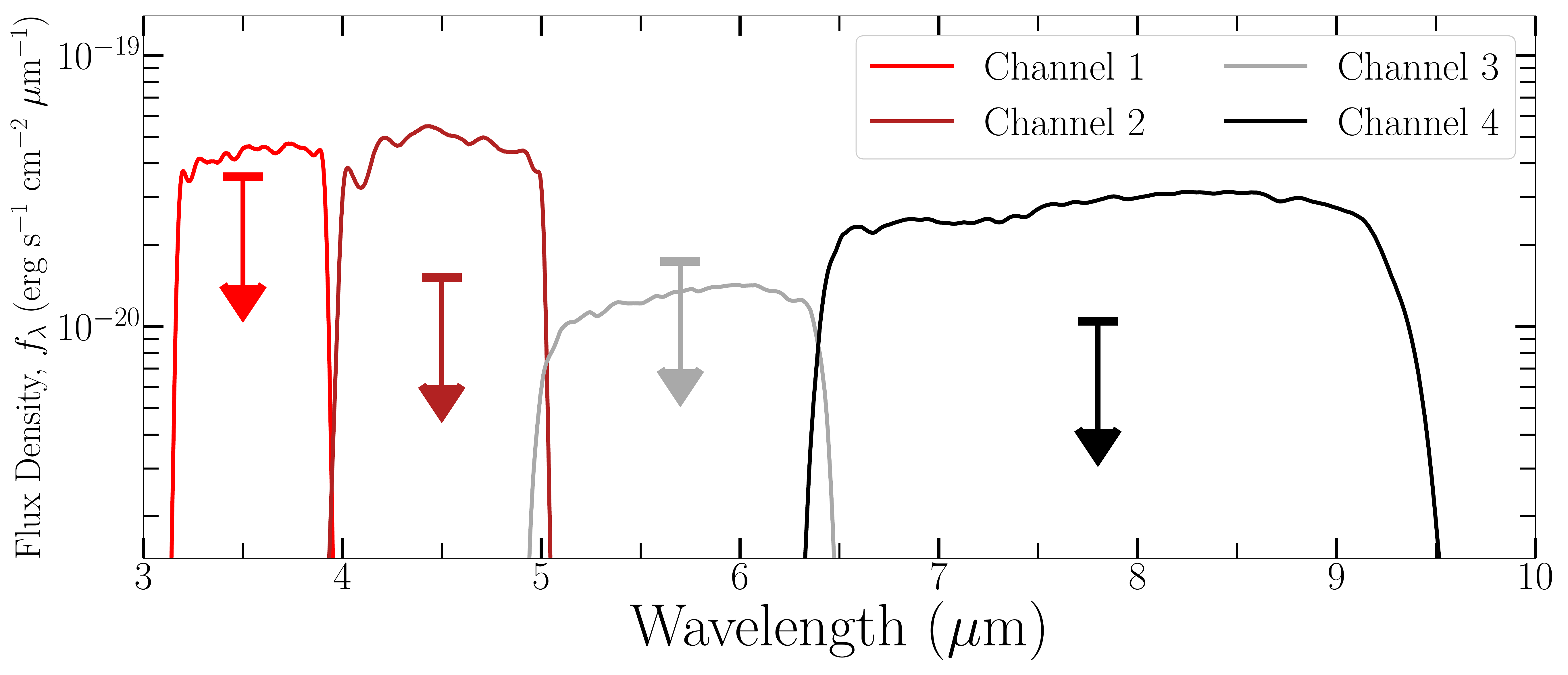} \caption{\emph{Spitzer} pre-explosion limits with respect to filter functions. \label{fig:spitzer_sed}}
\end{figure}

\newpage

\begin{deluxetable*}{ccccccc}[h!]
\tablecaption{\emph{HST} Pre-explosion Limits on Progenitor \label{tbl:hst_table}}
\tablecolumns{7}
\tablewidth{0.45\textwidth}
\tablehead{
\colhead{Instrument} & \colhead{Aperture} & \colhead{Filter} & \colhead{UT Date Obs.}& \colhead{Exp. Time} & \colhead{Proposal No.} & \colhead{3$\sigma$ Limit\tablenotemark{a}}\\
\colhead{} & \colhead{} & \colhead{} & \colhead{}& \colhead{(s)} & \colhead{} & \colhead{(mag)}
}
\startdata
WFPC2 & WF & F218W & 1999-02-02 & 1200 & 6358 & 21.2\\
WFPC2 & WF & F380W & 2008-01-04 & 1000 & 11171 & 25.2\\
WFPC2 & WF & F439W & 1993-12-31 -- 2008-01-04 & 60 -- 900 & 5195, 11171 & 26.6\\
WFC3 & UVIS & F475W & 2009-11-12 & 300-670 & 6358 & 28.2\\
WFPC2 & WF & F555W & 1993-12-31 -- 2008-01-04 & 10 -- 1000 & 5195, 5972, 9776, 10991, 11171, 11646 & 28.7\\
WFPC2 & WF & F702W & 1993-12-31 -- 2008-01-04 & 5 -- 600 & 5195, 11171 & 27.0\\
WFC3 & UVIS & F775W & 1999-02-02 & 1200 & 6358 & 25.2\\
WFPC2 & WF & F791W & 2008-01-04 & 500 & 11171 & 24.2\\
WFPC2 & WF & F814W & 1994-05-12 -- 1996-04-27 & 350 -- 2100 & 5972, 15133 & 26.6\\
WFC3 & IR & F160W & 2018-02-04 & 596 & 15133 & 24.3\\
\enddata
\tablenotetext{a}{All apparent magnitudes in Vega system.}
\end{deluxetable*}

\begin{deluxetable*}{ccccc}[h!]
\tablecaption{\emph{Spitzer} Pre-explosion Limits on Progenitor \label{tbl:spitzer_table}}
\tablecolumns{5}
\tablewidth{0.45\textwidth}
\tablehead{
\colhead{UT Date Obs. Range} & \colhead{Channel 1} & \colhead{Channel 2} & \colhead{Channel 3}& \colhead{Channel 4\tablenotemark{a}}
}
\startdata
2015-09-06 -- 2019-10-27 & 23.49~mag & 23.87~mag & 23.21~mag & 23.08~mag\\
\enddata
\tablenotetext{a}{All apparent magnitudes in AB system.}
\end{deluxetable*}

\begin{deluxetable*}{ccccc}[h!]
\tablecaption{X-ray Observations of SN~2019ehk \label{tab:xray_obs}}
\tablecolumns{6}
\tablewidth{0pt}
\tablehead{\colhead{MJD} & \colhead{Phase\tablenotemark{a}} & \colhead{Photon Index } & \colhead{0.3-10 keV Unabsorbed Flux} &  \colhead{Instrument} \\
\colhead{} & \colhead{(days)} & \colhead{($\Gamma$)} & \colhead{($10^{-12}$~erg~s$^{-1}$~cm$^{-2}$)} & \colhead{}}
\startdata
58604.61 & +2.81 & $0.1\pm0.4$ & $4.3^{+0.9}_{-0.8}$ &\textit{Swift-}XRT \\
58606.03 & +4.23 & $0.2\pm0.9$ & $1.3^{+0.9}_{-0.6}$&\textit{Swift-}XRT \\
58607.56 & +5.76 & -- & $<0.7\tablenotemark{b}$&\textit{Swift-}XRT \\
58612.71 & +10.91 & -- & $<0.9$ &\textit{Swift-}XRT \\
58619.64 & +17.84 & -- & $<1.6$ &\textit{Swift-}XRT \\
58624.56 & +22.76 &-- & $0.8$&\textit{Swift-}XRT \\
58629.30 & +27.50 &-- & $<0.7$ & \textit{Swift-}XRT \\
58894.00 & +292.2 &-- & $<1.1\times10^{-2}$ & \textit{Chandra} \\
\enddata
\tablenotetext{a}{Relative to explosion (MJD 58601.8).}
\tablenotetext{b}{Flux calibration performed assuming same spectral parameters inferred at $t=4.2$ d.}

\end{deluxetable*}

\begin{table*}
\centering
    \caption{VLA radio observations of SN~2019ehk.}
    \label{Tab:radio}
    \begin{tabular}{ccccc}
    \hline
    \hline
  Start Date & Time\footnote{Relative to second B maximum (MJD 58615.156)} & Frequency & Bandwidth & Flux Density\footnote{\label{Tab_radio_1}Upper-limits are quoted at $3\sigma$.} \\
(UT) & (days) & (GHz) & (GHz) & ($\mu$Jy/beam)\\
    \hline
2019-05-29 & 30 & 6.05 & 2.048 & $\leq27$\\
2019-06-18 & 51 & 6.05 & 2.048 & $\leq24.8$\\
2019-07-15 & 78 & 6.10 & 2.048 & $\leq28$\\
2019-08-29 & 122 & 6.10 & 2.048 & $\leq21$\\
2019-12-04 & 220 & 6.05 & 2.048 & $\leq880$\\
\hline
\end{tabular}
\end{table*}

\begin{deluxetable*}{ccccccc}\label{tbl:shocktable}
\tablecaption{Shock Cooling Models}
\tablecolumns{7}
\tablewidth{0.45\textwidth}
\tablehead{
\colhead{Model} & \colhead{Phase Range} & \colhead{$E(B-V)_{host}$} & \colhead{$R_e$} & \colhead{$M_e$} & \colhead{$v_{e}$} & \colhead{$t_{off}$}\\
\colhead{} & \colhead{} & \colhead{} & \colhead{$\Rsun$} & \colhead{$[\times 10^{-2}] \ \Msun$} & \colhead{$[\times 10^3] \ \kms$} & \colhead{days}
}
\startdata
\cite{nakar14} & $t<2$ & 0.47 & $110 \pm 50$ & $0.9 \pm 0.6$ & 12.0 & -- \\
\cite{nakar14} & $t<6$ & 0.47 & $105 \pm 27$ & $10.4 \pm 3.3$ & 12.0 & -- \\
\cite{piro15} & $t<2$ & 0.47 & $174.1^{+3.1}_{-4.4}$ & $0.51^{+0.1}_{-0.1}$ & $9.5 \pm 0.3$ & $0.01^{+0.01}_{-0.00}$ \\
\cite{piro15} & $t<6$ & 0.47 & $208.2^{+5.3}_{-6.5}$ & $7.2^{+1.1}_{-1.1}$ & $7.9 \pm 0.20$ & $0.01^{+0.01}_{-0.00}$ \\
\cite{sapir17} [n=3/2] & $t<2$ & 0.47 & $7.2^{+2.9}_{-2.9}$ & $20.2^{+14.1}_{-6.3}$ & $13.0^{+1.6}_{-0.7}$ & $0.17^{+0.2}_{-0.1}$\\
\cite{sapir17} [n=3/2] & $t<6$ & 0.47 & $\sim 30$ & $\sim 30$ & $\sim 12$ & --\\
\cite{sapir17} [n=3] & $t<2$ & 0.47 & $7.6^{+4.3}_{-3.0}$ & $83.3^{+17.0}_{-20.2}$ & $20.6^{+7.9}_{-3.2}$ & $0.3^{+0.1}_{-0.1}$\\
\cite{sapir17} [n=3] & $t<6$ & 0.47 & $\sim 43$ & $\sim 120$ & $\sim 19$ & --\\
\enddata
\end{deluxetable*}

\begin{deluxetable*}{ccccccccccc}\label{tbl:he_star_table}
\tablecaption{Helium Star Models}
\tablecolumns{11}
\tablewidth{0.45\textwidth}
\tablehead{
\colhead{Model} & \colhead{$M_{\rm i}$} & \colhead{$M_{\rm f}$} & \colhead{$M_{\rm He}$} & \colhead{$M_{\rm C/O}$} & \colhead{$L_{\rm f}$} & \colhead{$T_{\rm eff}$} & \colhead{$Y_{\rm s}$} &  \colhead{End Point} & \colhead{$T_{\rm max}$} & \colhead{Comments}\\
\colhead{} & \colhead{($\Msun$)} & \colhead{($\Msun$)} & \colhead{($\Msun$)} & \colhead{($\Msun$)}& \colhead{($\Lsun$)} & \colhead{(K)} & \colhead{} & \colhead{} & \colhead{($10^9$~K)} & \colhead{}
}
\startdata
\#1 & 3.00 & 2.61 & 1.10 & 1.51 & 4.50 & 6552 & 0.98 & O-burning & 2.0 & Single He-star, WR $\dot{M}$ \\
\#2 & 3.00 & 1.77 & 0.20 & 1.57 & 4.50 & 64094 & 0.97 & O-burning & 1.9 & Single He-star, artificial removal of He envelope \\
\#3 & 2.70 & 2.61 & 1.20 & 1.50 & 4.54 & 10625 & 0.98 & ONeMg Core & 1.2 & Single He-star, No $\dot{M}$ \\
\#4 & 2.70 & 1.75 & 0.34 & 1.41 & 4.46 & 6428 & 0.98 & O-burning & 1.9 & Single He-star, artificial removal of He envelope \\
\#5 & 2.70 & 1.50 & 0.11 & 1.38 & 4.41 & 16856 & 0.96 & ONeMg Core & 1.2 & Single He-star, artificial removal of He envelope \\
\#6 & 2.70 & 1.41 & 0.05 & 1.36 & 4.71 & 14486 & 0.65 & ONeMg Core & 1.1 & Single He-star, artificial removal of He envelope \\
\#7 & 3.00 & 1.89 & 0.46 & 1.43 & 3.49 & 8226 & 0.94 & O-burning & 2.0 & Binary w/ 1.4~$\Msun$ NS companion ($P_{\rm i} = 150d)$ \\
\#8 & 3.00 & 1.78 & 0.35 & 1.43 & 4.41 & 12436 & 0.98 & O-burning & 1.8 & Binary w/ 1.4~$\Msun$ NS companion ($P_{\rm i} = 50d)$ \\
\enddata
\tablecomments{L in log space. $Y_s$ is the surface helium mass fraction. Model luminosity and temperature presented in Figure \ref{fig:binary}: black polygons for O-burning end state and red polygons for a ONeMg core. }
\end{deluxetable*}

\begin{deluxetable*}{ccccccccccc}
\tablecaption{Binary Progenitor Models from \cite{yoon17} \label{tbl:y17_table}}
\tablecolumns{11}
\tablewidth{0.45\textwidth}
\tablehead{
\colhead{$M_{\textrm{p}}$} & \colhead{$M_{\rm p}$} & \colhead{$M_{\rm f}$} & \colhead{$L_{\rm f}$} & \colhead{$R_{\rm f}$}& \colhead{$T_{\rm eff}$} & \colhead{$H_{\rm env}$} & \colhead{$M_{\rm H}$} & \colhead{$M_{\rm He}$} & \colhead{$\dot{M}$} & \colhead{SN}\\
\colhead{} & \colhead{($\Msun$)} & \colhead{($\Msun$)} & \colhead{($\Lsun$)} & \colhead{($\Rsun$)}& \colhead{(K)} & \colhead{} & \colhead{($\Msun$)} & \colhead{($\Msun$)} & \colhead{($\Msun$ yr$^{-1}$)} & \colhead{}
}
\startdata
Sm13p50 & 13 & 3.88 & 4.82 & 6.50 & 4.56 & 0.00 & 0.00 & 1.63 & -5.65 & Ib \\
Sm13p50 & 13 & 3.96 & 4.84 & 6.20 & 4.58 & 0.00 & 0.00 & 1.65 & -5.63 & Ib \\
Sm16p50 & 16 & 4.99 & 5.05 & 4.90 & 4.68 & 0.00 & 0.00 & 1.66 & -5.35 & Ib \\
Sm16p300 & 16 & 5.01 & 5.06 & 5.10 & 4.67 & 0.00 & 0.00 & 1.65 & -5.34 & Ib \\
Sm16p1700 & 16 & 6.08 & 5.14 & 3.20 & 4.79 & 0.02 & 0.00 & 2.25 & -5.27 & IIb (BSG) \\
Sm18p50 & 18 & 5.44 & 5.10 & 2.10 & 4.88 & 0.00 & 0.00 & 1.57 & -5.29 & Ib \\
Sm18p500 & 18 & 5.55 & 5.10 & 1.90 & 4.90 & 0.00 & 0.00 & 1.61 & -5.29 & Ib \\
Sm18p2000 & 18 & 6.62 & 5.19 & 1.70 & 4.94 & 0.00 & 0.00 & 2.16 & -5.18 & Ib \\
Sm18p2200 & 18 & 7.04 & 5.16 & 1.70 & 4.93 & 0.08 & 0.01 & 2.53 & -5.36 & IIb (BSG) \\
\enddata
\tablecomments{L, T, $\dot{M}$ in log space.}
\end{deluxetable*}

\begin{deluxetable*}{ccccccccccc}[h!]
\tablecaption{BPASS Binary Progenitor Models \label{tbl:bp_table1}}
\tablecolumns{11}
\tablewidth{0.45\textwidth}
\tablehead{
\colhead{$M_i$} & \colhead{$L_f$} & \colhead{$T_f$} & \colhead{$R_f$}& \colhead{$M_{p,f}$} & \colhead{$M_{s,f}$} & \colhead{$M_{\rm H}$} & \colhead{$M_{\rm He}$} & \colhead{$M_{Ni}$} & \colhead{$M_{\rm ej}$} & \colhead{Delay Time}\\
\colhead{$\Msun$} & \colhead{$\Lsun$} & \colhead{K} & \colhead{$\Rsun$} & \colhead{$\Msun$}& \colhead{$\Msun$} & \colhead{$\Msun$} & \colhead{$\Msun$} & \colhead{$\Msun$} & \colhead{$\Msun$} & \colhead{yrs}
}
\startdata
9.50 & 4.50 & 4.42 & 8.77 & 1.69 & 3.87 & 0.000 & 0.22 & 0.006 & 1.76 & 7.49 \\
9.50 & 4.51 & 4.55 & 4.76 & 1.61 & 6.08 & 0.000 & 0.19 & 0.005 & 1.46 & 7.49 \\
10.00 & 4.51 & 4.46 & 7.35 & 1.65 & 1.00 & 0.000 & 0.19 & 0.008 & 1.61 & 7.46 \\
10.00 & 4.51 & 4.32 & 13.52 & 1.69 & 7.75 & 0.000 & 0.20 & 0.006 & 1.78 & 7.45 \\
10.00 & 4.52 & 4.63 & 3.28 & 1.63 & 6.60 & 0.000 & 0.18 & 0.005 & 1.54 & 7.46 \\
10.00 & 4.45 & 4.45 & 6.94 & 1.69 & 2.01 & 0.000 & 0.25 & 0.007 & 1.79 & 7.46 \\
10.00 & 4.29 & 4.38 & 7.97 & 1.59 & 3.06 & 0.000 & 0.12 & 0.006 & 1.40 & 7.46 \\
10.00 & 4.50 & 4.64 & 3.17 & 1.57 & 1.00 & 0.000 & 0.13 & 0.005 & 1.34 & 7.46 \\
10.00 & 4.46 & 4.32 & 12.82 & 1.72 & 3.04 & 0.000 & 0.20 & 0.006 & 1.90 & 7.45 \\
10.00 & 4.51 & 4.37 & 11.11 & 1.70 & 2.01 & 0.000 & 0.20 & 0.006 & 1.83 & 7.45 \\
10.00 & 4.54 & 4.34 & 13.14 & 1.74 & 1.00 & 0.000 & 0.24 & 0.006 & 1.98 & 7.45 \\
10.00 & 4.22 & 4.31 & 10.55 & 1.74 & 7.55 & 0.000 & 0.22 & 0.010 & 1.99 & 7.45 \\
10.00 & 4.51 & 4.31 & 14.67 & 1.70 & 9.03 & 0.000 & 0.21 & 0.004 & 1.83 & 7.45 \\
\enddata
\tablecomments{L, T and delay time in log space. Weak SN ($10^{50}$~erg), CO Core mass < 1.35~$\Msun$, $M_{p,f} > 1.5 ~ \Msun$ $M_{\rm ej} < 2 \ \Msun$, $M_{\rm H}< 0.01 \ \Msun$ , $M_{\rm He} > 0.1 \ \Msun$}
\end{deluxetable*}

\begin{deluxetable*}{cccccc}[h!]
\tablecaption{Optical Spectroscopy of SN~2019ehk \label{tab:spec_table}}
\tablecolumns{5}
\tablewidth{0.45\textwidth}
\tablehead{
\colhead{UT Date} & \colhead{MJD} &
\colhead{Phase\tablenotemark{a}} &
\colhead{Telescope} & \colhead{Instrument} & \colhead{Wavelength Range}\\
\colhead{} & \colhead{} & \colhead{(days)} & \colhead{} & \colhead{} & \colhead{(\AA)}
}
\startdata
2019-04-30 & 58603.3 & $-11.9$ & Shane & Kast & 4000--8600\AA \\
2019-05-01 & 58604.1 & $-11.1$ & Xinglong & BFOSC & 4200--8600\AA \\
2019-05-01 & 58604.2 & $-11.0$ & Shane & Kast & 3500--8200\AA \\
2019-05-02 & 58605.1 & $-10.1$ & Xinglong & BFOSC & 4200--8600\AA \\
2019-05-03 & 58606.2 & $-9.0$ & Shane & Kast & 3500--8600\AA \\
2019-05-04 & 58607.1 & $-8.1$ & Xinglong & BFOSC & 4200--8200\AA \\
2019-05-05 & 58608.1 & $-7.1$ & Xinglong & BFOSC & 4200--8200\AA \\
2019-05-05 & 58608.2 & $-7.0$ & Shane & Kast & 3500--11000\AA \\
2019-05-07 & 58610.1 & $-5.0$ & Faulkes North & FLOYDS & 3500--10000\AA \\
2019-05-07 & 58610.2 & $-5.1$ & Xinglong & BFOSC & 4200--8800\AA \\
2019-05-09 & 58612.1 & $-3.0$ & Xinglong & BFOSC & 4200--8800\AA \\
2019-05-11 & 58614.1 & $-1.0$ & SOAR & Goodman & 4000--9000\AA \\
2019-05-12 & 58615.1 & $0.0$ & NTT & EFOSC2 & 3600--9200\AA \\
2019-05-13 & 58616.1 & $+1.0$ & Shane & Kast & 3500--11000\AA \\
2019-05-18 & 58621.1 & $+6.0$ & LJT & YFOSC & 3500--8800\AA \\
2019-05-24 & 58627.1 & $+12.0$ & Faulkes North & FLOYDS & 4800--10000\AA \\
2019-05-28 & 58631.1 & $+16.0$ & Faulkes North & FLOYDS & 3500--10000\AA \\
2019-06-03 & 58637.1 & $+22.0$ & MMT & Binospec & 4800--7500\AA \\
2019-06-05 & 58639.1 & $+24.0$ & Bok & B\&C & 4000--7800\AA \\
2019-06-05 & 58639.1 & $+24.0$ & Faulkes North & FLOYDS & 3500--10000\AA \\
2019-06-14 & 58648.1 & $+31.0$ & Shane & Kast & 3500--11000\AA \\
2019-06-21 & 58655.1 & $+38.0$ & SOAR & Triple Spec & 9000--25000\AA \\
2019-06-30 & 58664.1 & $+49.0$ & Keck I & LRIS & 3200--10800\AA \\
2019-07-06 & 58670.1 & $+55.0$ & Shane & Kast & 3500--11000\AA \\
2019-07-10 & 58674.1 & $+59.0$ & Shane & Kast & 3500--11000\AA \\
2019-07-10 & 58674.1 & $+59.0$ & Faulkes North & FLOYDS & 3500--10000\AA \\
2020-01-24 & 58872.1 & $+257.0$ & Keck I & LRIS & 5400--10200\AA \\
\enddata
\tablenotetext{a}{Relative to second $B$-band maximum (MJD 58615.156)}
\end{deluxetable*}

\begin{deluxetable}{cccccc}[h!]
\tablecaption{Optical Photometry of SN~2019ehk \label{tbl:phot_table}}
\tablecolumns{6}
\tablewidth{0.45\textwidth}
\tablehead{
\colhead{MJD} &
\colhead{Phase\tablenotemark{a}} &
\colhead{Filter} & \colhead{Magnitude} & \colhead{Uncertainty} & \colhead{Instrument}
}
\startdata
58603.18 & $-11.98$ & $u$ & 19.15 & 0.02 & Swope \\
58603.22 & $-11.93$  & $u$ & 18.69 & 0.06 & Swope \\
58608.13 & $-7.02 $ & $u$ & 20.00 & 0.06 & Swope \\
58616.18 & $+1.02 $ & $u$ & 20.71 & 0.20 & Swope \\
58675.00 & $+59.84$  & $u$ & 24.00 & 0.20 & Swope \\
58603.18 & $-11.97$  & $B$ & 17.63 & 0.01 & Swope \\
58603.23 & $-11.93$  & $B$ & 17.63 & 0.01 & Swope \\
58608.14 & $-7.02 $ & $B$ & 18.06 & 0.01 & Swope \\
58609.17 & $-5.98 $ & $B$ & 18.21 & 0.02 & Swope \\
58611.14 & $-4.02 $ & $B$ & 18.06 & 0.02 & Swope \\
58615.16 & $+0.00 $ & $B$ & 17.99 & 0.01 & Swope \\
58636.09 & $+20.94$  & $B$ & 20.10 & 0.02 & Swope \\
58642.09 & $+26.93$  & $B$ & 20.45 & 0.09 & Swope \\
58644.04 & $+28.89$  & $B$ & 20.41 & 0.05 & Swope \\
58658.06 & $+42.90$  & $B$ & 20.99 & 0.05 & Swope \\
58670.04 & $+54.89$  & $B$ & 21.40 & 0.07 & Swope \\
58603.18 & $-11.98$  & $V$ & 16.91 & 0.01 & Swope \\
58603.23 & $-11.93$  & $V$ & 16.91 & 0.01 & Swope \\
58608.14 & $-7.02 $ & $V$ & 16.89 & 0.01 & Swope \\
58609.17 & $-5.98 $ & $V$ & 16.96 & 0.01 & Swope \\
58615.15 & $-0.00 $ & $V$ & 16.56 & 0.01 & Swope \\
58616.19 & $+1.03 $ & $V$ & 16.53 & 0.01 & Swope \\
58617.08 & $+1.92 $ & $V$ & 16.58 & 0.01 & Swope \\
58631.13 & $+15.98$  & $V$ & 18.09 & 0.10 & Swope \\
58636.10 & $+20.94$  & $V$ & 18.19 & 0.01 & Swope \\
58642.09 & $+26.93$  & $V$ & 18.45 & 0.02 & Swope \\
58644.05 & $+28.90$  & $V$ & 18.48 & 0.02 & Swope \\
58658.05 & $+42.90$  & $V$ & 19.01 & 0.02 & Swope \\
58670.04 & $+54.88$  & $V$ & 19.63 & 0.02 & Swope \\
58691.96 & $+76.81$  & $V$ & 20.38 & 0.04 & Swope \\
58603.18 & $-11.98$  & $g$ & 17.20 & 0.01 & Swope \\
58603.22 & $-11.94$  & $g$ & 17.21 & 0.01 & Swope \\
58608.13 & $-7.02 $ & $g$ & 17.44 & 0.01 & Swope \\
58609.18 & $-5.98 $ & $g$ & 17.54 & 0.01 & Swope \\
58611.15 & $-4.00 $ & $g$ & 17.36 & 0.01 & Swope \\
58615.15 & $-0.01 $ & $g$ & 17.27 & 0.01 & Swope \\
58616.18 & $+1.02 $ & $g$ & 17.27 & 0.01 & Swope \\
58617.08 & $+1.93 $ & $g$ & 17.36 & 0.01 & Swope \\
58631.12 & $+15.97$  & $g$ & 18.88 & 0.10 & Swope \\
58636.11 & $+20.95$  & $g$ & 19.01 & 0.02 & Swope \\
58639.05 & $+23.89$  & $g$ & 19.28 & 0.02 & Swope \\
58642.10 & $+26.94$  & $g$ & 19.44 & 0.11 & Swope \\
58644.06 & $+28.91$  & $g$ & 19.38 & 0.03 & Swope \\
58658.04 & $+42.89$  & $g$ & 19.82 & 0.02 & Swope \\
58670.02 & $+54.87$  & $g$ & 20.35 & 0.03 & Swope \\
\enddata
\tablenotetext{a}{Relative to second $B$-band maximum (MJD 58615.156)}
\end{deluxetable}

\begin{deluxetable}{cccccc}[h!]
\tablecaption{Optical Photometry of SN~2019ehk (Cont.) \label{tab:phot_table2}}
\tablecolumns{6}
\tablewidth{0.45\textwidth}
\tablehead{
\colhead{MJD} &
\colhead{Phase\tablenotemark{a}} &
\colhead{Filter} & \colhead{Magnitude} & \colhead{Uncertainty} & \colhead{Instrument}
}
\startdata
58688.98 & $+73.83$  & $g$ & 20.78 & 0.05 & Swope \\
58690.98 & $+75.83$  & $g$ & 20.78 & 0.04 & Swope \\
58697.98 & $+82.82$  & $g$ & 21.02 & 0.07 & Swope \\
58603.17 & $-11.98$  & $r$ & 16.59 & 0.01 & Swope \\
58603.22 & $-11.94$  & $r$ & 16.60 & 0.01 & Swope \\
58608.13 & $-7.03 $ & $r$ & 16.45 & 0.01 & Swope \\
58609.18 & $-5.97 $ & $r$ & 16.48 & 0.01 & Swope \\
58615.14 & $-0.01 $ & $r$ & 16.01 & 0.01 & Swope \\
58616.18 & $+1.02 $ & $r$ & 15.94 & 0.01 & Swope \\
58617.08 & $+1.93 $ & $r$ & 15.94 & 0.01 & Swope \\
58631.12 & $+15.96$  & $r$ & 17.07 & 0.01 & Swope \\
58636.08 & $+20.93$  & $r$ & 17.26 & 0.01 & Swope \\
58636.11 & $+20.96$  & $r$ & 17.26 & 0.01 & Swope \\
58639.05 & $+23.90$  & $r$ & 17.41 & 0.01 & Swope \\
58642.10 & $+26.95$  & $r$ & 17.54 & 0.01 & Swope \\
58644.07 & $+28.91$  & $r$ & 17.57 & 0.01 & Swope \\
58658.04 & $+42.88$  & $r$ & 18.17 & 0.01 & Swope \\
58670.01 & $+54.86$  & $r$ & 18.84 & 0.01 & Swope \\
58688.97 & $+73.81$  & $r$ & 19.49 & 0.02 & Swope \\
58690.97 & $+75.81$  & $r$ & 19.48 & 0.02 & Swope \\
58696.97 & $+81.82$  & $r$ & 19.80 & 0.04 & Swope \\
58603.18 & $-11.98$  & $i$ & 16.48 & 0.01 & Swope \\
58603.22 & $-11.94$  & $i$ & 16.45 & 0.01 & Swope \\
58608.13 & $-7.02 $ & $i$ & 16.05 & 0.01 & Swope \\
58609.18 & $-5.97 $ & $i$ & 16.17 & 0.01 & Swope \\
58611.15 & $-4.00 $ & $i$ & 15.76 & 0.01 & Swope \\
58615.14 & $-0.01 $ & $i$ & 15.55 & 0.01 & Swope \\
58616.18 & $+1.02 $ & $i$ & 15.47 & 0.01 & Swope \\
58617.08 & $+1.93 $ & $i$ & 15.41 & 0.01 & Swope \\
58631.12 & $+15.97$  & $i$ & 16.42 & 0.01 & Swope \\
58636.11 & $+20.96$  & $i$ & 16.56 & 0.01 & Swope \\
58639.05 & $+23.89$  & $i$ & 16.67 & 0.01 & Swope \\
58642.10 & $+26.95$  & $i$ & 16.77 & 0.01 & Swope \\
58644.07 & $+28.91$  & $i$ & 16.79 & 0.01 & Swope \\
58658.04 & $+42.88$  & $i$ & 17.17 & 0.01 & Swope \\
58670.02 & $+54.86$  & $i$ & 17.53 & 0.01 & Swope \\
58688.98 & $+73.82$  & $i$ & 17.94 & 0.01 & Swope \\
58690.98 & $+75.82$  & $i$ & 17.82 & 0.01 & Swope \\
58697.97 & $+82.82$  & $i$ & 18.11 & 0.01 & Swope \\
58601.28 & $-13.88$  & $B$ & >20.12 & -- & Joel Shepherd \\
58603.30 & $-11.86$  & $B$ & 17.79 & 0.16 & Joel Shepherd \\
58601.28 & $-13.88$  & $V$ & >18.85 & -- & Joel Shepherd \\
58603.30 & $-11.86$  & $V$ & 16.84 & 0.17 & Joel Shepherd \\
58602.24 & $-12.92$  & $g$ & 18.78 & 0.43 & Joel Shepherd \\
58601.28 & $-13.88$  & $r$ & >18.36 & -- & Joel Shepherd \\
58603.30 & $-11.86$  & $r$ & 16.52 & 0.09 & Joel Shepherd \\
\enddata
\tablenotetext{a}{Relative to second $B$-band maximum (MJD 58615.156)}
\end{deluxetable}

\begin{deluxetable}{cccccc}[h!]
\tablecaption{Optical Photometry of SN~2019ehk (Cont.) \label{tab:phot_table3}}
\tablecolumns{6}
\tablewidth{0.45\textwidth}
\tablehead{
\colhead{MJD} &
\colhead{Phase\tablenotemark{a}} &
\colhead{Filter} & \colhead{Magnitude} & \colhead{Uncertainty} & \colhead{Instrument}
}
\startdata
58600.10 & $-15.06$ & $g$ & >20.48 & -- & ZTF \\
58609.21 & $-5.94$ & $g$ & 17.52 & 0.03 & ZTF \\
58612.25 & $-2.90$ & $g$ & 17.33 & 0.02 & ZTF \\
58619.25 & $+4.10$ & $g$ & 17.76 & 0.15 & ZTF \\
58628.19 & $+13.04$ & $g$ & 18.86 & 0.10 & ZTF \\
58633.23 & $+18.07$ & $g$ & 19.06 & 0.12 & ZTF \\
58636.25 & $+21.09$ & $g$ & 19.20 & 0.13 & ZTF \\
58642.20 & $+27.04$ & $g$ & 19.54 & 0.20 & ZTF \\
58658.21 & $+43.05$ & $g$ & 19.77 & 0.24 & ZTF \\
58661.23 & $+46.07$ & $g$ & 20.36 & 0.46 & ZTF \\
58606.21 & $-8.95$ & $r$ & 15.84 & 0.01 & ZTF \\
58612.21 & $-2.94$ & $r$ & 16.09 & 0.01 & ZTF \\
58619.19 & $+4.04$ & $r$ & 16.14 & 0.06 & ZTF \\
58628.30 & $+13.15$ & $r$ & 16.79 & 0.02 & ZTF \\
58633.20 & $+18.05$ & $r$ & 17.13 & 0.03 & ZTF \\
58636.21 & $+21.05$ & $r$ & 17.24 & 0.03 & ZTF \\
58639.18 & $+24.02$ & $r$ & 17.44 & 0.04 & ZTF \\
58642.22 & $+27.07$ & $r$ & 17.54 & 0.04 & ZTF \\
58646.23 & $+31.07$ & $r$ & 17.39 & 0.08 & ZTF \\
58649.22 & $+34.06$ & $r$ & 17.63 & 0.08 & ZTF \\
58652.28 & $+37.12$ & $r$ & 18.00 & 0.14 & ZTF \\
58658.18 & $+43.02$ & $r$ & 18.36 & 0.09 & ZTF \\
58661.20 & $+46.04$ & $r$ & 18.39 & 0.10 & ZTF \\
58606.21 & $-8.95$ & $r$ & 15.82 & 0.03 & ZTF \\
58612.21 & $-2.94$ & $r$ & 16.07 & 0.04 & ZTF \\
58619.19 & $+4.04$ & $r$ & 16.06 & 0.04 & ZTF \\
58633.20 & $+18.05$ & $r$ & 17.07 & 0.04 & ZTF \\
58611.90 & $-3.26$ & $B$ & 18.14 & 0.15 & Konkoly \\
58613.92 & $-1.24$ & $B$ & 18.07 & 0.10 & Konkoly \\
58638.90 & $+23.74$ & $B$ & 20.88 & 0.79 & Konkoly \\
58647.87 & $+32.71$ & $B$ & 21.27 & 0.12 & Konkoly \\
58611.90 & $-3.26$ & $V$ & 16.72 & 0.04 & Konkoly \\
58613.92 & $-1.24$ & $V$ & 16.65 & 0.03 & Konkoly \\
58638.90 & $+23.74$ & $V$ & 18.30 & 0.08 & Konkoly \\
58647.87 & $+32.71$ & $V$ & 18.53 & 0.23 & Konkoly \\
58649.90 & $+34.74$ & $V$ & 18.52 & 0.22 & Konkoly \\
58611.90 & $-3.26$ & $g$ & 17.29 & 0.06 & Konkoly \\
58613.92 & $-1.24$ & $g$ & 17.22 & 0.02 & Konkoly \\
58638.90 & $+23.74$ & $g$ & 19.33 & 0.12 & Konkoly \\
58649.90 & $+34.74$ & $g$ & 19.19 & 0.20 & Konkoly \\
58611.90 & $-3.26$ & $r$ & 16.10 & 0.02 & Konkoly \\
58613.92 & $-1.24$ & $r$ & 16.14 & 0.01 & Konkoly \\
58638.90 & $+23.74$ & $r$ & 17.58 & 0.03 & Konkoly \\
58647.87 & $+32.71$ & $r$ & 17.95 & 0.11 & Konkoly \\
58649.90 & $+34.74$ & $r$ & 17.97 & 0.06 & Konkoly \\
58611.90 & $-3.26 $& $i$ & 16.10 & 0.02 & Konkoly \\
58613.92 & $-1.24 $& $i$ & 16.14 & 0.01 & Konkoly \\
58638.90 & $+23.74$ & $i$ & 17.58 & 0.03 & Konkoly \\
\enddata
\tablenotetext{a}{Relative to second $B$-band maximum (MJD 58615.156)}
\end{deluxetable}

\begin{deluxetable}{cccccc}[h!]
\tablecaption{Optical Photometry of SN~2019ehk (Cont.) \label{tab:phot_table4}}
\tablecolumns{6}
\tablewidth{0.45\textwidth}
\tablehead{
\colhead{MJD} &
\colhead{Phase\tablenotemark{a}} &
\colhead{Filter} & \colhead{Magnitude} & \colhead{Uncertainty} & \colhead{Instrument}
}
\startdata
58647.87 & $+32.71$ & $i$ & 17.95 & 0.11 & Konkoly \\
58649.90 & $+34.74$ & $i$ & 17.97 & 0.06 & Konkoly \\
58611.90 & $-3.26 $& $z$ & 15.77 & 0.04 & Konkoly \\
58613.92 & $-1.24 $& $z$ & 15.71 & 0.02 & Konkoly \\
58638.90 & $+23.74$ & $z$ & 16.88 & 0.03 & Konkoly \\
58647.87 & $+32.71$ & $z$ & 17.02 & 0.04 & Konkoly \\
58573.46 & $-41.70$ & $o$ & >19.58 & -- & ATLAS \\
58577.44 & $-37.72$ & $o$ & >18.94 & -- & ATLAS \\
58581.44 & $-33.71$ & $o$ & >19.51 & -- & ATLAS \\
58585.43 & $-29.73$ & $o$ & >20.38 & -- & ATLAS \\
58589.40 & $-25.75$ & $o$ & >18.39 & -- & ATLAS \\
58593.47 & $-21.68$ & $o$ & >18.45 & -- & ATLAS \\
58595.41 & $-19.74$ & $o$ & >20.67 & -- & ATLAS \\
58597.41 & $-17.75$ & $o$ & >20.00 & -- & ATLAS \\
58599.42 & $-15.74$ & $o$ & >19.90 & -- & ATLAS \\
58605.40 & $-9.75 $& $o$ & 15.61 & 0.02 & ATLAS \\
58609.39 & $-5.76 $& $o$ & 16.28 & 0.03 & ATLAS \\
58613.35 & $-1.81 $& $o$ & 15.79 & 0.02 & ATLAS \\
58615.44 & $+0.28 $& $o$ & 15.70 & 0.07 & ATLAS \\
58623.30 & $+8.15 $& $o$ & 16.13 & 0.10 & ATLAS \\
58625.38 & $+10.22$ & $o$ & 16.29 & 0.02 & ATLAS \\
58627.34 & $+12.19$ & $o$ & 16.44 & 0.01 & ATLAS \\
58629.32 & $+14.16$ & $o$ & 16.58 & 0.02 & ATLAS \\
58631.36 & $+16.20$ & $o$ & 16.70 & 0.02 & ATLAS \\
58633.39 & $+18.24$ & $o$ & 16.77 & 0.06 & ATLAS \\
58641.31 & $+26.15$ & $o$ & 17.19 & 0.06 & ATLAS \\
58643.37 & $+28.22$ & $o$ & 17.07 & 0.10 & ATLAS \\
58647.30 & $+32.14$ & $o$ & 17.36 & 0.03 & ATLAS \\
58649.37 & $+34.21$ & $o$ & 17.47 & 0.13 & ATLAS \\
58653.27 & $+38.11$ & $o$ & 17.48 & 0.15 & ATLAS \\
58655.29 & $+40.13$ & $o$ & 17.60 & 0.14 & ATLAS \\
58659.30 & $+44.14$ & $o$ & 17.71 & 0.14 & ATLAS \\
58665.29 & $+50.13$ & $o$ & 18.05 & 0.10 & ATLAS \\
58671.26 & $+56.10$ & $o$ & 18.21 & 0.21 & ATLAS \\
58681.30 & $+66.14$ & $o$ & 18.61 & 1.96 & ATLAS \\
58689.27 & $+74.11$ & $o$ & 18.46 & 0.24 & ATLAS \\
58575.46 & $-39.70$ & $c$ & >19.89 & -- & ATLAS \\
58579.45 & $-35.70$ & $c$ & >20.29 & -- & ATLAS \\
58583.42 & $-31.73$ & $c$ & >19.00 & -- & ATLAS \\
58603.42 & $-11.73$ & $c$ & 16.77 & 0.04 & ATLAS \\
58607.41 & $-7.74 $& $c$ & 16.86 & 0.06 & ATLAS \\
58611.40 & $-3.75 $& $c$ & 16.72 & 0.03 & ATLAS \\
58635.35 & $+20.20$ & $c$ & 18.01 & 0.09 & ATLAS \\
58639.31 & $+24.15$ & $c$ & 18.11 & 0.09 & ATLAS \\
58663.26 & $+48.10$ & $c$ & 19.08 & 0.37 & ATLAS \\
58667.28 & $+52.12$ & $c$ & 19.15 & 0.31 & ATLAS \\
58699.26 & $+84.11$ & $c$ & 19.95 & 0.79 & ATLAS \\
\enddata
\tablenotetext{a}{Relative to second $B$-band maximum (MJD 58615.156)}
\end{deluxetable}

\begin{deluxetable}{cccccc}[h!]
\tablecaption{Optical Photometry of SN~2019ehk (Cont.) \label{tab:phot_table5}}
\tablecolumns{6}
\tablewidth{0.45\textwidth}
\tablehead{
\colhead{MJD} &
\colhead{Phase\tablenotemark{a}} &
\colhead{Filter} & \colhead{Magnitude} & \colhead{Uncertainty} & \colhead{Instrument}
}
\startdata
58603.61 & $-11.55$ & $u$ & 17.55 & 0.01 & LCO\\
58604.59 & $-10.56$ & $u$ & 17.46 & 0.02 & LCO\\
58605.25 & $-9.91 $& $u$ & 16.55 & 0.02 & LCO\\
58607.38 & $-7.78 $& $u$ & 17.70 & 0.06 & LCO\\
58607.54 & $-7.62 $& $u$ & 17.91 & 0.22 & LCO\\
58614.37 & $-0.78 $& $u$ & 18.66 & 0.10 & LCO\\
58603.61 & $-11.54$ & $B$ & 17.52 & 0.02 & LCO\\
58604.61 & $-10.55$ & $B$ & 17.55 & 0.02 & LCO\\
58605.25 & $-9.91 $& $B$ & 16.81 & 0.02 & LCO\\
58607.38 & $-7.77 $& $B$ & 17.57 & 0.03 & LCO\\
58607.54 & $-7.62 $& $B$ & 17.58 & 0.06 & LCO\\
58614.38 & $-0.78 $& $B$ & 17.95 & 0.02 & LCO\\
58615.34 & $+0.19 $& $B$ & 17.98 & 0.03 & LCO\\
58622.50 & $+7.35 $& $B$ & 18.82 & 0.05 & LCO\\
58626.24 & $+11.09$ & $B$ & 19.53 & 0.01 & LCO\\
58630.91 & $+15.76$ & $B$ & 19.88 & 0.03 & LCO\\
58635.87 & $+20.72$ & $B$ & 20.02 & 0.05 & LCO\\
58636.33 & $+21.17$ & $B$ & 20.19 & 0.07 & LCO\\
58641.31 & $+26.15$ & $B$ & 20.27 & 0.06 & LCO\\
58652.70 & $+37.54$ & $B$ & 20.68 & 0.32 & LCO\\
58657.51 & $+42.36$ & $B$ & 20.62 & 0.09 & LCO\\
58661.88 & $+46.72$ & $B$ & 21.05 & 0.09 & LCO\\
58667.45 & $+52.30$ & $B$ & 20.68 & 0.11 & LCO\\
58603.61 & $-11.54$ & $V$ & 17.04 & 0.02 & LCO\\
58604.61 & $-10.55$ & $V$ & 16.94 & 0.02 & LCO\\
58605.25 & $-9.90 $& $V$ & 16.14 & 0.02 & LCO\\
58607.38 & $-7.77 $& $V$ & 16.64 & 0.03 & LCO\\
58607.54 & $-7.61 $& $V$ & 16.74 & 0.04 & LCO\\
58614.38 & $-0.77 $& $V$ & 16.53 & 0.01 & LCO\\
58615.35 & $+0.19 $& $V$ & 16.52 & 0.01 & LCO\\
58622.51 & $+7.35 $& $V$ & 17.18 & 0.02 & LCO\\
58626.25 & $+11.09$ & $V$ & 17.49 & 0.02 & LCO\\
58630.92 & $+15.77$ & $V$ & 17.87 & 0.02 & LCO\\
58636.34 & $+21.18$ & $V$ & 18.10 & 0.06 & LCO\\
58641.32 & $+26.16$ & $V$ & 18.32 & 0.03 & LCO\\
58652.70 & $+37.55$ & $V$ & 18.91 & 0.01 & LCO\\
58657.52 & $+42.37$ & $V$ & 19.06 & 0.02 & LCO\\
58661.88 & $+46.73$ & $V$ & 19.29 & 0.04 & LCO\\
58667.46 & $+52.31$ & $V$ & 19.50 & 0.04 & LCO\\
58687.85 & $+72.69$ & $V$ & 20.18 & 0.16 & LCO\\
58603.62 & $-11.54$ & $g$ & 17.33 & 0.02 & LCO\\
58614.39 & $-0.77 $& $g$ & 17.34 & 0.01 & LCO\\
58615.35 & $+0.20 $& $g$ & 17.38 & 0.01 & LCO\\
58622.51 & $+7.36 $& $g$ & 18.35 & 0.02 & LCO\\
58652.71 & $+37.55$ & $g$ & 19.96 & 0.06 & LCO\\
58657.52 & $+42.37$ & $g$ & 20.15 & 0.05 & LCO\\
\enddata
\tablenotetext{a}{Relative to second $B$-band maximum (MJD 58615.156)}
\end{deluxetable}

\begin{deluxetable}{cccccc}[h!]
\tablecaption{Optical Photometry of SN~2019ehk (Cont.) \label{tab:phot_table6}}
\tablecolumns{6}
\tablewidth{0.45\textwidth}
\tablehead{
\colhead{MJD} &
\colhead{Phase\tablenotemark{a}} &
\colhead{Filter} & \colhead{Magnitude} & \colhead{Uncertainty} & \colhead{Instrument}
}
\startdata
58603.62 & $-11.54$  & $r$ & 16.92 & 0.01 & LCO\\
58604.61 & $-10.54$  & $r$ & 16.78 & 0.01 & LCO\\
58607.39 & $-7.77 $ & $r$ & 16.38 & 0.02 & LCO\\
58607.55 & $-7.61 $ & $r$ & 16.44 & 0.02 & LCO\\
58614.39 & $-0.77 $ & $r$ & 16.22 & 0.01 & LCO\\
58615.36 & $+0.20 $ & $r$ & 16.18 & 0.02 & LCO\\
58622.52 & $+7.36 $ & $r$ & 16.53 & 0.02 & LCO\\
58626.26 & $+11.10$  & $r$ & 16.87 & 0.02 & LCO\\
58636.35 & $+21.20$  & $r$ & 17.39 & 0.03 & LCO\\
58652.71 & $+37.56$  & $r$ & 18.14 & 0.02 & LCO\\
58657.53 & $+42.38$  & $r$ & 18.39 & 0.03 & LCO\\
58687.86 & $+72.70$  & $r$ & 19.64 & 0.02 & LCO\\
58603.62 & $-11.53$  & $i$ & 16.69 & 0.05 & LCO\\
58604.62 & $-10.54$  & $i$ & 16.47 & 0.01 & LCO\\
58605.26 & $-9.90 $ & $i$ & 15.69 & 0.01 & LCO\\
58607.55 & $-7.61 $ & $i$ & 16.00 & 0.03 & LCO\\
58614.39 & $-0.77 $ & $i$ & 15.66 & 0.02 & LCO\\
58615.36 & $+0.21 $ & $i$ & 15.60 & 0.02 & LCO\\
58622.52 & $+7.36 $ & $i$ & 15.84 & 0.02 & LCO\\
58626.26 & $+11.10$  & $i$ & 16.10 & 0.01 & LCO\\
58630.93 & $+15.78$  & $i$ & 16.35 & 0.01 & LCO\\
58636.35 & $+21.20$  & $i$ & 16.61 & 0.01 & LCO\\
58641.33 & $+26.18$  & $i$ & 16.84 & 0.03 & LCO\\
58652.72 & $+37.56$  & $i$ & 17.14 & 0.02 & LCO\\
58657.54 & $+42.38$  & $i$ & 17.32 & 0.02 & LCO\\
58667.48 & $+52.32$  & $i$ & 17.55 & 0.02 & LCO\\
58687.86 & $+72.71$  & $i$ & 17.92 & 0.02 & LCO\\
58605.25 & $-9.90 $ & $g$ & 16.39 & 0.02 & Thacher \\
58606.26 & $-8.89 $ & $g$ & 16.71 & 0.02 & Thacher \\
58607.24 & $-7.91 $ & $g$ & 17.17 & 0.02 & Thacher \\
58633.27 & $+18.11$  & $g$ & 19.19 & 0.07 & Thacher \\
58605.25 & $-9.90 $ & $r$ & 15.76 & 0.01 & Thacher \\
58606.26 & $-8.89 $ & $r$ & 15.99 & 0.01 & Thacher \\
58607.24 & $-7.91 $ & $r$ & 16.30 & 0.03 & Thacher \\
58632.18 & $+17.03$  & $r$ & 17.30 & 0.03 & Thacher \\
58633.27 & $+18.11$  & $r$ & 17.22 & 0.02 & Thacher \\
58634.18 & $+19.03$  & $r$ & 17.25 & 0.02 & Thacher \\
58640.18 & $+25.03$  & $r$ & 17.40 & 0.03 & Thacher \\
58641.19 & $+26.03$  & $r$ & 17.96 & 0.08 & Thacher \\
58642.30 & $+27.14$  & $r$ & 17.78 & 0.03 & Thacher \\
58643.19 & $+28.03$  & $r$ & 17.83 & 0.04 & Thacher \\
58644.19 & $+29.03$  & $r$ & 17.76 & 0.06 & Thacher \\
58645.31 & $+30.15$  & $r$ & 18.13 & 0.23 & Thacher \\
58647.19 & $+32.03$  & $r$ & 18.23 & 0.08 & Thacher \\
58654.21 & $+39.06$  & $r$ & 18.43 & 0.06 & Thacher \\
58658.21 & $+43.06$  & $r$ & 18.74 & 0.06 & Thacher \\
58662.21 & $+47.06$  & $r$ & 18.60 & 0.05 & Thacher \\
58663.21 & $+48.06$  & $r$ & 18.79 & 0.06 & Thacher \\
\enddata
\tablenotetext{a}{Relative to second $B$-band maximum (MJD 58615.156)}
\end{deluxetable}

\begin{deluxetable}{cccccc}[h!]
\tablecaption{Optical Photometry of SN~2019ehk (Cont.) \label{tab:phot_table7}}
\tablecolumns{6}
\tablewidth{0.45\textwidth}
\tablehead{
\colhead{MJD} &
\colhead{Phase\tablenotemark{a}} &
\colhead{Filter} & \colhead{Magnitude} & \colhead{Uncertainty} & \colhead{Instrument}
}
\startdata
58664.24 & $+49.08$ & $r$ & 18.65 & 0.07 & Thacher \\
58666.19 & $+51.03$ & $r$ & 18.87 & 0.07 & Thacher \\
58667.21 & $+52.06$ & $r$ & 18.56 & 0.07 & Thacher \\
58676.21 & $+61.06$ & $r$ & 19.21 & 0.16 & Thacher \\
58606.26 & $-8.89 $& $i$ & 15.69 & 0.01 & Thacher \\
58607.24 & $-7.91 $& $i$ & 16.01 & 0.01 & Thacher \\
58620.19 & $+5.03 $& $i$ & 15.59 & 0.02 & Thacher \\
58631.18 & $+16.02$ & $i$ & 16.53 & 0.02 & Thacher \\
58632.18 & $+17.03$ & $i$ & 16.22 & 0.14 & Thacher \\
58633.27 & $+18.11$ & $i$ & 16.50 & 0.02 & Thacher \\
58634.18 & $+19.03$ & $i$ & 16.60 & 0.02 & Thacher \\
58640.18 & $+25.03$ & $i$ & 17.03 & 0.03 & Thacher \\
58641.19 & $+26.03$ & $i$ & 16.98 & 0.03 & Thacher \\
58642.30 & $+27.14$ & $i$ & 16.92 & 0.03 & Thacher \\
58643.19 & $+28.03$ & $i$ & 17.05 & 0.03 & Thacher \\
58644.19 & $+29.03$ & $i$ & 16.92 & 0.04 & Thacher \\
58646.19 & $+31.03$ & $i$ & 16.99 & 0.04 & Thacher \\
58647.19 & $+32.03$ & $i$ & 16.88 & 0.06 & Thacher \\
58650.21 & $+35.06$ & $i$ & 17.08 & 0.04 & Thacher \\
58658.21 & $+43.06$ & $i$ & 17.43 & 0.03 & Thacher \\
58662.21 & $+47.06$ & $i$ & 17.34 & 0.03 & Thacher \\
58663.21 & $+48.06$ & $i$ & 17.38 & 0.03 & Thacher \\
58664.24 & $+49.08$ & $i$ & 17.42 & 0.05 & Thacher \\
58666.19 & $+51.03$ & $i$ & 17.48 & 0.03 & Thacher \\
58667.21 & $+52.06$ & $i$ & 17.54 & 0.04 & Thacher \\
58668.21 & $+53.06$ & $i$ & 17.52 & 0.04 & Thacher \\
58669.21 & $+54.06$ & $i$ & 17.57 & 0.04 & Thacher \\
58670.21 & $+55.06$ & $i$ & 17.61 & 0.04 & Thacher \\
58672.19 & $+57.03$ & $i$ & 17.75 & 0.08 & Thacher \\
58674.21 & $+59.05$ & $i$ & 17.64 & 0.06 & Thacher \\
58675.21 & $+60.06$ & $i$ & 17.74 & 0.06 & Thacher \\
58676.21 & $+61.06$ & $i$ & 17.73 & 0.08 & Thacher \\
58605.25 & $-9.90 $& $z$ & 15.53 & 0.02 & Thacher \\
58606.26 & $-8.89 $& $z$ & 15.66 & 0.02 & Thacher \\
58607.24 & $-7.91 $& $z$ & 15.97 & 0.02 & Thacher \\
58620.19 & $+5.03 $& $z$ & 16.25 & 0.06 & Thacher \\
58631.18 & $+16.02$ & $z$ & 16.07 & 0.03 & Thacher \\
58632.18 & $+17.03$ & $z$ & 16.13 & 0.04 & Thacher \\
58633.27 & $+18.11$ & $z$ & 16.07 & 0.03 & Thacher \\
58634.18 & $+19.03$ & $z$ & 16.10 & 0.03 & Thacher \\
58640.18 & $+25.03$ & $z$ & 16.37 & 0.04 & Thacher \\
58641.19 & $+26.03$ & $z$ & 16.46 & 0.06 & Thacher \\
58642.30 & $+27.14$ & $z$ & 16.31 & 0.04 & Thacher \\
58643.19 & $+28.03$ & $z$ & 16.29 & 0.03 & Thacher \\
58644.19 & $+29.03$ & $z$ & 16.27 & 0.04 & Thacher \\
58646.19 & $+31.03$ & $z $& 16.36 & 0.06 & Thacher \\
58647.19 & $+32.03$ & $z $& 16.25 & 0.03 & Thacher \\
58650.21 & $+35.06$ & $z $& 16.31 & 0.04 & Thacher \\
58658.21 & $+43.06$ & $z $& 16.47 & 0.03 & Thacher \\
58662.21 & $+47.06$ & $z $& 16.62 & 0.04 & Thacher \\
\enddata
\tablenotetext{a}{Relative to second $B$-band maximum (MJD 58615.156)}
\end{deluxetable}

\begin{deluxetable}{cccccc}[h!]
\tablecaption{Optical Photometry of SN~2019ehk (Cont.) \label{tab:phot_table8}}
\tablecolumns{6}
\tablewidth{0.45\textwidth}
\tablehead{
\colhead{MJD} &
\colhead{Phase\tablenotemark{a}} &
\colhead{Filter} & \colhead{Magnitude} & \colhead{Uncertainty} & \colhead{Instrument}
}
\startdata
58663.21 & $+48.06$ & $z $& 16.65 & 0.03 & Thacher \\
58664.24 & $+49.08$ & $z $& 16.61 & 0.10 & Thacher \\
58666.19 & $+51.03$ & $z $& 16.74 & 0.04 & Thacher \\
58667.21 & $+52.06$ & $z $& 16.77 & 0.05 & Thacher \\
58668.21 & $+53.06$ & $z $& 16.79 & 0.05 & Thacher \\
58669.21 & $+54.06$ & $z $& 17.02 & 0.07 & Thacher \\
58670.21 & $+55.06$ & $z $& 16.84 & 0.05 & Thacher \\
58672.19 & $+57.03$ & $z $& 16.91 & 0.08 & Thacher \\
58674.21 & $+59.05$ & $z $& 16.92 & 0.08 & Thacher \\
58675.21 & $+60.06$ & $z $& 16.84 & 0.07 & Thacher \\
58604.61 & $-10.55$ & $B$ & 17.10 & 0.10 & \emph{Swift} \\
58606.04 & $-9.12$ & $B$ & 17.31 & 0.11 & \emph{Swift} \\
58607.56 & $-7.60$ & $B$ & 18.07 & 0.17 & \emph{Swift} \\
58612.70 & $-2.45$ & $B$ & -- & 0.00 & \emph{Swift} \\
58619.64 & +4.49 & $B$ & 18.25 & 0.00 & \emph{Swift} \\
58624.57 & +9.41 & $B$ & 19.03 & 0.35 & \emph{Swift} \\
58629.14 & +13.99 & $B$ & 19.33 & 0.00 & \emph{Swift} \\
58604.61 & $-10.55$ & $V$ & 16.26 & 0.11 & \emph{Swift} \\
58606.04 & $-9.12$ & $V$ & 16.11 & 0.13 & \emph{Swift} \\
58607.56 & $-7.60$ & $V$ & 16.80 & 0.14 & \emph{Swift} \\
58612.70 & $-2.45$ & $V$ & -- & 0.00 & \emph{Swift} \\
58619.64 & +4.49 & $V$ & -- & 0.00 & \emph{Swift} \\
58624.57 & +9.41 & $V$ & 17.17 & 0.19 & \emph{Swift} \\
58629.14 & +13.99 & $V$ & 17.72 & 0.25 & \emph{Swift} \\
58604.61 & $-10.55$ & $U$ & 17.05 & 0.12 & \emph{Swift} \\
58606.04 & $-9.12$ & $U$ & 17.32 & 0.13 & \emph{Swift} \\
58607.56 & $-7.60$ & $U$ & 18.57 & 0.27 & \emph{Swift} \\
58612.70 & $-2.45$ & $U$ & -- & 0.00 & \emph{Swift} \\
58619.64 & +4.49 & $U$ & 18.93 & 0.00 & \emph{Swift} \\
58624.57 & +9.41 & $U$ & 18.92 & 0.00 & \emph{Swift} \\
58629.14 & +13.99 & $U$ & 19.09 & 0.00 & \emph{Swift} \\
58604.61 & $-10.55$ & $W1$ & 17.80 & 0.15 & \emph{Swift} \\
58606.04 & $-9.12$ & $W1$ & 18.15 & 0.17 & \emph{Swift} \\
58607.56 & $-7.60$ & $W1$ & 19.27 & 0.00 & \emph{Swift} \\
58612.70 & $-2.45$ & $W1$ & 18.96 & 0.27 & \emph{Swift} \\
58619.64 & +4.49 & $W1$ & 19.15 & 0.00 & \emph{Swift} \\
58624.57 & +9.41 & $W1$ & 19.15 & 0.00 & \emph{Swift} \\
58629.14 & +13.99 & $W1$ & 19.15 & 0.00 & \emph{Swift} \\
58604.61 & $-10.55$ & $W2$ & 18.91 & 0.22 & \emph{Swift} \\
58606.04 & $-9.12$ & $W2$ & 19.28 & 0.00 & \emph{Swift} \\
58607.56 & $-7.60$ & $W2$ & 19.67 & 0.00 & \emph{Swift} \\
58612.70 & $-2.45$ & $W2$ & 19.38 & 0.00 & \emph{Swift} \\
58619.64 & +4.49 & $W2$ & -- & 0.00 & \emph{Swift} \\
58624.57 & +9.41 & $W2$ & 19.55 & 0.00 & \emph{Swift} \\
58629.14 & +13.99 & $W2$ & 19.75 & 0.00 & \emph{Swift} \\
58604.61 & $-10.55$ & $M2$ & 19.24 & 0.24 & \emph{Swift} \\
58606.04 & $-9.12$ & $M2$ & 19.23 & 0.00 & \emph{Swift} \\
58607.56 & $-7.60$ & $M2$ & 19.75 & 0.33 & \emph{Swift} \\
58612.70 & $-2.45$ & $M2$ & 19.25 & 0.00 & \emph{Swift} \\
58619.64 & +4.49 & $M2$ & -- & 0.00 & \emph{Swift} \\
58624.57 & +9.41 & $M2$ & 19.71 & 0.00 & \emph{Swift} \\
58629.14 & +13.99 & $M2$ & 19.83 & 0.34 & \emph{Swift} \\
\enddata
\tablenotetext{a}{Relative to second $B$-band maximum (MJD 58615.156)}
\end{deluxetable}


\end{document}